\documentclass[twoside,12pt]{article}
\usepackage{amssymb}
\usepackage{graphicx}
\usepackage{mcite}

\begin{document}

\title{ \vspace{1cm} Effective interactions and operators in no-core shell model}
\author{ I. Stetcu$^{1,2}$ and J. Rotureau$^{3,4}$\\
$^1$Theoretical Division, Los Alamos National Laboratory, \\
Los Alamos, New Mexico 87545, USA \\
$^2$Department of Physics, University of Washington, Seattle, WA, 98125, USA\\
$^3$Fundamental Physics, Chalmers University of Technology, \\
412 96 G\"oteborg, Sweden \\
$^4$Department of Physics, University of Arizona, Tucson, AZ 85721, USA }

\maketitle

\hspace*{7.in}
\vspace*{-1.5in}
\rotatebox{90}{%
\fbox{\parbox[t]{1.2in}{
LA-UR-12-21687}}}

\begin{abstract}
Solutions to the nuclear many-body problem rely on effective interactions, and in general effective operators,  to take into account effects not included in calculations. These include effects due to the truncation to finite model spaces where a numerical calculation is tractable, as well as physical terms not included in the description  in the first place. In the no-core shell model (NCSM) framework, we discuss two approaches to the effective interactions based on (i) unitary transformations and (ii) effective field theory (EFT) principles. Starting from a given Hamiltonian, the unitary transformation approach is designed to take into account effects induced by the truncation to finite model spaces in which a numerical calculation is performed. This approach was widely applied to the description of nuclear properties of light nuclei; we review the theory and present representative results. In the EFT approach, a Hamiltonian is always constructed in a truncated model space  according to the symmetries of the underlying theory, making use of power counting to limit the number of interactions included in the calculations. Hence, physical terms not explicitly included in the calculation are treated on the same footing with the truncation to a finite model space. In this approach, we review results for both nuclear and trapped atomic systems, for which the effective theories are formally  similar, albeit describing different underlying physics. Finally, the application of the EFT method of constructing effective interactions to Gamow shell model is briefly discussed.
\end{abstract}


\section{Introduction}

It is generally accepted that the nuclear problem is pathologically complicated. On one hand, the inter-nucleon interaction is non-central, non-local and unconstrained at short distances, and, on the other hand, the complexity of solving the many-body Schr\"odinger equation increases dramatically with the number of nucleons. Phenomenological \cite{Wiringa:1995,QMC_rev} and one-boson exchange models \cite{cdbonn} of the inter-nucleon interactions have provided guidance and have proved successful in applications to light nuclei. However, a deeper understanding of the interactions between nucleons has been achieved using effective field theories (EFTs) \cite{weinberg1990,*weinberg1991,*ordonez1992,eftreview}, which provide interactions consistent with the symmetries of the underlying theory of the strong interactions, QCD. EFT has several advantages: eliminates the model dependence (\textit{e.g.}, there are an infinite number of nucleon-nucleon interactions which are phase-shift equivalent, but many-body solutions depend upon the particular choice), explains naturally the hierarchy of the nuclear interactions, provides a framework for error estimation, and can mitigate shortcomings with the description of low-momentum observables, such as multipole transitions.

For light nuclei, several methods are now available for solving the nuclear few-body problem, such as Green's function Monte Carlo (GFMC) \cite{Carlson:1986,*Carlson:1997qn,Schiavilla:1993tk}, Fadeev \cite{Nogga199719}, Fadeev-Ya\-ku\-bov\-sky \cite{Yakubovsky:1966ue,*Gloeckle:1993mx,*Kamada1992205,*PhysRevLett.85.944}, (effective-interaction) Hyperspherical Harmonics \cite{Kievsky:1992um,Barnea:2000be}, and the no-core shell model (NCSM) \cite{Navratil:2000gs,Navratil:2009ut}, and they all agree within the statistical or numerical errors inherent to each method \cite{benchmark:4he_gs}. Coupled-cluster (CC) expansion has been applied to the description of (nearly) closed-shell nuclei, from $^4$He to $^{40}$Ca \cite{PhysRevLett.92.132501,*Hagen2007169,*PhysRevC.76.044305,PhysRevC.82.034330}. In most cases, the theoretical description of the experimental data is excellent. Among the few-body methods, while subject to greater errors, the NCSM, CC, and the auxiliarly-field difusion quantum Monte Carlo (AFDMC) \cite{Gandolfi:2006bt,*PhysRevLett.99.022507}, are the only methods with the potential to be extended to heavier nuclei with fewer restrictions. Thus, NCSM can handle both local and non-local interactions unlike GFMC that has difficulties in the presence of strong non-localities. AFDMC has been restricted to more schematic interactions like Argonne $v_6$, and to date only an approximate inclusion of three-body forces has been possible in CC. It is also conceivable that  a lattice approach like the one in Ref. \cite{latticeEFT} could be applied to heavy $N=Z$ nuclei.

NCSM is a direct numerical diagonalization approach, in which the Schr\"odinger equation is solved by expanding the system wave function in a many-body basis. The many-body basis is truncated to a finite size, so that a numerical diagonalization can be performed. This is similar to the phenomenological shell model, shortly described in Sec. \ref{Stetcu_Sec:phSM}, which deals with even more drastic truncations of the available model spaces.  However, in contrast with the phenomenological version, in NCSM one starts with a given Hamiltonian (which can contain two- and three-body interactions) defined in an infinite Hilbert space and one derives an effective interaction in a given model space via a unitary transformation. We review the traditional derivation for effective interactions, based on the  Okubo-Lee-Suzuki transformation \cite{Okubo:1954,DaProvidencia:1964,Suzuki:1980,*Suzuki:1994ok,Suzuki:1982}, as well as effective operators, and present selected representative results in Sec. \ref{Stetcu_Sec:NCSM}. 

Applications of few-body methods like GFMC and NCSM has allowed tremendous progress in the last decade  in our understanding of how nuclear structure arises from the properties of the interactions among nucleons inside a nucleus. Often, these interactions are modeled only in terms of \textit{ad hoc} potentials, which are not necessarily rooted in QCD. EFT provides the framework to construct nuclear potentials that respect the symmetries of QCD and produce observables in a systematic and controlled expansion in a small parameter \cite{weinberg1990,*weinberg1991,*ordonez1992,eftreview}. In such an approach, the shape of the nuclear interactions, including the range of pion exchange, is restricted and the complicated short-range physics is encoded into a number low-energy constants, which, so far have been fitted to data. However, given the advances in lattice QCD simulations \cite{lattQCD}, it will be possible in the near future to directly compute these from the underlying theory. 

While the EFTs can shed light on the properties of the inter-nucleon interactions, the solutions to many-body problems, especially those involving a large number of nucleons, rely on the increasing computational power, supported by the development of sophisticated numerical algorithms. In general, the development of many-body methods are considered independent from the derivation of the interactions. To some extent, this is a justified approach especially if one needs to asses the reliability of the many-body method, so that in conventional approaches the two issues have been considered decoupled: the effective interaction has been seen as merely an input to the many-body codes, with minimal mixing between the two worlds, required for the determination of the three-nucleon interaction parameters. 

With the development of EFT, it has been recognized that one has never access to the ``full'' Hilbert space associated with the quantum system. It becomes therefore necessary to truncate the Hilbert space in order to exclude those states that could not be constrained experimentally or theoretically. In other words, the interactions are only defined in the context of a model space, and often independent of whether the excluded space physics is known or not. Much in the same way one is forced to restrict the calculation to a tractable model space in NCSM. This has motivated the development of a hybrid approach, in which the interactions are derived using the principles of EFT directly in NCSM model spaces. In this paper, we review the efforts to combine the power of EFTs with the NCSM, with the goal of not only extending the benefits of QCD compatible solutions to a larger class of nuclei, but also of eliminating the model dependence and mitigating some shortcomings of the conventional NCSM. Reference \cite{Stetcu:2006ey} was a first attempt at generating effective interactions in NCSM restricted spaces, by fitting two- and three-body parameters directly to levels in light nuclei. An alternate method of fitting the leading order (LO) low-energy constants for the two-body interaction was presented in Ref. \cite{Stetcu:063613}, where a physical trap was added. The method has been later extended in order to include corrections beyond leading order, with applications to few-body systems of cold atoms \cite{NCSMeft_trap_2b,NCSMeft_trap_34atoms} and nucleons \cite{NCSMeft_trap_23nucl}. We present in detail the two methods in Sec. \ref{Stetcu_Sec:NCSMasEFT}. Furthermore, the approach is general and has been recently applied to the derivation of effective interactions used in Gamow shell model calculations, as illustrated in Sec. \ref{eft_gsm}.  We conclude and discuss further applications in Sec. \ref{Stetcu_Sec:concl}.

\section{Phenomenological shell model}
\label{Stetcu_Sec:phSM}

Although the main interest of the present paper is NCSM, we find useful to briefly review the  conventional shell model, which is the phenomenological method of choice for the description of medium nuclei. To some extent, the theoretical derivation of the effective interaction is even more challenging than in the NCSM, because a much larger amount of correlations have to be included in the effective Hamiltonian. 

In the phenomenological shell model, \textit{all} the correlations between a reduced number of nucleons are included. The tradeoff is that most of the nucleons are considered inert. Thus, the main assumption is that the wave function factorizes in a component describing a closed shell nucleus, and a few valence nucleons:

\begin{equation}
|\Psi^A\rangle \simeq |\Psi^{core}\rangle |\Psi^{valence}\rangle,
\label{eq:factwf}
\end{equation}
where $|\Psi^A\rangle$ is the wave function of the $A$-nucleon system, while $|\Psi^{core}\rangle$ and $|\Psi^{valence}\rangle$ describe the core and valence nucleon wavefunctions respectively. A necessary condition for the factorization (\ref{eq:factwf}) to work is that the core subcomponent of the system have no low-lying excited states. While this is condition is satisfied for p-shell nuclei, because the core nucleus $^4$He has no low-lying excited states below 20 MeV, the condition could be badly violated for other cases. For example, $^{16}$O, the core for so-called sd-shell nuclei (mass number between 20 and 40), has low-lying excited states around 4 MeV. In such cases, in order to keep the simple factorization of the wave function (\ref{eq:factwf}), one needs to include in the effective operators the effects of the nucleons being excited out of the closed-shell core, process called core polarization \cite{bertsch_core_pol}. Assuming that one can separate the nuclear Hamiltonian in core and valence contributions

\begin{equation}
H=H_{core}+H_{val},
\end{equation}
the eigenvalue equations decouple in a part that involves only the core (and is assumed solved)

\begin{equation}
H_{core}|\Psi^{core}\rangle = E_{core}|\Psi^{core}\rangle,
\end{equation}
and a part involving only the valence space contribution 
\begin{equation}
H_{val}|\Psi^{val}\rangle = E_{val}|\Psi^{val}\rangle,
\label{eq:valeig}
\end{equation}
which constitutes the main object of investigation in the phenomenological shell model.

Even if we assume the wave-function factorization produces small enough errors, one needs further simplifications since the valence particle wavefunction $|\Psi^{val}\rangle$ given by Eq. (\ref{eq:valeig}) allows particles to scatter into an infinite number of states above the core. In order to make calculations tractable, in the phenomenological approach one considers that the valence space spans only a small enough number of states. For example, for p-shell nuclei, a good description of a large number of low-lying states can be obtained if the lowest p-shell single particle states are accessible to the valence nucleons. The tradeoff is that only a subset of low-lying states can be described in such a restricted space, even if all missing correlations are included in the effective operators. Thus, for even-$A$ nuclei one can only describe positive-parity states, while for odd-$A$ nuclei only negative-parity states are accessible. In order to describe the remaining states, one has to work in valence spaces that include multiple shells.


In order to derive the interaction in the restricted model space, we consider two interacting particles. Let 
$|\tau\rangle$ be the wave function so that the Schr\"odinger equation is
\begin{equation}
(H_0+V)|\tau\rangle = E_\tau|\tau\rangle,
\end{equation}
\noindent
where $H_0$ is the kinetic energy term, and $V$ can be, in principle, the bare nucleon-nucleon interaction.
The last equation transforms into two equations if we multiply on the left
by $P$ and $Q$, the projectors
inside and outside the valence space respectively (by definition, the projector $P$ projects into the valence space, i.e., $P|\tau\rangle = |\Psi^{val}\rangle$):
\begin{equation}
(-E_\tau +H_0 +PVP)P|\tau\rangle = -PVQ|\tau\rangle,
\end{equation}
\begin{equation}
(-E_\tau +H_0 +QVQ)Q|\tau\rangle = -QVP|\tau\rangle.
\end{equation}
Solving the latter for the wave function outside the model space $Q|\tau\rangle$, and
plugging into the former, one obtains a Schr\" odinger equation inside
the valence space \cite{ring}:
\begin{equation}
(H_0+PV^{E_\tau}_{eff}P)P|\tau\rangle = E_\tau P|\tau\rangle,
\end{equation}
with the effective nucleon-nucleon interaction in the valence space given
by
\begin{equation}
V^{E_\tau}_{eff}=V+VQ\frac{1}{E_{\tau}-H_0-QVQ}QV.
\end{equation}
This can be rewritten as an integral equation for $V^{E_\tau}_{eff}$ \cite{ring};
furthermore, in the particular case when the $Q$ operator corresponds
to the two-particle excited states outside the model space
(which brings the largest contribution), 
one can show that the equation for $V_{eff}$ is similar to the equation for
Br\"ueckner $G$-matrix \cite{kuo74}. Therefore, computation of effective 
interaction reduces to calculation of the $G$-matrix.

In general, the effective interaction obtained through the procedure outlined
above does not provide good description of
the low-lying states. Therefore, one further adjusts the two-body interaction
using a fitting procedure so that one obtains a correct description of ground
and excited states for a large number of nuclei. Very often, even if one starts with a realistic two-body potential, for a better description of the experimental data, one can completely ignore the analytical form
of the nucleon-nucleon potential $V(\vec{r}_i,\vec{r}_j)$. Thus, the
effective TBME approximation consists in assuming that each matrix element is assumed a
parameter that can be adjusted in order to obtain the experimental spectra of
nuclei, and the $G$-matrix is usually the first step of the iteration. 
Because it is analytically convenient, one usually chooses 
single-particle states corresponding to the harmonic
oscillator. This has no relevance for diagonalization, but it becomes important
for calculating transition strengths when one has to evaluate matrix elements of
different operators, as the reduced matrix elements of the corresponding
operator depend upon the radial structure of the single-particle wave function. In general, however, in the phenomenological shell model, no diagrammatic basis for generating the effective operator exists. Empirical renormalization of transition operators must also be introduced, obscuring the underlying physics and undercutting the phenomenological shell model as a predictive tool.

However, the phenomenological shell model is considered a success. In the full $sd$ space the parameters are the three single-particle energies (the
spin-orbit interaction removes the degeneracy) $\varepsilon_{1s_{1/2}}$,
$\varepsilon_{0d_{3/2}}$ and $\varepsilon_{0d_{5/2}}$ plus 63 TBME. The Wildenthal ``USD'' interaction
\cite{wildenthal} is fitted to reproduce 447 ground and excited states of $sd$ shell nuclei $(A=17-40)$. In the space spanned by the $1p_{1/2}-1p_{3/2}-0f_{5/2}-0f_{7/2}$ single particle states outside a $^{40}$Ca inert core, the monopole-modified Kuo-Brown ``KB3'' interaction \cite{KB3,*KB3-2} is very successful in reproducing experimental features; this interaction was derived starting from the $G$-matrix as first approximation, and then fitting the matrix elements to describe experimental data. Further refinement of the existing interactions to include more experimental information in the fitting procedure have allowed even more precise description of nuclear properties both in the sd \cite{PhysRevC.74.034315} and pf \cite{PhysRevC.65.061301} shells. A more systematic approach to the derivation of effective interactions \cite{NCSMcore} and operators \cite{NCSMcore_op} in one single shell has been implemented within the NCSM framework, and will be discussed in Sec. \ref{Stetcu_sec:core}.

Phenomenological shell-model inspired methods have been applied to the description of trapped gases, even though the model spaces are not so drastically restricted \cite{Johnson2010:pei}. Given the fact that the off-shell properties of the NN interactions cannot be constrained experimentally, Johnson has proposed a unitary transformation that preserves the on-shell properties, but is fitted, similar to the phenomenological interactions, to give the correct binding energies for select few-body systems, thus minimizing the expectation value of the induced few-body forces. Using this method, several few-body systems not included in the fit have been predicted with good accuracy in small model spaces \cite{Johnson2010:pei}. Thus, it is possible that such an approach could be used for the derivation of effective interactions in small model spaces for heavy nuclei, although its predictability should be properly investigated.

\section{No-core shell model}
\label{Stetcu_Sec:NCSM}

NCSM is a few- and many-body technique specialized in solving the Schr\"odinger equation by direct diagonalization in a restricted basis set constructed with harmonic oscillator (HO) wave functions. The method has its roots in the phenomenological shell model described in the previous section. The advantage of this approach, however, is that the results are independent upon any parameters intrinsic to the calculation, such as the HO frequency of the single particle states used to construct the many-body basis, even though in practice this is achieved only for small systems.

In NCSM all the nucleons are allowed to interact. Formally, the non-relativistic intrinsic Hamiltonian describing the system of $A$ protons and neutrons writes
\begin{equation}
 H_{int}=\frac{1}{A}\sum_{i<j}\frac{(\vec p_i-\vec p_j)^2}{2 M_N}+\sum_{i>j=1}^AV_{ij}^{NN}+
 \sum_{i>j>k=1}^AV_{ijk}^{NNN}+...,
 \label{Stetcu_eq:ham}
\end{equation}
where $V^{NN}_{ij}$ is the NN interaction, which depends only on the relative coordinates between the $i$ and $j$ particle, $V_{ijk}^{NNN}$ the NNN interaction, $\vec p_i$ the momentum of particle $i$, and $M_N$ the nucleon mass. Usually, $V^{NN}$ is fitted with high accuracy low-energy observables (phase-shifts and de\-u\-te\-ron properties), while the NNN interactions are adjusted to reproduce properties of the three-body system (tritium, nu\-cle\-on\--de\-uteron scattering). This Hamiltonian can be numerically diagonalized in an appropriate basis. 

There are two equivalent approaches to solving the many-body problem, depending on how the basis is set up. In the first approach, one considers the relative (or Jacobi) coordinates, defined, for example, in terms of differences between the CM positions of sub-clusters within the $A$-body system: 

\begin{eqnarray}
\vec{\xi}_1&=&\sqrt{\frac{1}{2}}\left(\vec{r}_1-\vec{r}_2\right),
\nonumber\\
\vec{\xi}_2&=&\sqrt{\frac{2}{3}}
\left[\frac{1}{2}\left(\vec{r}_1+\vec{r}_2\right)-\vec{r}_3\right],
\nonumber\\
\vdots && 
\nonumber\\
\vec{\xi}_{A-1}&=& \sqrt{\frac{A-1}{A}}
\left[\frac{1}{A-1}\left(\vec{r}_1+\vec{r}_2+\cdots+\vec{r}_{A-1}\right)
-\vec{r}_{A}\right].
\label{Stetcu_eq:JacobiCoord}
\end{eqnarray}
Subsequently, the internal coordinates are described with HO wave functions \cite{Navratil:1999pw}. For example, the basis states for the three-nucleon system are
\begin{equation}
{\cal A}\left\{\left[\phi_{n_1 l_1}(\vec{\xi}_1) \otimes 
\phi_{n_2 l_2}(\vec{\xi}_2) \right]_{L} 
|(\frac{1}{2} \frac{1}{2} )s_2 \frac{1}{2};S\rangle
|(\frac{1}{2} \frac{1}{2} )t_2 \frac{1}{2};T\rangle
\right\},
\label{Stetcu_eq:basis3b}
\end{equation} 
which have the spatial part constructed using HO wavefunctions in 
$\vec{\xi}_1$ and $\vec{\xi}_2$
with quantum numbers  $n_1$, $l_1$ and 
$n_2$, $l_2$, respectively, 
with the angular momentum coupled to $L$, while the spin (isospin) part 
is constructed by coupling 
first two 
spins (isospins) $s=1/2$ ($t=1/2$) into spin (isospin) $s_2$  ($t_2$) 
and then a third spin (isospin) $s=1/2$ ($t=1/2$)
to total spin $S$ (isospin $T$).
In Eq.~(\ref{Stetcu_eq:basis3b}), ${\cal A}$ stands for the operator that 
antisymmetrizes the three-particle wavefunction.  
Details on the construction of a fully antisymmetrized basis can be found 
in Ref. \cite{Navratil:1999pw}, which include generalization to more than three particles. Because in this method only relative coordinates are involved, the solutions are translationally invariant. However, due to the complexity of the anti\--symmetriza\-tion, this approach is not efficient for nuclei with $A>4$.

For heavier nuclei, it is more efficient to work with a Slater determinant (SD) basis, in which antisymmetric many-body states are constructed from single HO basis states. In such an approach, center-of-mass (CM) excitations can mix with the internal degrees of freedom. However, if one performs a truncation on the number of energy quanta above the non-interacting minimum configuration, the CM wave function factorizes exactly, so that the results are free of spurious modes. In particular, by adding 
$$\beta H_{HO}^{CM} =\beta \left(\frac{\vec P_{CM}^2}{2M_NA}+\frac{1}{2}AM_N\omega^2\vec R_{CM}^2-\frac{3}{2}\omega\right)$$ 
to the Hamiltonian (\ref{Stetcu_eq:ham}), with $\omega$ the HO frequency and $\beta$ a numerical parameter of order 10, one can ensure that for the low-lying states, the CM is in the lowest HO. Hence, for operators depending only upon the intrinsic coordinates, the matrix elements are free of CM contributions. While in this case the anti-symmetrization is trivial, the size of the many-body model space increases very quickly with the number of particles and the number of single-particle states. Even so, the diagonalization in a SD basis remains the method of choice for $A>4$, and a lot of effort is dedicated toward the development of algorithms to handle larger and larger dimensions.


In principle, NCSM provides exact solutions to the Hamiltonian in Eq. (\ref{Stetcu_eq:ham}). In either the trans\-la\-ti\-o\-na\-lly-invariant or the SD approaches, for a complete calculation, an infinite number of HO states should be included in calculations, and therefore the dimension of the many-body basis is infinite. Since the NCSM involves a numerical diagonalization, the many-body basis has to be truncated, and hence a method to include correlations left out by truncation is required. In NCSM, the truncation is determined by $N_{max}$, which is the number of $\omega$ excitations above the minimum non-interacting configuration. Thus, the model space is defined by $N_{max}$ and the frequency $\omega$. Note that if the truncation in the SD basis, given by $N_{max}$, is related to the truncation in relative coordinates, so that both include the same number of excitations on top of the minimum configuration, calculations by the two means produce the same results for the same truncations. 

The intrinsic properties of the $A$-body system are not affected by the addition of a Hamiltonian term that depends only on the CM coordinates. Thus, following Lipkin's idea \cite{Lipkin_CM}, we add a HO CM Hamiltonian, so that the new Hamiltonian reads

\begin{equation}
  H_{A}=H_{int}+H_{HO}^{CM}= \sum_{i=1}^A h_i 
  +\sum_{i>j=1}^A \left(V_{ij}-\frac{M_N\omega^2}{2A}(\vec r_i -\vec r_j)^2\right)
   + \sum_{i>j>k=1}^AV_{ijk}^{NNN}+...,
   \label{Stetcu_eq:Ham_Abody}
\end{equation}
where $h_i$ stands for a single-particle HO Hamiltonian of frequency $\omega$. As one subtracts the CM term in the final many-body calculation, it does not introduce any net influence on the converged intrinsic properties of the many-body calculation. Furthermore, this addition and subtraction does not affect our exact treatment of the CM motion. However, this procedure is an essential step in the derivation of the effective interaction within the NCSM approach. 

\subsection{Unitary transformation method for effective interactions}
\label{Stetcu_sec:ut}

In most of the approaches to solving the nuclear many-body problem, the inter-nucleon interactions are considered input. Early applications of NCSM have used as input phenomenological NN interactions, such as the local Argonne $v_{18}$ \cite{Wiringa:1995,QMC_rev} or the non-local charge-dependent (CD) Bonn interaction \cite{cdbonn}. In such interactions, the short-range part has been parametrized as a repulsive code of the order of 1 GeV.  As a consequence, even low-lying states have considerable high-momentum components, making the convergence in finite model spaces very difficult. The unitary transformation approach takes into account the eliminated states to produce an effective interaction for the truncated model space. Any truncation induces up to $A$-body interactions, so an effective interaction that reproduces exactly the low-lying spectrum is as difficult to derive as solving the $A$-body problem. Hence, the so called cluster approximation has been derived, where the interaction is obtained for $a<A$ particles, and the result used in the $A$-body problem. However, in finite models spaces, the addition of the CM procedure discussed at the end of the previous section introduces a pseudo-dependence upon the HO frequency $\omega$, and the cluster approximation described below will sense this dependence. In the largest model spaces, important observables manifest a relatively weak dependence of the frequency $\omega$ and the model space size. In the following, as it is customary in NCSM applications, we make the distinction usual distinction in the traditional NCSM between the ``bare" interaction, which is the original interaction whose full space full Hilbert space solution we seek, and the ``effective'' interaction, which is derived in a small model space. We will see that in the EFT inspired approach discussed later that such a distinction is artificial.

Perhaps the most general derivation of the unitary transformation approach was presented in Refs. \cite{Navratil:1993plb,Navratil:1993NPA}, which we closely follow in this paper. 

As discussed previously, for a given Hamiltonian $H_{int}$, in particular given by Eq. (\ref{Stetcu_eq:ham}), the goal is to solve the Schr\"odinger equation

\begin{equation}
H_{int}|\Phi\rangle=E_\Phi|\Phi\rangle
\label{Stetcu_eq:goal}
\end{equation}
in a finite model space, by direct numerical diagonalization. Hence, in the NCSM approach one divides the full Hilbert space into a model space, with associated projection operator $P$ ($P^2=P$), and a complementary,
excluded space, with the associated projection operator $Q$ ($Q^2=Q$, $P+Q=1$, and $PQ=QP=0$). In principle, the excluded space $Q$ is infinite, but in practical realizations is taken to be as large as possible and still be numerically tractable. In addition, we introduce a similarity transformation operator $X$ (not necessary unitary) with the goal to perform many-body calculations in the model space $P$, using a transformed Hamiltonian
$\mathcal{H}$,
\begin{equation}
{\mathcal{H}}=XH_{int}X^{-1},
\label{transfHam}
\end{equation}
so that a finite subset of eigenvalues of the initial Hamiltonian
$H_{int}$ in Eq. (\ref{Stetcu_eq:ham}) are reproduced. This is a general
approach, which can be applied to non-Hermitian Hamiltonian operators 
that can arise, for example, in the context of boson mappings.

In order to derive the conditions to be imposed on the 
transformation operator $X$, we revisit the Feshbach projection formalism applied to the Schr\"odinger
equation
\begin{equation}
{\mathcal H}|\Psi\rangle=E_\Psi |\Psi\rangle.
\label{schr0}
\end{equation}
We note that, for the same energy eigenvalue, the wavefunctions $|\Phi\rangle$ and $\Psi\rangle$ are connected via the unitary transformation
\begin{equation}
|\Psi\rangle=X|\Phi\rangle.
\label{Stetcu_eq:eigvs}
\end{equation}
In general for non-Hermitian Hamiltonians the left and
right eigenvectors are not related simply by a Hermitian conjugation. Nevertheless, we
have the freedom to choose a normalization so that 
$\langle\tilde\Psi_E|\Psi_E\rangle=1$, where $\langle \tilde\Psi_E|$ is
the left eigenvector corresponding to the eigenvalue $E_\Psi$.
From Eq. (\ref{schr0}), we obtain immediately two coupled equations for the components of the wave function in the models space $P$,
\begin{equation}
P\mathcal{H} P|\Psi\rangle +  P\mathcal{H} Q|\Psi\rangle = E_\Psi P|\Psi\rangle,
\label{Stetcu_eq:Ppsi}
\end{equation} 
and in the excluded space
\begin{equation}
Q\mathcal{H} P|\Psi\rangle +  Q\mathcal{H} Q|\Psi\rangle = E_\Psi Q|\Psi\rangle.
\label{Stetcu_eq:Qpsi}
\end{equation} 
Solving formally for $Q|\Psi\rangle$ in Eq. (\ref{Stetcu_eq:Qpsi})
\begin{equation}
Q|\Psi\rangle= \frac{1}{E_\Psi-Q{\mathcal H}Q}Q{\mathcal H}P|\Psi\rangle,
\label{QrightEV}
\end{equation}
and inserting the result into Eq. (\ref{Stetcu_eq:Ppsi}), one obtains immediately that the effective Hamiltonian in the model space can be expressed as
\begin{equation}
H_{eff}=P{\mathcal H}P+P{\mathcal H}Q\frac{1}{E_\Psi-Q{\mathcal H}Q}Q{\mathcal H}P,
\label{Heff0}
\end{equation}
which is manifestly energy and state dependent. While approaches in which the energy dependence have been proposed, but their application has been limited to two- and three-particle systems \cite{Haxton:2001,*Haxton:2002kb,*Haxton:2008,luu:103202}. Using Eq. (\ref{Heff0}), we can eliminate at once the state dependence and energy dependence if we  impose \textit{one} of the following
decoupling conditions
\begin{equation}
Q{\mathcal{H}}P=0,\label{decouplQP}
\end{equation}
or
\begin{equation}
P{\mathcal{H}}Q=0.\label{decouplPQ}
\end{equation}
We note, however, that the former condition also ensures that the
$Q$-space component of the wave function $|\Psi\rangle$ vanishes,
although this is not true for its complementary left eigenstate.
Moreover, as it will become clear in the derivation of the effective
operators below, \textit{both} conditions have to be satisfied so that
one obtains energy-independent effective operators corresponding to
other observables besides the Hamiltonian.

In a consistent approach general operators $O$ are transformed by the
same transformation operator $X$, $\mathcal{O}=XOX^{-1}$. In this case, one needs to compute a matrix element of the form
$\langle \tilde \Phi |{\mathcal O}|\Psi\rangle$, where $\langle \tilde
\Phi|$ corresponds possibly to another left eigenvector of
${\mathcal H}$.
Similar to Eq.  (\ref{QrightEV}), the $Q$-component of the left eigenstate
$\langle\tilde \Phi|$ can be written
\begin{equation}
\langle\tilde\Phi|Q=\langle\tilde\Phi|P{\mathcal H} Q
\frac{1}{E_\Phi-Q{\mathcal H}Q}.
\label{QleftEV}
\end{equation}
one can immediately extract the expression for the effective operator in the
model space $P$
\begin{eqnarray}
\lefteqn{{\mathcal O}_{eff}=P{\mathcal O}P+P{\mathcal H}Q
\frac{1}{E_\Phi-Q{\mathcal H}Q}Q{\mathcal O}P
+P{\mathcal O}Q \frac{1}{E_\Psi-Q{\mathcal H}Q}Q{\cal H}P\nonumber} \\
& &
+P{\mathcal H}Q\frac{1}{E_\Phi-Q{\mathcal H}Q}Q{\mathcal O}Q
\frac{1}{E_\Psi-Q{\mathcal H}Q}Q{\cal H}P.
\end{eqnarray}
As advertised, in order to obtain an energy-independent expression
for a general effective operator one needs to construct the
transformation operator $X$ so that \textit{both} decoupling 
conditions (\ref{decouplQP}) and (\ref{decouplPQ}) are satisfied.
Consequently, both left and right $P$ eigenstates of the
transformed Hamiltonian ${\mathcal H}$ have components only in the
model space. Additional subtleties exist within this effective
operator approach \cite{ViazVary}.

In order to obtain the transformation operator $X$, let us consider first an operator $\hat \omega$ that satisfies the following condition:
\begin{equation}
\hat \omega = Q\hat \omega P.
\end{equation}
In terms of $\hat \omega$, let us further assume that the transformation operator $X$ can be written as
\begin{equation}
X\equiv\exp(-\hat\omega)=P+Q-\hat\omega,
 \label{ansatz}
\end{equation}
so that the decoupling condition (\ref{decouplPQ}) is satisfied automatically. The remaining decoupling condition (\ref{decouplQP}) provides a quadratic equation that determines the auxiliary operator $\hat \omega$:
\begin{equation}
  Q{\mathcal H}P=QHP-Q\hat \omega H P +Q H\hat\omega P -\hat\omega H\hat\omega=0.
  \label{Stetcu_eq:W}
\end{equation}
Finally, we obtain at once the expressions for the effective Hamiltonian in the model space
\begin{equation}
  {\mathcal H}_{eff}=PHP+PH\hat\omega P,
  \label{Stetcu_eq:nonHermH}
\end{equation}
and similarly for the effective operator
\begin{equation}
  {\mathcal O}_{eff}=POP+PO\hat\omega P.
  \label{Stetcu_eq:nonHermO}
\end{equation}
Manifestly, the Hamiltonian (\ref{Stetcu_eq:nonHermH}) is non-Hermitian, even if the original Hamiltonian is Hermitian, and the effective operators given by Eq. (\ref{Stetcu_eq:nonHermO}) also change symmetry properties under a Hermitian conjugation. 

As noted before, it is desirable to obtain Hermitian Hamiltonians in the model space, as they are easier to handle numerically. Thus, assuming that $X=\exp(G)$, where $G$ is anti-Hermitian ($G^\dagger=-G$), one obtains a considerably more complicated equation that determines $G$ (see Eq. (2.19) in Ref. \cite{Suzuki:1982}). However, one can show that one can derive a relation between $G$ and $\hat\omega$ \cite{Suzuki:1982}
\begin{equation}
G=\mathrm{arctanh}(\hat\omega-\hat\omega^\dagger)=\sum_{n=0}^\infty (-1)^n \frac{1}{2n+1}\hat\omega(\hat\omega^\dagger\hat\omega)^n-\mathrm{h.c.},
\end{equation}
where we have used the property $\hat\omega^k=0$, for $k>1$. Using $G$ given by the previous equation, we obtain the unitary transformation that ensures the decoupling of the model and excluded spaces 
\begin{equation}
X=(1+\hat\omega-\hat\omega^\dagger)(1+\hat\omega\hat\omega^\dagger+\hat\omega^\dagger\hat\omega).
\end{equation}
If we calculate $Q\mathcal{H}P$ we obtain for $\hat\omega^\dagger$ an equation which is exactly the conjugate of (\ref{Stetcu_eq:W}). Finally, because $\hat\omega^\dagger \hat \omega$ and $\hat \omega\hat\omega^\dagger$ act only in the model space $P$ and in the excluded space $Q$ respectively, the expression for the effective Hamiltonian in the model space is

\begin{equation}
 {\mathcal H}_{eff}=\frac{P+\hat\omega^\dagger}{\sqrt{P+\hat\omega^\dagger\hat\omega}}
                H\frac{P+\hat\omega}{\sqrt{P+\hat\omega^\dagger\hat\omega}},
\label{Stetcu_eq:effHam}
\end{equation}
while the expression for general effective operators is similar
\begin{equation}
  {\mathcal O}_{eff}=\frac{P+\hat\omega^\dagger}{\sqrt{P+\hat\omega^\dagger\hat\omega}}
                O\frac{P+\hat\omega}{\sqrt{P+\hat\omega^\dagger\hat\omega}}.
\label{Stetcu_eq:effOp}
\end{equation}

The construction method we have presented here is not unique. In Ref. \cite{Navratil:1993plb}, the derivation is a bit more involved, using an additional operator. However, the end result is the same, with the effective operators in the model space given by Eqs. (\ref{Stetcu_eq:effHam}) and (\ref{Stetcu_eq:effOp}). We should, however, point out that our derivation does not exclude the existence of a different unitary transformation that would ensure the same decoupling between the model and excluded space.

The fact that one usually chooses to work with the Hermitian Hamiltonian (\ref{Stetcu_eq:effHam}) is a matter of convenience. As discusses above, there exists at least one similarity transformation, Eq. (\ref{ansatz}), that ensures the decoupling condition, and there could be more. Nevertheless, as long as the transformations are computed exactly, they all produce the same spectrum and properties of the  system. The wave functions will be different, but this is not alarming since they are not observables. As we will see in the following, however, one always uses an approximation, so that the decoupling condition is only approximate, and the exact decoupling is achieved only in a large model space limit. It is therefore conceivable that one transformation could be more suitable than another in the sense of faster convergence to ``full" Hilbert space results. Where relevant, we will discuss further implications of the particular choice of transformation. Hitherto there is no approximation in our derivation.

A great deal of effort has been directed toward the calculation of the operator $\hat\omega$ \cite{Suzuki:1980,*Suzuki:1994ok,Suzuki:1982,Krenciglowa,Navratil:1996_effint,Navratil:1998_pshell}, as it is an essential step in the derivation of the model space effective operators. Two iterative solutions have been devised:  one that converges to the states with the largest  $P$-space components and is equivalent to the solution of Krenciglowa and Kuo \cite{Krenciglowa}, and another which converges to states lying closest to a chosen parameter appearing in the iteration procedure \cite{Suzuki:1980,*Suzuki:1994ok,Suzuki:1982}. A more efficient method was introduced later by Navr\'atil and Barrett \cite{Navratil:1996_effint,Navratil:1998_pshell}. It relies on the fact that the components of the exact eigenvectors in the complementary space are mapped into the model
space. Thus, going back Eq. (\ref{Stetcu_eq:eigvs}) and using the definition of the similarity transformation operator (\ref{ansatz}) we obtain
\begin{equation}
|\Psi\rangle =  P|\Phi\rangle + \hat\omega |\Phi\rangle,
\label{Stetcu_eq:mapEV}
\end{equation}
taking also into account the decoupling condition (\ref{decouplQP}). Equation (\ref{Stetcu_eq:mapEV}) represents the formal proof that $\hat\omega$ maps back from the model space the excluded component of the wave function. Hence, if we choose a set $\mathcal{K}$ of \textit{exact} eigenvectors of the Hamiltonian in the full space, one obtains immediately that the matrix elements of the operator $\hat\omega$ are given by \cite{Navratil:2000gs,Navratil:1996_effint,Navratil:1998_pshell}
\begin{equation}
\langle\alpha_{Q}|\hat\omega|\alpha_{P}\rangle=
\sum_{k\in{\mathcal{K}}}\langle\alpha_{Q}|\Psi_k\rangle
\langle{\Psi_k}|\alpha_{P}\rangle^{-1},\label{Stetcu_eq:omega}
\end{equation}
where $|\alpha_{P}\rangle$ and $|\alpha_{Q}\rangle$ are the basis states
of the $P$ and $Q$ spaces, respectively.  The dimension of the subspace ${\mathcal{K}}$ is equal
with the dimension of the model space $P$ and we assume that the overlap matrix between the chosen eigenvectors and the basis states in the model space is non singular. In the next subsection, we
will present a practical implementation of Eq. (\ref{Stetcu_eq:omega}).

\subsection{The cluster approximation}
\label{Stetcu_sec:cluster}

A closer examination of Eq. (\ref{Stetcu_eq:omega}) reveals that the computation of the matrix elements of the operator $\hat\omega$ requires the knowledge of a set of $A$-body eigenvectors, which is in itself the goal of calculation. Moreover, it shows that the effective Hamiltonian is a $A$-body operator. Hence, an exact calculation of the unitary transformation is as difficult as obtaining the many-body solution. The practical solution is the so-called cluster approximation, in which the unitary transformation operator is calculated for $a<A$ particles, and an effective $a$-body interaction in the model space is used to solve the full $A$-body problem. Only the two- and three-body cluster approximation has been implemented up to now. 

Let us consider for the moment that one can neglect for now three- and higher-body forces and let us consider the $a$-body cluster approximation. In this case, from the $A$-body Hamiltonian in Eq. (\ref{Stetcu_eq:Ham_Abody}) one considers the $a$-body problem 
\begin{equation}
h^{(a)}=\sum_{i=1}^a h_i+\sum_{i>j=1}^a\left(V_{ij}-\frac{M_N\omega^2}{2A}(\vec r_i - \vec r_j)^2\right).
\end{equation}
The $a$-body Hamiltonian $h_{a}$ is further separated into a relative contribution and a pair CM contribution, so that the previous equation can be cast as
\begin{equation}
h^{(a)}=h_{rel}^{(a)}+H_{a}^{CM},
\end{equation}
with the CM of the $a$-body system decoupling from the relative Hamiltonian given by
\begin{equation}
h_{rel}^{(a)}=\frac{1}{a}\sum_{i<j=1}^a\left[\frac{(\vec p_i-\vec p_j)^2}{2M_N}+\frac{1}{2}M_N\omega^2\left(1-\frac{2}{A}\right)(\vec r_i-\vec r_j)^2\right]+\sum_{i<j}^{a}V_{ij}.
\label{Stetcu_eq:hrelA}
\end{equation}
We assume that Eq. (\ref{Stetcu_eq:hrelA}) can be solved in a space large enough space that could be considered the full Hilbert space (although in practice is always truncated), obtaining a set of eigenvectors $|\Psi^{(a)}_k\rangle$, which in turn determines by means of Eq. (\ref{Stetcu_eq:omega}) the operator $\hat \omega^{(a)}$. Hence, the effective $a$-body interaction in the model space will be given by the analogous of Eq. (\ref{Stetcu_eq:effHam})
\begin{equation}
V^{(a)}=\frac{P_a+\hat{\omega}^{(a)\dagger}}{\sqrt{P_{a}+\hat\omega^{(a)\dagger}\hat\omega^{(a)}}}
                h^{(a)}\frac{P+\hat\omega^{(a)}}{\sqrt{P_a+\hat\omega^{(a)\dagger}\hat\omega^{(a)}}}-\sum_{i=1}^a h_i,
\end{equation}
so that the effective Hamiltonian for the $A$-body problem becomes
\begin{equation}
H_{eff}=\sum_{i=1}^A h_i+\frac{\left(\begin{array}{c}
A \\
2 \end{array}\right)}{\left(\begin{array}{c}
A \\
a \end{array} \right) \left( \begin{array}{c}
a \\
2 \end{array}\right)}\sum_{i_1<i_2\ldots i_a=1}^A V^{(a)}_{i_1i_2\ldots i_a}
\end{equation}
There is no summation over $a$.
Note that the decoupling conditions (\ref{decouplQP}) and (\ref{decouplPQ}) are now valid in the $a$-body space and not the full $A$-body space. 

In particular, in the lowest approximation, i.e., the two-body cluster approximation, the relative two-body Hamiltonian (\ref{Stetcu_eq:hrelA}) writes

\begin{equation}
h^{(2)}_{rel}=\frac{p^2}{2M_n}+\frac{1}{2}M_n\omega^2r^2+V_{12}(\sqrt{2}r)-\frac{M_N\omega^2}{2A} r^2,
\label{Stetcu_eq:2brelH}
\end{equation}
where $\vec p = (\vec p_1-\vec p_2)/\sqrt{2}$ and $\vec r=(\vec r_1 - \vec r_2)/\sqrt{2}$ (we have also assumed a local two-body interaction). 
By solving the two-body Schr\"odinger equation using the previous Hamiltonian with high accuracy in a large model space, one can construct by means of Eq. (\ref{Stetcu_eq:omega}) the approximate operator $\hat\omega^{(2)}$ and explicitly an effective interaction in a chosen model space. We note that the use of the HO piece is essential. Thus, it not only provides a term that depends on the particle number, but also ensures that \textit{all} the eigenstates of Hamiltonian (\ref{Stetcu_eq:2brelH}) are bound, and hence the HO basis is appropriate for the eigenvector expansion.  Finally, it is important to note that, at the two-body cluster level, the unitary transformation accommodates mostly the short-range correlations, while the long-range any many-body correlations are included by increasing the model space or the cluster level.

By construction, there are two limits in which the exact solutions are recovered:
\begin{itemize}
\item[(i)] working in the same cluster approximation, but increasing the size of the model space approaching the full Hilbert space;
\item[(ii)] keeping the size of the model space constant, but increasing $a$ up to $A$.
\end{itemize} 
In practical applications, the cluster approximation was never applied for $a>3$, and one must rely on property (i) in order to obtain solutions that do not depend upon parameters of the calculation. Thus, the dependence on the HO frequency and model space, introduced by the cluster approximation, is minimized by increasing the model space. Because many-body terms in the interaction are neglected in this approximation, one looses the variational character of the calculation and it is not unusual to have a convergence to the exact eigenenergy from below. However, the advantage is that one usually obtains a much faster convergence than if one uses the ``bare'' interaction, which does preserve the variational character. 

\begin{figure}
\centering{\includegraphics[clip,scale=0.56,angle=-90]{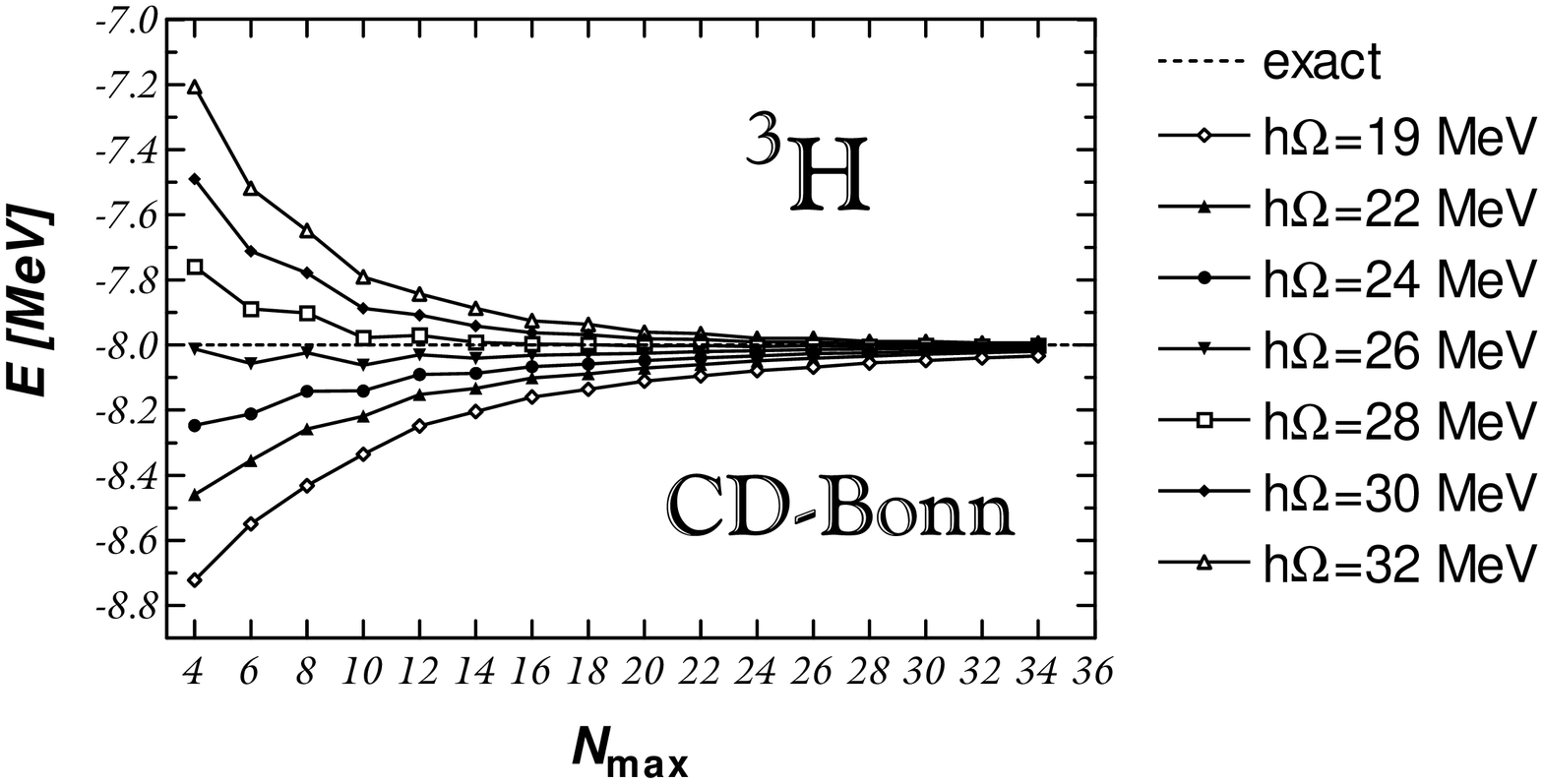}}
\caption{The ground-state energy of the triton as a function of the shell model truncation parameter $N_{max}$ using an effective interaction derived in the two-body cluster approximation. We present several HO frequencies: 19 MeV (open diamonds),  22 MeV (full up triangles), 24 MeV (circles), 26 MeV (full down triangles), 28 MeV (open squares), 30 MeV (full diamonds), and 32 MeV (open triangles), respectively.  The dashed line is the exact ground state for the CD Bonn interaction, calculated independently in a Fadeev approach \protect\cite{cdbonn}. Figure taken from Ref. \protect\cite{Navratil:1999pw}, courtesy P. Navr\' atil.}
\label{Stetcu_fig:tritonEn}
\end{figure}

In Fig. \ref{Stetcu_fig:tritonEn} we illustrate the convergence properties of the effective interaction in the two-body cluster approximation for $^3$H. In this case, the starting Hamiltonian contains, in addition to the kinetic energy, only the non-local NN potential CD Bonn \cite{cdbonn}. In small model spaces, the energy can deviate quite significantly from the exact solution obtained in a charge-dependent Fadeev calculation in 34 channels \cite{cdbonn}. However, increasing the model space to large enough values, the ground-state energy converges to the Fadeev approach (note that because of missing three-body interactions, the calculated value misses the experiment). The reason is that by increasing the model space, the neglected contributions from higher order clusters become smaller and smaller.  In fact, it was demonstrated analytically that, if one starts with a ``bare" NN interaction, in the large model space limit, the two-body cluster effective interaction is exact to second order in perturbation theory \cite{Barnea:clusterExpansion}. Additionally, one often can find a special frequency for which the rate of convergence is very large for some states, as shown in Fig. \ref{Stetcu_fig:tritonEn} for $\omega=26$ MeV. In this case, the contribution from neglected three-body clusters is small. This usually happens when the length associated with the HO frequency is close to the size of the respective state.

\begin{figure}
\centering{\includegraphics*[scale=0.55]{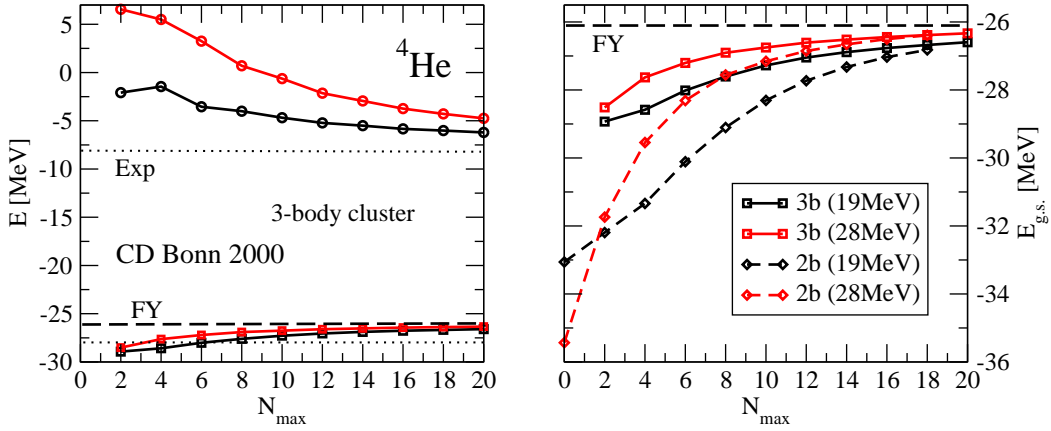}}
\caption{The ground- and first excited-state energies of $^4$He as a function of the model space truncation parameter $N_{max}$ in the three-body cluster approximation (left panel), and comparison of the convergence rates of the ground-state energy for the two- and three-body cluster approximations (right panel) for two different HO energies . The dashed line is the exact ground-state energy for the CD Bonn potential used in this investigation, while the dotted lines represent the experimental ground- and first excited-state energies. Figure from Ref. \protect\cite{nonHermOp}.}
\label{Stetcu_fig:he4En}
\end{figure}

In order to demonstrate the effect of the higher cluster approximation, in Fig. \ref{Stetcu_fig:he4En} we compare the convergence rates of the energies of the two lowest spin and isospin zero states in $^4$He,  obtained using effective interactions calculated in two- and three-body cluster approximations respectively. Like in the previous example, the same charge-dependent CD Bonn potential was employed. As expected, a higher-body cluster approximation 
includes more correlations in the interactions,
and the convergence is faster. This is illustrated in the right panel
of Fig. \ref{Stetcu_fig:he4En}, where we plot the ground-state 
energy dependence on $N_{max}$
obtained by computing the effective interaction using both the two- and 
three-body cluster approximations. The rate of convergence
is faster in the three-body cluster approximation for both HO energies
chosen for this example. Like for the $^3$H energy shown in Fig. \ref{Stetcu_fig:tritonEn}, HO frequencies for which the HO length is of the size of the states considered show faster convergence to the exact solution. The latter differs from the experimental value mostly because we neglect three-body interactions. Unlike for the ground state, the first $0^+$ excited-state energy has a faster convergence rate for $\omega = 19$ MeV. However, this state has a much more slower convergence rate than the ground state, and even in the largest model spaces the results are quite sensitive to the choice of the HO energy parameter.

The NCSM approach described here has been applied successfully to the description of a large number of properties of low-lying states in light nuclei \cite{Navratil:2000gs,Hayes:2003ni,Caurier:2005rb,Nogga:2005hp,Navratil:2007,Stetcu:polarizHe,forssen:2p,PhysRevLett.106.202502}, nuclear reaction observables \cite{Quaglioni:2007eg,PhysRevC.79.044606,*PhysRevC.82.034609,*PhysRevC.83.044609,*PhysRevLett.108.042503} and even to predict properties relevant to the physics beyond the Standard Model \cite{Stetcu:2008vt,*PhysRevC.84.065501}. We will review select applications below.

The cluster approximation employed in these calculations is not justified \textit{a priori}.  Mintkevich and Barnea have demonstrated that the higher order clusters are negligible in the limit of the model space approaching the ``full'' Hilbert space \cite{Barnea:clusterExpansion}, as expected. They have also presented arguments suggesting that, at least for a system of bosons, the induced many-body terms fall much faster than the effective two-body interactions as the model space is increased to the full space. Nevertheless, there is no clear limit when the model space is large enough to ensure that the neglected many-body terms are small. All the results, however, suggest that, as long as the procedure is used to converge with increasing the size of the model space, one can obtain the exact result for a given Hamiltonian, as demonstrated in a benchmark calculation of the $^4$He ground-state properties by several few-body methods \cite{benchmark:4he_gs}. Sometimes one can rely on extrapolation methods to inferrer the converged value \cite{PhysRevC.77.024301}, although it might not be obvious how large the model space has to be in order to predict the correct value.

Several other methods based on unitary transformations have been proposed in the last decade to derive interactions better suitable for many-body calculations \cite{Feldmeier:1997zh,Bogner:2001yi,*Bogner:2002yw,Bogner:2006pc}, either by means of exact \cite{PhysRevLett.103.082501,PhysRevLett.107.072501,PhysRevC.83.034301} or approximate \cite{Coraggio:2008in,PhysRevC.83.064317} methods. In this case, the goal is to ``soften'' the interactions, so that even initially the model and excluded spaces are not as strongly coupled as when using ``bare" interactions. The similarity renormalization group (SRG) in particular provides an elegant way to  accomplish the decoupling of the low- and high-momentum spaces, and still preserve the accuracy of the initial interactions. But such approaches suffer from the same shortcoming as the unitary transformation in NCSM: the procedure induces up to $A$-body interactions and out of practical considerations one needs to truncate the interactions to a level where calculations can be performed, hitherto by including up to three-body induced (and genuine) interactions. However, as in the case of the cluster approximation for the unitary transformation in NCSM, there is no proof that the neglected terms are small necessarily. For light nuclei has been found that the contribution of four- and higher-body induced interactions is negligible \cite{PhysRevLett.103.082501,PhysRevLett.107.072501,PhysRevC.83.034301}, but calculations of medium-mass nuclei have suggested that, although the hierarchy of the many-body forces seems to be preserved, the neglected terms become increasingly important \cite{PhysRevLett.107.072501,PhysRevC.83.034301,Roth:2011fk}. Finally, it is important to note that in the calculations involving SRG and similar approaches, one usually uses inter-nucleon interactions that are derived from EFT \cite{N3LO,vanKolck:1994,*Epelbaum:2002}. Such interactions have intrinsic systematic errors that come from the neglected terms at the order of the expansion. And while one can argue that the similarity/unitary transformations preserve the accuracy of the initial EFT Hamiltonian, they also induce terms not included in the EFT expansion, thus potentially enhancing those contributions. Hence, the net result can be an enhancement of the terms that are small (to some order) and, possibly, a loss of predictability.


\subsection{Effective operators in NCSM}
\label{Stetcu_sec:eff_op}

In a consistent approach, the wave functions obtained using a transformed Hamiltonian could be used to compute other observables only if the operators associated with those observables are transformed similarly to the Hamiltonian. In Sec. \ref{Stetcu_sec:ut} we derived a unitary transformation that depends only on the Hamiltonian, ensuring that the effective operators are energy independent. The advantage of the unitary transformation approach is that any general operator preserves its symmetry properties. 

Of extreme importance are the electromagnetic properties, which are well known experimentally. In the phenomenological shell model, a long-standing problem is the relatively large effective charges that were found essential in the overall description of the transition strengths. They arise from the truncation of the model space and previous attempts at describing them in perturbation theory have proved unsuccessful \cite{Osnes}. In NCSM, Navr\'atil et. al have reported effective charges consistent with the phenomenological values \cite{Navratil:1996jq}. A fundamental derivation within the NCSM of the phenomenological effective charges represents, in addition to consistency, a strong motivation to pursue the renormalization of general operators.

Because the implementation of the unitary transformation for general operators is a non-trivial task \cite{Stetcu:2004bk,Stetcu:2004wh,Stetcu:2006zn}, for non-scalar operators it was done somewhat later than the introduction of the unitary transformation in NCSM \cite{Navratil:1996_effint,Navratil:1998_pshell}. The difficulty stems from the fact that, for a general non-scalar operator of rank $\Delta J$ and $\Delta T$, $O^{(\Delta J,\Delta T)}$, the calculation of effective operator matrix elements require in general the use of transformations in different channels, i.e., Eq. (\ref{Stetcu_eq:effOp}) becomes

 \begin{equation}
  {\mathcal O}_{eff}^{(\Delta J\Delta T)}=\frac{P_{JT}+\hat\omega^\dagger_{JT}}{\sqrt{P_{JT}+\hat\omega^\dagger_{JT}\hat\omega_{JT}}}
                O^{(\Delta J,\Delta T)}\frac{P'_{J'T'}+\hat\omega_{J'T'}}{\sqrt{P'_{J'T'}+\hat\omega^\dagger_{J'T'}\hat\omega_{J'T'}}},
\label{Stetcu_eq:effOpJT}
\end{equation}
where $J,\:T,\:J',\:T'$ fulfill the usual angular momentum rules $|J-J'|\leq \Delta J$ and  $|T-T'|\leq \Delta T$. We emphasize that the transformation operator $\hat\omega_{JT}$ is determined by the decoupling conditions imposed on the Hamiltonian. 

As in the case of effective interactions, the effective operators will acquire $A$-body contributions, even if the original operator is one-body. However, given that the transformation can be calculated only approximately, higher order terms are neglected. In particular, given the complexity of the procedure, the effective operators have been implemented only in the two-body cluster approximation \cite{Stetcu:2004bk,Stetcu:2004wh,Stetcu:2006zn}. 

For the one-body operators special care has to be taken in order to eliminate higher order contributions to the effective operators. Thus, in the two-body cluster, the unitary transformation writes

\begin{equation}
X_2\approx\exp(- G_{2}),
\end{equation}
where $G_{2}=\sum_{ij}\left(\arctan(\hat\omega_{ij}-\hat\omega_{ij}^\dagger)\right)$ with $\hat\omega_{ij}$ calculated at the two-body level. Using the operator identity 

\begin{equation}
\exp(-G_2)O\exp(G_{2})=O + [O,G_2]+ \frac{1}{2!}\left[ [O,G_2],G_2\right]+\ldots
\end{equation}
one can obtain immediately for a general one-body operator $O^{(1)}=\sum_i O_i$,  the expression of the effective operator in the two-body cluster approximation

\begin{equation}
P_2\mathcal{O}^{(1)}_{eff}P_2=\sum_i O_i + P_2\sum_{i,j}\left( e^{-G_{ij}}(O_i + O_j )e^{G_{ij}}-(O_i+O_j)\right)P_2,
\label{Stetcu_eq:one_body_eff}
\end{equation}
where we have retained only one- and two-body terms in the expression. 

Similarly one obtains the expression for a general two-body operator $O^{(2)}=\sum_{i,j}O_{ij}$
\begin{equation}
P_2\mathcal{O}^{(2)}_{eff}P_2= P_2\sum_{i,j} e^{-G_{ij}}O_{ij}e^{G_{ij}}P_2.
\label{Stetcu_eq:two_body_eff}
\end{equation}
Combining Eqs. (\ref{Stetcu_eq:one_body_eff}) and (\ref{Stetcu_eq:two_body_eff}) for the special case of a one- plus two-body Hamiltonian, we recover the expression of the effective Hamiltonian (\ref{Stetcu_eq:hrelA}) in the two-body cluster approximation.

\subsection{Selected results}

NCSM has been successful in describing a large number of nuclear properties, from energy levels to reaction observables. While the main scope of the paper is to discuss the effective interactions used within the NCSM framework, we find appropriate to illustrate with a few applications to nuclear systems how well this approach works, especially in the case of nuclear spectra. The interested reader can find more examples in a recent review of the latest NCSM calculations \cite{Navratil:2009ut}.

We have already discussed the two- and three-body cluster approximations for the effective interactions in the case of triton and alpha particle. However, because the main goal was a demonstration of the efficacy of the effective interactions in numerically accessible model spaces, as well as a comparison of the two- and three-body approximations, the starting Hamiltonian included two-body terms only. As a result, we have seen that the ground-state energies do not converge to the experimental values in either of the cases. In addition, precise calculations have shown that two-body forces only cannot account for the experimental ordering of some states in a number of p-shell nuclei \cite{PhysRevLett.88.152502} and in general predict binding energies that are too small \cite{QMC_rev,Navratil:1998_pshell,PhysRevC.65.054003}. Today it is generally accepted that, in order to obtain a good agreement with the experimental data it is important to include three-body forces in the calculations. Experience with phenomenological forces \cite{QMC_rev}, as well as rigorous analytical proof in the case of EFT interactions \cite{Weinberg1992}, has shown that four- and higher-body forces are highly suppressed, although it is conceivable that for some processes involving large enough momenta they could become significant.

Early light nuclei calculations have been based on phenomenological two-body forces, which provide an excellent description of the NN scattering data up the pion threshold, augmented by three-body interactions. In general, the phenomenological two- and three-body interactions provide very good agreement for a large number of low-energy observables \cite{QMC_rev}, but in this approach the relation to the theory of strong interactions, QCD, is completely lost. The EFT interactions, and in particular chiral interactions, have emerged with the potential to provide a unified approach to the derivation of two- and many-body interactions. The symmetries of QCD, such as the approximate chiral symmetry of QCD, play a central role in the derivation. In the EFT approach, one starts with the most general Hamiltonian with the appropriate degrees of freedom and symmetries \cite{weinberg1990,*weinberg1991,*ordonez1992,eftreview}. Interactions among nucleons consist of pion exchanges and contact interactions, which model short-range dynamics (e.g., exchange of heavier mesons) and at very low energies even pion exchange can be treated as short ranged, so that the theory contains only contact interactions. Because the number of terms allowed by the symmetry is infinite, it is important to order these terms so that observables
can be calculated in an expansion in powers of a small parameter, given by the ratio of the relevant momentum $Q$ divided by $M_{QCD}\sim 1$ GeV. This is called power counting. A truncation of this expansion at any given order must respect renormalization-group (RG) invariance except for small errors contained in higher orders. In nuclear physics, the leading order (LO) terms must contain non-perturbative physics to obtain nuclear bound states and resonances, while subleading-order corrections, if truly corrections, should be treated in perturbation theory. The interested reader can find more details in other publications (see, e.g., \cite{eftreview}).

\begin{figure}
\centering\includegraphics[clip,scale=0.4]{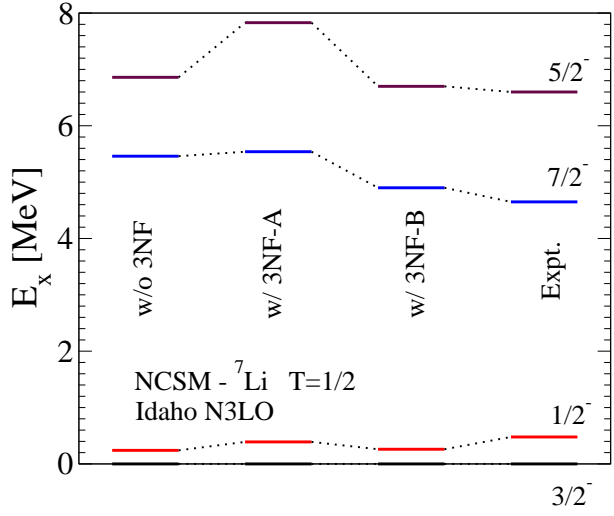}
\caption{Excitation energy of the lowest states of $^7$Li using two-body interactions only as well as two- plus three-nucleon interactions (3NF), with two determinations A and B of the low-energy constants $c_E$ and $c_D$, compared with the experimental data. Figure taken from Ref. \protect \cite{Nogga:2005hp}, courtesy A. Nogga.}
\label{Stetcu_fig:Li7ex}
\end{figure}

In most present applications one uses the so-called  ``Weinberg power counting," in which all the irreducible diagrams are summed to infinite order. Although this approach is disputed \cite{Kaplan1998}, it is practical. Furthermore, there is indication that for small enough cutoffs (about 500 MeV), necessary for the renormalization of the Lipmann-Schwinger equation, the shortcomings of this approach can be minimized \cite{Fleming2000}. In Ref. \cite{Nogga:2005hp}, Nogga et. al. have used the two-body Idaho-N3LO interaction \cite{N3LO}, derived in the ``Weinberg" scheme, as an input to NCSM in order to compute low-energy properties of $^7$Li. In addition to the two-body forces, with contributions up to $(Q/M_{QCD})^4$, the three-body interactions appearing at the $(Q/M_{QCD})^3$ order \cite{vanKolck:1994,*Epelbaum:2002} have been included. The later contain, in addition to the two pion-exchange term, two three-body contact interactions with or without one-pion exchange involving two undetermined low-energy constants, $c_D$ and $c_E$ respectively, that need to be adjusted to experimental data. Two determinations have been done for these strengths: (A) a fit to the triton binding energy and and the N-deuteron doublet scattering length and (B) a fit to the $^3$H and $^4$He binding energies. Three sets of calculations have been performed in NCSM: one in which only two-body forces have been included and two in which, in addition to the two-body interactions, the two determinations for the low-energy constants (A) and (B) in front of the contact terms have been used. The effective interactions derived within the three-body cluster approximation described in Sec. \ref{Stetcu_sec:ut} were then used to predict the low-lying states in $^7$Li. The best agreement of the predicted binding energy with experiment has been found for the case (A) (about 38 MeV). As shown in Fig. \protect\ref{Stetcu_fig:Li7ex}, all three Hamiltonians used as input predict the correct ordering of the states. However, some splittings between low-lying states are significantly influenced by the choice of the starting Hamiltonian. In particular, the splitting between 7/2$^-$ and 5/2$^{-}$ is best described by choice (B), in which the binding energy is only 36.7 MeV (experimentally, the $^7$Li binding energy is 39.2 MeV). While these results might be viewed as a undesirable model dependence, we point out that small deviations are expected due to the exclusion of higher order terms and should be reduced even further when higher order are included in the initial interactions.

\begin{figure}
\centering\includegraphics[clip,scale=0.35]{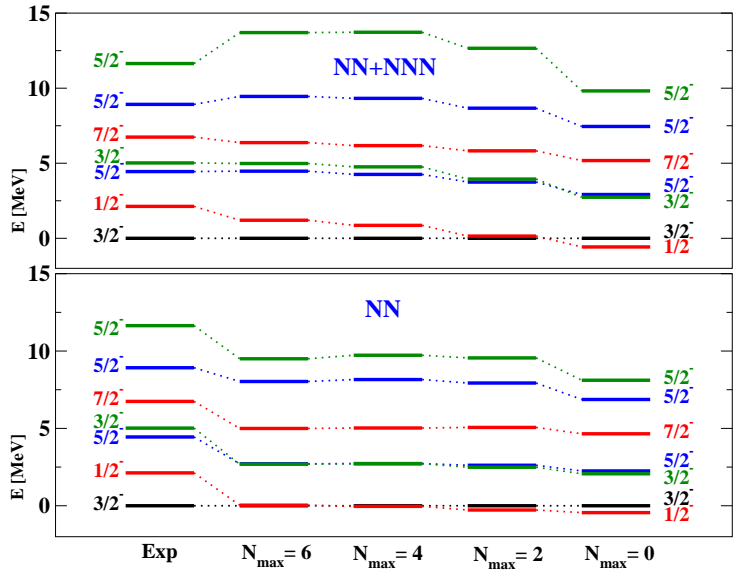}
\caption{Theoretical and experimental natural parity excitation spectra of $^{11}$B. Results with NN and NN+NNN interactions are presented as a function of the model space truncation $N_{max}$ for the HO frequency of 15 MeV. Figure taken from Ref. \protect\cite{Navratil:2007}, courtesy P. Navr\'atil.}
\label{Stetcu_fig:11Bex}
\end{figure}

The freedom of choice regarding the fitting procedure of the low-energy constants has been investigated further in Ref. \cite{Navratil:2007}, where $c_D$ and $c_E$ have been determined so that the binding energies of the $A=3$ systems ($^3$H and $^3$He) reproduce the experimental values. In Fig. \ref{Stetcu_fig:11Bex} we present the prediction of the excitation spectra in $^{11}$B for a particular choice of the combination of the two low-energy constants that also reproduces the binding energy for the alpha particle. We observe the good convergence properties with the model space truncation, especially for the low-lying states. This figure also illustrates the overall improvement in the theoretical spectra when three-body interactions are included. In particular note the lowest 1/2$^-$ and 3/2$^-$ states, whose converged values are basically degenerated when only two-body interactions are included, while the addition of the three-body terms produces a splitting that is consistent with the experimental data. The same holds for the (3/2$^-$, $5/2^-$) doublet. In addition, a large number of other energy levels and electromagnetic properties of select p-shell nuclei have been investigated and found in reasonable agreement with experiment \cite{Navratil:2007}.

\begin{figure}
\centering\includegraphics[clip,scale=0.5]{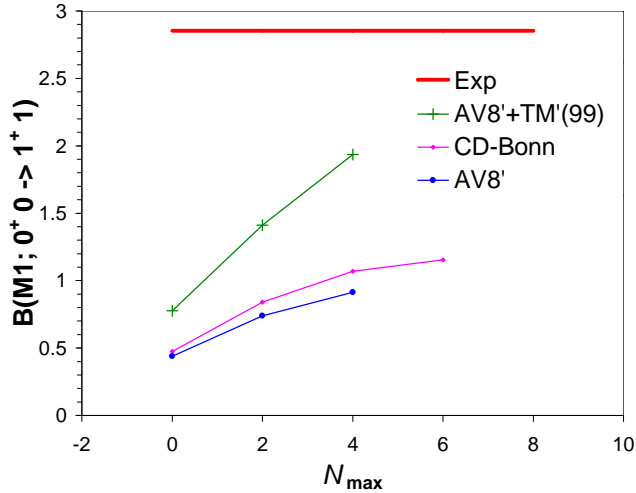}
\caption{B(M1), in $\mu_ N^2$, as a function of model-space truncation $N_{max}$, for the the $0^+ 0\rightarrow 1^+ 1$ transition in $^{12}$C. Figure taken from Ref. \protect\cite{Hayes:2003ni}, courtesy A.C. Hayes. }
\label{Stetcu_fig:c12bm1}
\end{figure}

As expected, the inclusion of three-body interactions has an important effect for the description of other observables. This was nicely demonstrated in Ref. \cite{Hayes:2003ni}, where Hayes et. al. compared the convergence pattern of inelastic electron scattering to the 15.11 MeV state of $^{12}$C, muon capture to the 
ground state of $^{12}$B, and neutrino scattering to the ground  state of $^{12}$N, explicitly showing how the theoretical predictions improve with the inclusion of three-body forces. Thus, as shown in Fig. \ref{Stetcu_fig:c12bm1}, if only two-body forces are included in the Hamiltonian, the M1 strength converges to about one third the experimental value. However, the introduction of the phenomenological Tucson-Melbourne three-body interaction (TM$^\prime$(99)) \cite{TMprime99} significantly increases the strength, although the results are far from being convergent in model spaces accessible at the time (note the large model space dependence in this case). The improvement is  associated with the improved strength of the spin-orbit splitting when genuine three-body interactions are included.

Except for energy, most of the calculations (including those in Fig. \ref{Stetcu_fig:c12bm1}) involving observables did not take into account the renormalization of general operators, as it is required for consistency. Before 2004, calculations involving effective operators have been scarce and limited to scalar operators \cite{Navratil:2000gs,benchmark:4he_gs}, which have the same transformation properties as the Hamiltonian. The first implementation of the effective operator approach discussed in Sec. \ref{Stetcu_sec:eff_op} to general non-scalar operators has been introduced in Ref. \cite{Stetcu:2004wh}. Like for the Hamiltonian, the renormalization was implemented in relative coordinates, as long as the operators could be written in relative coordinates \cite{Stetcu:2004wh}. As a first test, the quadrupole moment of the deuteron has been calculated within the two-body approximation, which is exact for the two-body system. Calculations in a $4\hbar\omega$ model space yielded 0.179 $e$ fm$^2$ when the bare operator was used, while the value of 0.270 $e$ fm$^2$, was obtained using the corresponding effective operator in the same model space. The later is in excellent agreement with independent calculation of the deuteron quadrupole moment for wave functions obtained with the same two-body potential. Applying the same method to more than three particles has yielded a different result. Thus, a very small renormalization has been obtained even in small model spaces for B(E2). For example, in $^6$Li, if a bare E2 operator is used  B(E2; $3^+1 \to 1^+1$)=2.647 $e^2$ fm$^4$, while the same quantity in the same model space gives 2.784  $e^2$ fm$^4$ if the renormalization procedure is performed \cite{Stetcu:2004wh}. Both calculations have used the Argonne V8$^\prime$ NN potential. Using the CD-Bonn 2000 NN interaction, the same observable calculated using the bare operator in $10\hbar\omega$ yields 10.221 $e^2$ fm$^4$, expected to be comparable with the results obtained with Argonne V8$^\prime$. Overall, the difference between the bare operator results in the $2\hbar\omega$ and $10\hbar\omega$ model spaces, coupled with the small renormalization at the two-body cluster level, indicate sizable effective many-body effects needed to correct the $2\hbar\omega$ B(E2) value. 

\begin{figure}
\centering\includegraphics[clip,scale=0.8]{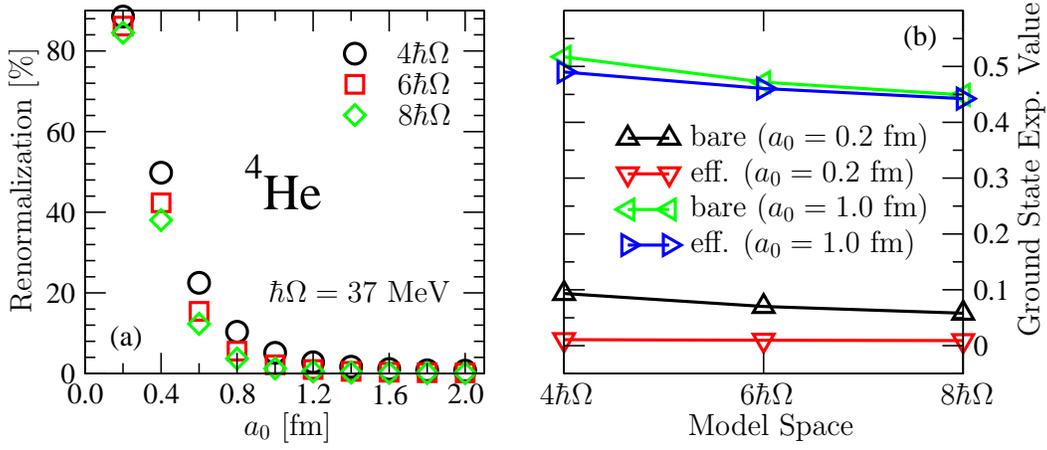}
\caption{Left panel: renormalization of the ground state expectation value of the relative Gaussian operator using realistic wave functions for $^{4}$He, as a function of the range of the operator for $4\hbar\Omega$ (circles), $6\hbar\omega$ (squares), and $8\hbar\omega$ (diamonds). Right panel: expectation values in different model spaces for Gaussian operators of selected ranges. Figure from Ref. \protect\cite{Stetcu:2004wh}.}
\label{Stetcu_fig:gaussianOp}
\end{figure}

In order to better understand the renormalization behavior in the case of general operators, a test case has been devised in Ref. \cite{Stetcu:2004wh}, where a two-body Gaussian scalar operator of variable range $a_0$ has been considered. Mathematically, the operator writes
\begin{equation}
O(\vec r_1,\vec r_2)=\frac{1}{\pi^{3/2}a^3_0}\exp\left(-\frac{(\vec{r}_1-\vec{r}_2)^2}{a_0^2}\right).
\label{gaussopdef}
\end{equation}
While clearly not a realistic observable, this operator can be used to observe the behavior of the renormalization operator with the range. The left panel of Fig. \ref{Stetcu_fig:gaussianOp} illustrates this dependence, showing a clear contrast between the renormalization of the operators of short range, whose values change by up to 90\% for the shortest ranges considered, and  that of the long-range operators that present little or no renormalization. The right-hand panel demonstrates that, in the case of a short range operator ($a_0=0.2$ fm), the renormalization produces observables that are model space independent to a large extent, while the results obtained with the bare operator vary significantly with the cutoff. In contrast, for $a_0=1$ fm, there is little contrast between the expectation value obtained with either the effective or bare operator, even though, as expected, the renormalization is more seizable in the smaller model space. Hence, one can immediately draw the conclusion that the unitary transformation at the two-body cluster level strongly (and effectively) renormalizes short-range operators, while long-range operators are largely unaffected by the procedure. This further reinforces the assumption that the two-body cluster approximation renormalizes the short-range part of the interaction, while the long-range and the many-body effects have to be incorporated by increasing the model space. Obviously, the long-range operators are going to require large spaces to converge, as they are sensitive to the correct description of the asymptotic of the wave function. The same behavior has been confirmed within a SRG renormalization of general operators in Ref. \cite{PhysRevC.82.054001}.

\begin{figure}
\centering\includegraphics[clip,scale=0.65]{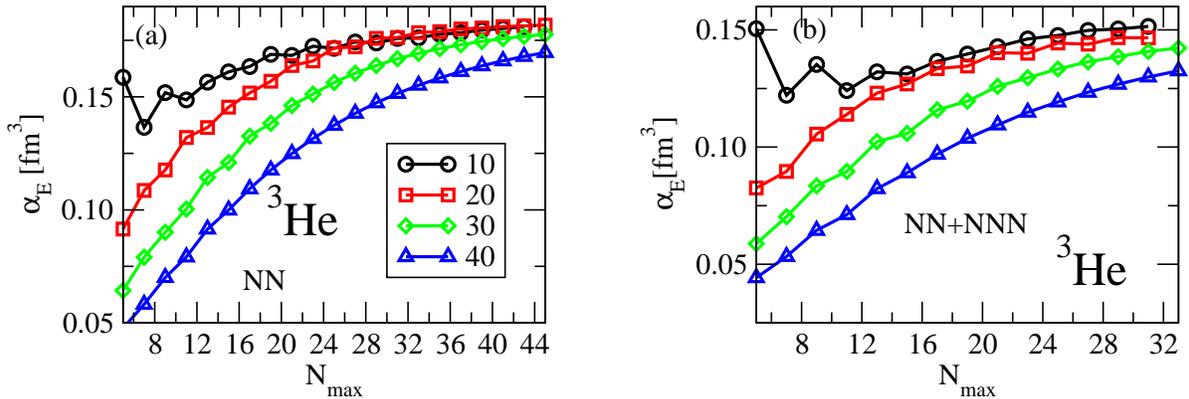}
\caption{The running of the $^3$He electric polarizability with $N_{max}$. In the left panel we show the results including two-body forces only, while in the right panel we include both two- and three-body interactions, thus obtaining an excellent description of the ground-state energy. The polarizability results are in good agreement with previous calculations. Figure from Ref. \protect\cite{Stetcu:polarizHe}.}
\label{Stetcu_fig:alphaE}
\end{figure}

In very few cases, it is possible to increase the model space enough so that reasonable convergence can be achieved, even if the operators are not consistently renormalized. This is mostly the case of the $A=3$ system, where in relative coordinates one can increase $N_{max}$ to large values. As an example, we illustrate in Fig. \ref{Stetcu_fig:alphaE}, the dependence of the $N_{max}$ for the electric polarizability $\alpha_E$ defined by

\begin{equation}
\alpha_{\rm E} = 2 \alpha \; \sum_{N \neq 0} \; \frac{| \langle N |  D_z | 0 \rangle |^2}{E_N - E_0} \, ,
\label{alphaE_def}
\end{equation}
where $\alpha$ is the fine-structure constant, $E_0$ is the energy of the ground-state $|0\rangle$, $E_N$ is the 
energy of the N${\underline{th}}$ excited state, $|N\rangle$ (all of which are in the continuum for few-nucleon systems), and $D_z$ is the component  of the (non-relativistic, in our case) electric-dipole operator in the $\hat{z}$ direction, which generates the transition between those  states. Two model Hamiltonians have been used in this calculation: one which include only EFT derived NN interactions at next-to-next-to-next-to-next order \cite{N3LO}, and one where the additional three-body interactions in the local form \cite{Navratil_FBS} ensure a correct description of the binding energy.  As shown, results obtained using different frequencies eventually converge to the same value, even if the operator has not been transformed in the same way as the Hamiltonian. In fact, as Fig. \ref{Stetcu_fig:gaussianOp} demonstrates, for a long range operator like the electric dipole, there is little to gain by applying the unitary transformation. Finally, a lower bound for the error can be estimated by using the small variation of the observable in the largest model space for different frequencies.

In most of the cases, however, it is impossible to follow the same kind of convergence for all the observables. An assessment regarding the convergence of a certain observable is often performed by taking into account the change with respect the size of the model spaces. Extrapolation procedures, based on simple assumption of an exponential dependence on $N_{max}$, have been devised for both energy levels \cite{PhysRevC.77.024301} and observables \cite{PhysRevC.79.021303}. The fitting involved in the procedure allows another estimation of the error for each observable considered. However, it should be noted that the error strongly depends on the particular form chosen for the dependence of $N_{max}$, and, assuming  the form to be correct, whether $N_{max}$ is large enough so that one can reliably extract the parameters that provide the asymptotic values. Furthermore, like in the cases where the error is assessed from the variation in the largest model spaces, this represents just a lower bound. Systematic errors that can arise, for example, from the omission of relevant terms in the Hamiltonian are harder to assess. In principle, EFT provides the framework for error estimate and systematic improvement, even though it is still difficult to estimate the errors in systems with more than two particles.

\subsection{Single-shell effective interactions and operators in NCSM}
\label{Stetcu_sec:core}

The successful application of NCSM to the description of properties of light nuclei motivates the effort to attempt an \textit{ab initio} description of nuclear properties in medium mass nuclei. In the past years, several many-body methods, such as the importance truncated no-core shell model \cite{Roth:2007sv}, the \textit{ab initio} shell model with a core \cite{NCSMcore,NCSMcore_op,Navratil:1996jq}, group-theory approaches \cite{PhysRevLett.98.162503,*PhysRevC.76.014315,*Dytrych:2008}, and others \cite{Roth:2011fk}, have been developed with the goal of extending the NCSM description to medium-mass nuclei. In the following we discuss the NCSM with a core because the goal of the current paper is a discussion of derivation of effective interactions rather than many-body truncation methods on which Refs. \cite{Roth:2011fk,Roth:2007sv} are based. In addition, this method is naturally connected with the phenomenological shell model discussed in Sec. \ref{Stetcu_Sec:phSM}. We note that, for simplicity, only two-body interactions have been included in the starting Hamiltonian. Hence, a direct comparison with the experimental data is a bit problematic and  in order to mitigate this shortcoming the INOY (inside nonlocal outside Yukawa) interaction \cite{INOY:2003,*INOY:2004} has been employed. This NN interaction minimizes the need for three-body forces in the three-body systems, reproducing the NN data with the same accuracy as the other NN potentials. Even for $^6$Li, INOY interaction provides a reasonable description of the experimental data, as illustrated in Fig. 1 of Ref. \cite{NCSMcore}. However, for a complete \textit{ab initio} description, three-body interactions should be included in future development.

The approach discussed in Sec. \ref{Stetcu_Sec:phSM} has the advantage that, even for nuclei as heavy as mid $pf$ shell (e.g., $^{56}$Fe), one can perform a full space calculation with modest computing power for today standards and still obtain an excellent agreement with the experiment. The reason is a strong restriction of the model space to only one HO shell as well as a limitation of the active nucleons to a small number.  The drawback is the phenomenological fitting of the single shell interaction used in the calculation. NCSM provides the opportunity to derive such an interaction from an underlying NN (and NNN) interaction, with the rigorous inclusion of a large amount of many-body effects (e.g., core polarization). 

While the method has been introduced by Navr\'atil et. al. in 1996 in order to the derive electromagnetic effective charges in one shell \cite{Navratil:1996jq}, a comprehensive study of the NCSM-derived single-shell effective interactions and operators has been performed only more recently by Lisetskiy et. al. \cite{NCSMcore,NCSMcore_op}. The method is based on a unitary transformation approach performed in two steps. The first step is identical to the usual NCSM approach (illustrated in the previous section), allowing the calculation of converged $A$-body states in large model spaces. In the second step, an $A$-body transformation to a $0\hbar\omega$ model space is performed, and in this model space all $A$-body correlations between nucleons are included. However, because most of the nucleons in $0\hbar\omega$ will be confined to the closed shell, the many-body Hamiltonian will contain lower rank interactions. 

In order to better understand this method, let us take $^6$Li and $^6$He. In this case, after the first step, a six-body solution can be obtained in a large model space. Since the eigenfunctions have been calculated in the large mode space, using Eq. (\ref{Stetcu_eq:omega}) one can calculate the six-body operator $\hat \omega$ and hence the effective interaction in $0\hbar\omega$. Like for the two-body cluster for the deuteron, the six-body effective interaction is exact in this case. However, in $0\hbar\omega$ model space, for $^6$Li and $^6$He one has to confine two protons and two neutrons to the 0s shell (forming a ``core"), allowing only two nucleons move in the p-shell, which in analogy with the phenomenological shell model, we label as ``valence." Thus, the only interactions allowed in the model space are one- (between the valence nucleons and core) and two-body interactions (between the valence nucleons), even after the exact six-body cluster calculation. The same procedure can be applied to $^5$He and $^5$Li, which provide the equivalent of single-particle energies in the phenomenological shell model, as well as to $^4$He, which provides the reference energy of the closed shell nucleus. 

\begin{figure}
\centering\includegraphics[clip,scale=0.2]{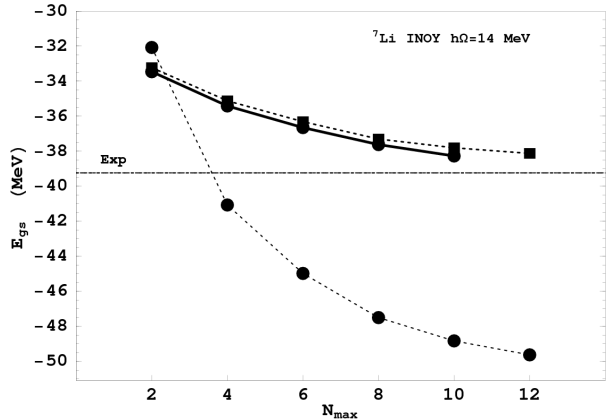}
\caption{The ground-state energy of $^7$Li calculated in the two-body valence cluster approximation discussed in the text (filled circles connected with full line), which include exactly up to six-body correlations. $N_{max}$ represents the truncation of the model space in which four-, five- and six-body solutions have been obtained. Figure from Ref. \protect\cite{NCSMcore}.}
\label{Stetcu_fig:7Licore}
\end{figure}

Combining the information from $A=4, 5$ and 6 nuclei, taking special care as to not double-count contributions, one can construct a Hamiltonian in the six-body approximation that can be further used in order to predict the $^7$Li ground-state spectrum in $0\hbar\omega$. We show the ground-state energy (marked with square symbols) for a fixed HO frequency in Fig. \ref{Stetcu_fig:7Licore}. The convergence is displayed as a function of the model space used in the first transformation in order to obtain four-, five- and six-body solutions. The  ground-state results in the conventional NCSM approach, in which the seven-body calculation is performed in larger and larger model spaces, is displayed with full circles connected by a full line. The agreement between the two methods is remarkable. Note that in this approach, the core contribution as well as one- and two-body terms become $A$-dependent, because the HO CM term added to the Hamiltonian, as discussed in Sec. \ref{Stetcu_Sec:NCSM}, corresponds to the targeted nucleus (in this case $^7$Li). In contrast, while in the Wildenthal interaction \cite{wildenthal} the two-body matrix elements include a mass dependence, the core and the single particle energies are fixed in phenomenologically derived interactions. A similar approach within the NCSM method, marked by full circles connected by dotted lines in Fig. \ref{Stetcu_fig:7Licore}, shows the need for stronger and stronger repulsive three-body forces with increasing the underlying model space, used to derive the effective interaction. Furthermore, although the effect is smaller, the $A$-dependent core and interactions, also provide better description of the excited state energies \cite{NCSMcore}.

Despite the remarkable the agreement between the two theoretical predictions in Fig. \ref{Stetcu_fig:7Licore}, one can expect an increasing importance of the neglected many-body forces, caused by the drastic reduction of the model space to one single major shell. This is indeed observed for $A>7$, where single-shell calculations suggest that three- and higher-body interactions play an increasingly important role, and even the partial inclusion of induced three-body forces considerably improve the description \cite{NCSMcore}. 

One of the puzzles of the phenomenological shell model is the good description of the observed electromagnetic transition strengths, using the full-space transition operators complemented by a simple rescaling of the electromagnetic charges. In order to investigate the renormalization properties of the quadrupole operator, the NCSM approach to the single-shell interaction was also extended to operators \cite{NCSMcore_op}. In this framework, higher-body correlations (up to six bodies, in this case) have been included in the procedure, like in the case of the Hamiltonian. Five-body renormalization of the $^5$Li and $^5$He systems yields the one-body contribution, while the two-body part is obtained from the renormalization of $^6$Li. Calculations of different matrix elements in the six-body system have shown that the one-body term contributes significantly more than half to the total strength. In the case of isoscalar transitions, the two-body terms become important, thus suggesting a sensitivity to higher-body correlations in general. In contrast, for isovector transitions, the two-body contributions remain relatively small. 

The NCSM approach to single-shell interactions and operators could be further extended to different directions. First, for a consistent derivation of the interactions, it is important to include genuine three-body interactions in the initial Hamiltonian used to obtain the initial NCSM solutions in large model spaces. Second, the same approach can be extended to more challenging mass regions, like the sd shell, where the applicability of the traditional NCSM is very limited. In this case, it is conceivable that an implementation of the unitary transformation proposed by Johnson \cite{Johnson2010:pei}, tested for trapped cold atoms, and mentioned in Sec. \ref{Stetcu_Sec:phSM}, could improve the description of a large number of observables. However, considerable computational power is still required in order to obtain converged solutions for nuclei with one and two nucleons above the closed shell. 

\section{NCSM as an effective theory}
\label{Stetcu_Sec:NCSMasEFT}

In the previous section we have discussed a derivation of the effective interactions in truncated model spaces based on the use of a carefully designed unitary transportation. The starting point was a given ``bare" interaction, so that the results depend in general upon the model Hamiltonian chosen. In addition, because it is practically impossible to compute all the induced interactions, which contain up to $A$-body terms, one is forced to adopt the cluster approximation that neglects interactions of rank greater than a chosen order. Finally, the description of low-momentum observables requires large model spaces, which makes the application of the method to heavy nuclei extremely challenging.

In EFTs, the restriction to a model space generates all the interactions allowed by the underlying symmetries \cite{kaplan}. In the traditional approach in a continuum basis, the particle momenta are limited within the restricted space, one can treat short-distance interactions in a derivative expansion, similar to the multipole expansion in classical electrodynamics. The coefficients of this expansion, called low-energy constants (LECs), carry information about the details of the short-range dynamics. LECs change with the model space in such a way that low-energy observables remain (approximatively) independent upon the size of the model space.   

Because the derivation of inter-nucleon interactions is based on the symmetries of the QCD Lagrangean, the  EFT approach provides a modern understanding of the nuclear forces at low energies \cite{eftreview}, even in the absence of exact solutions from QCD. In particular, the appearance of light pions is the immediate consequence of the small and spontaneous breaking of chiral symmetry. Hence, it is not difficult to write the most general Hamiltonian with the appropriate degrees of freedom and symmetries \cite{weinberg1990,*weinberg1991,*ordonez1992}, which is a generalization to systems with more than one nucleon of the Lagrangian used in chiral perturbation theory. Interactions among nucleons consist of pion exchanges and contact interactions, which model the short-range dynamics like, for example, the exchange of heavier mesons. For phenomena involving momenta much lower than the pion mass, even pion exchange can be treated as short ranged, leaving only contact interactions (and their derivatives) in the theory \cite{pionless,*pionless2}.

Including an infinite number of interactions is clearly not practical. However, they can be organized as an expansion of the relevant momentum scale over the nucleon mass, $Q/M_N$.  Such an organization, called power counting, allows for a consistent truncation of the relevant terms in the nuclear Hamiltonian. If desired, the precision can be systematically improved by adding higher order terms. This approach had mostly been applied in particle physics to systems where unitarity could be accounted for perturbatively. In nuclear physics, the leading order (LO) must contain non-perturbative physics to generate nuclear bound states and resonances, while the subleading-order corrections, should be treated in perturbation theory. 

If the power counting of the theory with pions is debated and still being explored \cite{nogga}, the power counting is well established for the theory in which the pions are integrated out (pionless theory) \cite{pionless,*pionless2}. Thus, at low enough energies, P and higher partial waves can be neglected at low orders in the two-body sector. Given that the S-wave two-nucleon scattering length is much larger than the range of the nuclear force, set by the pion mass, one can formulate an interparticle potential as a series of contact interactions with an increasing number of derivatives \cite{pionless,*pionless2,Bedaque:1998kg,*Bedaque:1998km,*Platter:fk}. Applications of the pionless theory have demonstrated excellent results results for $^2$H \cite{rupak}, $^3$H \cite{triton_eft,*triton_eft2,*triton_eft3,*PhysRevC.69.034010}, and even the ground state of $^4$He \cite{Platter:2004zs}. With trivial modifications, this EFT has also proved useful for atomic and molecular systems with large scattering lengths \cite{hammerrev}. While continuum momentum-space calculations are  considerably  simplified in this EFT, they are still quite involved beyond the four-body system. Attempts have been made to derive bulk properties of matter using a spatial lattice \cite{lattice}, but the limit of applicability of the pionless theory (which should break down if the momenta involved approach the pion mass) with increasing density is at present unknown.

In this section, we formulate the NCSM as an effective theory. In such an approach, the interaction preserves the form in each model space, as dictated by the power counting. Truncation to a certain order involves including all interactions up to a certain rank, hence justifying the cluster approximation. The effective operators describing interactions with external probes can be consistently treated within the same framework.

Because the contact interactions are singular, an ultraviolet (UV) momentum cutoff $\Lambda$ has to be introduced in order to solve the Schr\"odinger equation. This is natural in a continuum basis formulation of EFT, but in NCSM the cutoff has to be formulated in terms of the HO parameters. Thus, since $(N_{max}+3/2)\omega$ is the maximum energy allowed in the truncated model space, we define $\Lambda$ as the momentum associated with this energy in the relative coordinate of two particles:
\begin{equation}
\Lambda=\sqrt{M_N (N_{max}+3/2)\omega}.
\label{Stetcu_eq:Lambda_def}
\end{equation}
Additionally, the HO frequency sets the spacing between HO levels and provides an infrared energy cutoff $\hbar\omega$ or, equivalently, a momentum cutoff $\lambda = \sqrt{M_N\hbar\omega}$. Other authors have defined the infrared momentum which corresponds to the maximal radial extent needed to encompass the many-body system to be described \cite{PhysRevC.82.034330,PhysRevC.83.034301}, with demonstrated benefits \cite{Coon:2012ab}. While it would be interesting to revisit the calculations of Ref. \cite{Stetcu:2006ey} using the more recently proposed infrared cutoff definition, in this paper we restrict ourselves to $\lambda = \sqrt{M_N\hbar\omega}$, which was initially proposed. 

A model space is defined by the two cutoffs. In the following, we will consider two types of applications. In one type, the system is placed on a trap, and therefore the infrared cutoff is physics and the solutions depend explicitly upon $\lambda$. In the second type of applications, we consider untrapped few-nucleon systems. In this case, since $\Lambda = \lambda \sqrt{Nmax + 3/2}$, the running of the observables with $\Lambda$ cannot be obtained by increasing $\hbar\omega$ arbitrarily in a fixed-$N_{max}$ model space, as this procedure increases the infrared cutoff as well, introducing additional errors. Instead, we verify explicitly that cutoff dependences decrease with increasing $\Lambda$ and decreasing $\lambda$, and we remove the influence of the infrared cutoff by extrapolating to the continuum limit, where $\hbar\omega\to 0$ with $N_{max} \to \infty$ so that $\Lambda$ is fixed. Traditional shell-model calculations use larger values for the HO frequencies, of the order of $41/A^{1/3}$ MeV, but in this approach we are interested in a small infrared cutoff limit, which removes the HO frequency dependence.

The EFT method of deriving effective interactions can be applied to either pionfull or pionless EFTs. For simplicity, all the applications to nuclear systems and cold atoms have been limited to pionless EFT. In the following we discuss various applications to nuclear and atomic systems.

\subsection{Light nuclei in EFT framework}
\label{Stetcu_Sec:EFT_NCSM}

Attempts to introduce EFT methods into the shell model date back to 2000  \cite{haxtonSMasEF,Haxton:2001,*Haxton:2002kb,*Haxton:2008}. Such an approach in based, like the unitary transformation method, on underlying NN (and NNN) interactions, thus suffering from the same model dependence like the conventional NCSM. While initially only a contact-gradient expansion modeled after EFT has been employed in the derivation of the effective interactions \cite{haxtonSMasEF}, the approach was later supplemented by the addition of an exact summation of the relative kinetic energy, which accounts for the long-range behavior coming from the excluded space. The result is an energy-dependent effective interaction, which, to our knowledge, has been applied only to the description of the deuteron and three-particle systems \cite{Haxton:2001,*Haxton:2002kb,*Haxton:2008,luu:103202}. In this section, however, we present an approach based entirely on EFT principles, in which the Hamiltonian is defined according to power counting directly in the model space, without knowledge of the excluded space whose contribution is encoded in the LECs.

In LO pionless theory, the Hamiltonian is composed of the relative kinetic energy and two two-body contact interactions in the $^3S_1$ and $^1S_0$ NN channels \cite{pionless,*pionless2}, with corresponding parameters $C_0^1$ and $C_0^0$, and one three-body contact interaction that appears in the NNN $S_{1/2}$ channel \cite{triton_eft,*triton_eft2,*triton_eft3,*PhysRevC.69.034010}, with the corresponding strength parameter $D_0$:
\begin{eqnarray}
\lefteqn{H_{int}=\frac{1}{2M_NA}\sum_{i,j}(\vec p_i-\vec p_j)^2 \nonumber }\\
& &+C_0^0(\omega,N_{max})\sum_{[i<j]^0}\delta(\vec r_i-\vec r_j)+C_0^1 (\omega,N_{max})\sum_{[i<j]^1}\delta(\vec r_i-\vec r_j) \nonumber \\ 
& & +D_0(\omega,N_{max})\sum_{[i<j<k]}\delta(\vec r_i-\vec r_j)\delta(\vec r_j-\vec r_k),
\label{ham}
\end{eqnarray}
where $[i<j]^s$ denotes pairs of particles in the $S$-wave NN channel of spin $s$ and $[i<j<k]$ triplets of particles in the spin-$1/2$ $S$-wave 3N channel. The LECs depend upon the truncation $N_{max}$ as well as the HO frequency, and, in order to achieve RG they have to be adjusted in each model space so that one preserves the physical observables. The form of the interaction us fixed (i.e., matrix elements of the contact interactions in all model spaces), which represents a clear departure from the unitary transformation approach, in which the structure of the interaction changes from one model space to the other. In the continuum basis approach, the two-body parameters are fixed to two-body observables, like the deuteron binding energy in the $^3S_1$ channel, and two-body scattering phaseshifts. However, because the NCSM basis states are built using bound states only, the direct connection with the continuum observables is difficult. Only $C_0^1(\omega,N_{max})$ can be directly fixed in each model space to reproduces the deuteron binding energy, so alternate few-body observables have to be considered in order to determine $C_0^0(\omega,N_{max})$ and $D_0(\omega,N_{max})$. Hence, the $^3$H and $^4$He binding energies have been used to simultaneously fix the remaining LECs in each model space.

\begin{figure}
\centering
\includegraphics[scale=0.82]{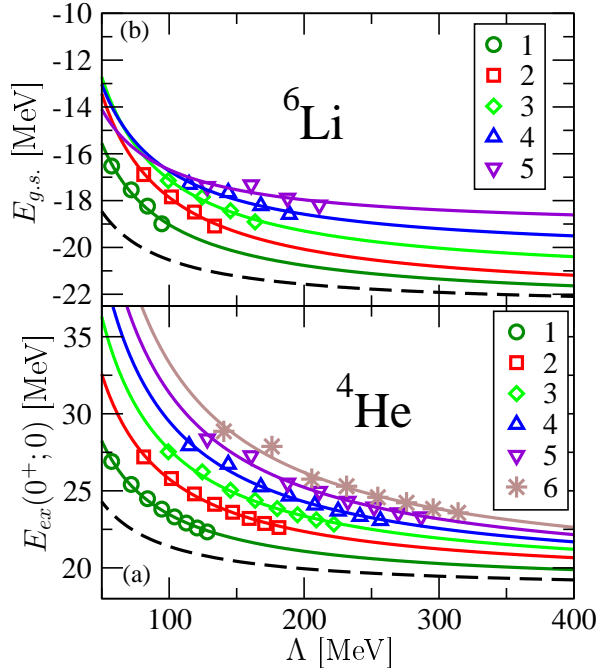}
\caption{Running with the ultraviolet cutoff for two observables: the first $(0^+;0)$ excited state in $^4$He (lower panel) and the $^6$Li ground-state energy (upper panel). The discrete points represent calculations for different frequencies, denoted in the legend in MeV, while the continuous lines represent a fit to a linear dependence of $1/\Lambda$ for fixed $\omega$. The dashed line represents the $\omega\to 0$ limit.}
\label{Stetcu_fig:lightnuclei}       
\end{figure}

With the coupling constants thus determined, one can solve the many-body problem, predict other states, and compare against the experimental data. Thus, in Fig. \ref{Stetcu_fig:lightnuclei}, the energy of the first $(J^\pi;T)=(0^+;0)$ state in $^4$He (lower panel) and the ground-state energy of $^6$Li are plotted vs. the UV cutoff, for fixed IR cutoff. As discussed earlier, we are interested in the limit of large $\Lambda$ and small $\lambda$. Very large $N_{max}$ calculations are prohibitive even for few particles, so one needs to rely on the extrapolation of the results obtained at accessible $N_{max}$. We assume that the energy of the state runs linearly with the inverse of the ultraviolet cutoff for a fixed infrared cutoff, $E(\Lambda,\omega) =E_0(\omega)+A(\omega)/\Lambda$. This particular choice, marked by the continuous lines in Fig. \ref{Stetcu_fig:lightnuclei}, is motivated by the same type of LO running of the bound-state energy in the two-body sector in the continuum \cite{pionless,*pionless2,rupak}. The IR dependence is removed by taking numerically the limit $\omega\to 0$ for a fixed $\Lambda$  (the dashed line in Fig. \ref{Stetcu_fig:lightnuclei}). Following this procedure, the predicted value for the the first excited state in $^4$He is within $10\%$ of the experiment. This represents an excellent agreement which can be easily explained by the proximity of that state to the four-nucleon continuum threshold, well within the regime of applicability of the pionless EFT. The $^6$Li binding energy is about 70\% of experiment \cite{Stetcu:2006ey}. 

One-photon exchanges between nucleons are non-perturbative only for momenta below $\alpha M_N\approx 7$ MeV, where $\alpha$ is the fine structure constant. Because the typical momenta of the bound states is considerably larger than $7$ MeV, the Coulomb interactions do not appear explicitly in LO. We used the observed $^4$He binding energy as a
fitting parameter; the difference between that and a Coulomb-corrected value is a higher-order effect. The Coulomb contribution, however, grows with the square of number of protons and so increases for heavier nuclei. One even expects an improvement of the $^6$Li results when Coulomb is included explicitly, since the fit to $^4$He assumes that the Coulomb repulsion is divided among all nucleon pairs. Hence, Coulomb effects are growing with the square of the mass number rather than the square of the number of protons. The effect, however, yields an underbinding of 2 MeV for $^6$Li, well within the expected size for subleading corrections.

Although the LO results are not as precise as the ones obtained from phenomenological potentials, they are consistent with QCD, with the information about QCD contained in the parameters. Moreover, the results are improvable order by order and adding next-to-leading order terms and beyond should improve the precision, as it happens in continuum calculations. Neglecting operators beyond the LO is expected to produce an error of about 30\%, so that the $^6$Li result was considered the first successful application of the pionless EFT to $A>4$ nuclei. Unlike for the calculations in the continuum, that there is an additional source of errors: the assumption that for fixed $\omega$ the energy runs like $1/\Lambda$, although a softer running cannot be excluded. Thus adding an extra $\log(a_0\Lambda)/\Lambda$ term leaves the $^4$He result virtually unchanged, but moves the result for the ground-state of $^6$Li within 15\% of the experimental result. However, more investigations are necessary in order to pin down the running with the ultraviolet cutoff.

Before a definitive conclusion can be drawn with respect to the applicability of the pionless EFT to the description of $^6$Li, one needs to demonstrate systematic improvement when operators beyond LO are included in the calculation. In the approach presented so far, i.e., a fit of LECs to few-body levels, the determination of coupling constants becomes quickly impracticable. It is, therefore, desirable, to identify a method to connect the two-body contact parameters to the scattering (or lattice) data. This is indeed possible if one places the interacting nucleons in a Harmonic trap, relating the \textit{bound-state} energy levels to the scattering phaseshifts \cite{HT}, very similar to the L\"uscher's formula on the lattice\cite{luscher}. We devote the rest of this section to the description of such an approach and review its applications to atomic and nuclear systems.

\subsection{Trapped atomic systems}

The area of ultra cold atomic systems has been the subject of great experimental progress in the past decades. The use of magnetic fields to create Feshbach resonances in cold, trapped atomic systems has opened up the possibility of dialing two-atom scattering lengths to values much larger than the typical range of the van der Waals potential, thus creating systems whose properties are universal, i.e., do not depend upon the details of the interaction. A ground-breaking achievement \cite{atomexpt,*stoferle:030401} is the ability to further confine just a few atoms in nearly isolated sites of optical lattices formed by laser beams. At low temperatures, the lattice sites may be considered as HO traps. The atomic systems with large scattering length display close similarities to nuclear physics systems at low energy. Hence, such atomic systems can be studied with techniques developed in nuclear physics and, conversely, provide an excellent testing ground the development of for few- and many-body methods that can be further applied, with little or no change, to the description of nuclear systems at low energies. In the following we show applications of EFT to two- and few-fermion system in a HO trap, and then further extend the approach to nuclear systems.

\subsubsection{Two-fermion in a trap in a EFT approach}
\label{Stetcu_sec:two_ferms_trap}

We consider a non-relativistic system of two spin 1/2 particles of reduced mass $\mu$ that interact
with each other in the S-wave only, although the extension to other partial waves is straight forward. In free space, at sufficiently low-energy, {\it{i.e.}} for $k$ much smaller than the inverse of the range of the interaction $R$, the phase shift $\delta_0(k)$ is given by the effective range expansion (ERE) \cite{beth}:
\begin{equation}
k \, \cot \delta_0(k) =-\frac{1}{a_2}+\frac{1}{2}r_2 k^{2}+ \ldots,
\label{ERE}
\end{equation}
with $a_2$, $r_2$, $\ldots$ , respectively, the scattering length, effective range, and higher ERE parameters not shown explicitly. The ERE is an expansion in powers of $kR$, which, for $k\ll 1/R$, is a small parameter. Generically, the sizes of ERE parameters are set by $R$, for example $|r_2| \sim R$, although the most interesting are the fine-tuned cases in which the scattering length is much larger than the range of the interaction. For such systems, a bound (virtually bound) state exists close to the threshold, largely independent upon the details of the potential.

The ERE produces model-independent results. Thus, if one needs to describe the phaseshift with a certain precision, the ERE can be truncated and the system described by a finite number of parameters. The precision can be always improved by adding higher orders in the ERE. Potentials that generate the same values for this finite number of ERE parameters cannot be distinguished at the fixed precision level and all generate the same wavefunction for distances beyond the range of the force, $r \gtrsim R$.

Unfortunately, ERE cannot be used to characterize systems with more than two particles. For this, it is necessary to go back to the notion of potential. For particles interacting in free space, it has been demonstrated \cite{pionless} that the ERE expansion (\ref{ERE}) can be reproduced  at each power of $kR$ by a potential $V$ constructed within  EFT as a Taylor series in momentum space. In coordinate space, this expansion for the potential writes:
\begin{eqnarray}
V(\vec{r}\,', \vec{r})
&=&C_0 \delta(\vec{r}\,') \delta(\vec{r})
  -C_2\left\{\left[\nabla\,'^2\delta(\vec{r}\,')\right] \delta(\vec{r})
    +\delta(\vec{r}\,') \left[\nabla^2\delta(\vec{r})\right]\right\}
         \nonumber\\
&&+ C_4 \left\{\left[\nabla\,'^4\delta(\vec{r}\,')\right] \delta(\vec{r})
	     +\delta(\vec{r}\,') \left[\nabla^4\delta(\vec{r})\right]
  + 2 \left[\nabla\,'^2\delta(\vec{r}\,')\right]
        \left[\nabla^2\delta(\vec{r})\right] \right\}   
+\ldots \label{taylorcoord}
\end{eqnarray}
where $C_0$, $C_2$, 
and $C_4$ are parameters,
and ``\ldots'' denote interactions that contribute at higher orders. The contact interactions (and their derivatives) in Eq. (\ref{taylorcoord}) are singular so that, when one solves the two-body problem with $V$, an ultraviolet (UV) cutoff $\Lambda$ has to be introduced. The parameters $C_i$ depend upon $\Lambda$ in such a way that observables are UV cutoff independent, as required by RG invariance. In the generic situation {\it{i.e.}} when the size of the ERE parameters are set by $R$, one can simply treat the whole potential in perturbation theory. However, in the more interesting cases when $|a_2|\gg R$, like in nuclear physics, the $C_0$ term in Eq. (\ref{taylorcoord}) needs to be solved exactly, while the remaining terms can still be accounted for in perturbation
theory \cite{pionless}. (Note that for the description of resonances, the first two terms would have to be treated to all orders.)

Let us assume that the same system of two particles is now confined by a HO potential characterized by the frequency $\omega$. In the zero-range interaction approximation, the energy eigenstates $\epsilon$ (in units of $\omega$) of this system are related to the S-wave scattering phase shift  in free space $\delta_{0}(k)$ by the transcendental equation \cite{NCSMeft_trap_2b,HT,DFT_sr,mehen}

\begin{equation}
\frac{\rm{\Gamma}\left(\frac{3}{4}-\frac{\epsilon}{2}\right)}
      {\rm{\Gamma}\left(\frac{1}{4}-\frac{\epsilon}{2}\right) }
=-\frac{b k}{2} \cot\delta_0(k),
\label{busch1}
\end{equation}
where $b=1/\sqrt{\mu\omega}$ is defined in terms of the reduced mass and $k=\sqrt{2\mu\omega\varepsilon}$ is the relative momentum of the two particles. Given that the trapped system has only bound states, Eq. (\ref{busch1}) provides a connection between observables in the trapped system and observables in the free space. This connection allows us to develop a EFT approach in which two-body parameters in the expansion (\ref{taylorcoord}) can be fixed by observables in the two-body system alone, as long as one traps the system. Theoretically, one can always trap a many-body system by adding a CM Hamiltonian, as described in Sec. \ref{Stetcu_Sec:NCSM}. However, a lot of experimental effort is directed toward trapped atomic systems, and, inspired by these developments, we first consider fermionic systems in external HO traps. For atomic systems the trap is physics, but when we apply the same methods to the description of self-bound nuclear systems, we remove the trap and take the continuum limit. This approach will be discussed later. For now, we concentrate on the description of trapped spin 1/2 fermions interacting via S-wave interactions. We will consider a large range of two-body scattering lengths $a_2$, as experimentally one can tune it over a large range of values, including around the Feshbach resonance. We also allow for other ERE parameters, but consider those set by the range of the interaction.

Using Eq. (\ref{ERE}) the spectrum of the 
two-fermion system (\ref{busch1})
in the trap can then be written as a function of the ERE parameters as
\begin{equation}
\frac{ \rm{\Gamma}\left(\frac{3}{4}-\frac{\epsilon}{2}\right)}
      {\rm{\Gamma}\left(\frac{1}{4}-\frac{\epsilon}{2}\right)} 
=\frac{b}{2 a_2} \left (1-\frac{a_2r_2k^2}{2}+\ldots \right )
\label{busch2}
\end{equation}
In quantum mechanics, the two-fermion system in the harmonic trap is described by the Hamiltonian 
\begin{equation}
H= -\frac{1}{2\mu}\nabla^2 + \frac{1}{2}\mu\omega r^2 + V,
\label{hamil}
\end{equation}
where the two-body interaction $V$ is constructed in such a way so that the spectrum of $H$ is indeed given by Eq. (\ref{busch2}). In the following, we use EFT principles to construct such an interaction at low energies.

Since here we are interested in systems with  large scattering length, corrections beyond 
the lowest order given by $C_0$ in Eq. (\ref{taylorcoord}) are treated as perturbations. For that we
 write the Hamiltonian (\ref{hamil}) for 
the relative motion as
\begin{equation}
H=H^{(0)}+V^{(1)}+V^{(2)}+\ldots,
\end{equation}
with the wavefunction and energy decomposed accordingly,
\begin{equation}
|\psi\rangle =|\psi^{(0)}\rangle +|\psi^{(1)}\rangle +|\psi^{(2)}\rangle 
+\ldots,
\label{wf}
\end{equation}
and 
\begin{equation}
E=E^{(0)}+E^{(1)}+E^{(2)}+\ldots
\label{E}
\end{equation}
This generic expansion is valid for all eigenvalues and eigenvectors, as long as the physics of the state is dominated by physics beyond the range of the interaction.
The superscript $^{(n)}$ corresponds to the order in perturbation theory of 
the different terms. The natural basis to solve this two-body problem is the set of HO eigenfunctions $\phi_{nlm}(\vec{r})$  with the corresponding eigenenergies  $E_{n}=(N+3/2)  \omega\equiv (2n+l+3/2) \omega$ where $N$ is the total number of quanta 
and $n$ the radial quantum number. 
Since we consider only interaction in the S-wave, the radial part of the solution of Eq. (\ref{hamil}) can be expanded
with the radial S-wave HO wavefunction 
\begin{eqnarray}
\phi_{n}(r)&=&\pi^{-3/4} b^{-3/2}
\left[L_n^{(1/2)}\left(0\right)\right]^{-1/2}
e^{-r^2/2b^2}
L_n^{(1/2)}\left(r^2/b^2\right), \label{ho_wf}
\end{eqnarray}
 where $L_n^{(\alpha)}$ is the generalized Laguerre polynomial.
 In this basis,
 the singularity of the potential (\ref{taylorcoord}) 
can be removed by taking into account a finite number of shells below the highest energy shell characterized by $n=n_{max}$. As discussed in Sec. \ref{Stetcu_Sec:EFT_NCSM}, this corresponds to having a UV momentum cutoff,
\begin{equation}
\Lambda=\frac{1}{b}\sqrt{2N_2^{max}+3}.
\label{Lambda}
\end{equation} 
with $N_2^{max}=2n_{max}$. Equation (\ref{Lambda}) is the equivalent of Eq. (\ref{Stetcu_eq:Lambda_def}), but it is more general as it allows for different particle masses. For an eigenstate $E_{\nu}$, the corresponding wavefunction $\psi^{(\nu)}(r)$ of the two-particle  system in the trap is expanded in this finite HO basis,
\begin{equation}
\psi^{(\nu)}(r)= \sum_{n=0}^{n_{max}} c^{(\nu)}_{n} \phi_{n}(r),
\label{expansion}
\end{equation}
with the coefficients $c^{(\nu)}_{n}$ to be determined.  

The  eigenenergies (\ref{E}) will depend on both $N_{max}$ and $\omega$, $E=E(N_{max},\omega)$.
Since $N_{max}$ is arbitrary, the energies should not depend sensitively on $N_{max}$. Although this cannot be achieved in general, it can for the shallow levels of interest, i.e., those which are dominated by the physics at distances $r>R$. As we show in the following,
this is accomplished
by allowing the $C_i^{(\nu)}$ to depend on both 
$N_{max}$ and $\omega$, $C_i^{(\nu)}=C_i^{(\nu)}(N_{max},\omega)$.
Nevertheless, at any order a residual  $N_{max}$ dependence 
introduces an error 
in the calculation of shallow levels,
which should be proportional to powers of $1/\Lambda$.
At the end of the calculation we want to take $N_{max}$
sufficiently large, $\Lambda \sim 1/R$,
so that the truncation error is smaller than the error proportional to powers of $R$ arizing from the truncation of Eq. (\ref{taylorcoord}).

The leading-order component $H^{(0)}$ of the  Hamiltonian $H$ (\ref{hamil}) is 
\begin{equation}
H^{(0)}=-\frac{1}{2\mu}\nabla^2 + \frac{1}{2}\mu\omega^2r^2
        + C_0^{(0)} \delta(\vec{r}),
\label{LOpot}
\end{equation}
with corresponding wavefunctions at LO solutions of the Schr\"odinger equation
\begin{equation}
\left(H^{(0)}-E^{(0)}\right)\psi^{(0)}(\vec{r})=0.
\label{sch0}
\end{equation}
By inserting Eq. (\ref{expansion}) into Eq. (\ref{sch0}) and using Eq. (\ref{ho_wf}) one can derive a relationship between the LO energy $E^{(0)}$ of any state and the coefficient $C^{0}_0(n_{max},\omega)$, that is \cite{Stetcu:063613,NCSMeft_trap_2b}
\begin{equation}
\frac{1}{C_0^{(0)}(n_{max},\omega)} = \frac{\mu}{2 \pi^{3/2} b  } 
\sum_{n=0}^{n_{max}} 
\frac{L_n^{(1/2)}(0)}{\frac{E^{(0)}}{2  \omega}- (n+\frac{3}{4})}. \label{C0_LO}
\end{equation}

In a given model space, $C_0^{(0)}(n_{max},\omega)$ can be adjusted to reproduce one energy of the two-body system. Although we can choose any energy of the trapped two-body system, since we are interested in low-energy properties, we can always fix the ground-state energy in the trap
\begin{equation}
E^{(0)}_0=E_0(\omega),
\end {equation}
so that $C_0^{(0)}(n_{max},\omega)$ is determined from Eq. (\ref{C0_LO}). In particular, at unitarity ($b/a_2=0$), if we choose to fix the ground state energy $E_0(\omega)= \omega /2$, the summation over $n$ in Eq. (\ref{C0_LO}) can be performed analytically \cite{alhassid2008}:
\begin{equation}
C_0^{(0)}(n_{max},\omega)=-\frac{\pi^2b}{2\mu}\frac{\Gamma\left(n_{max}+1\right)}{\Gamma\left(n_{max}+3/2\right)},
\end{equation}
so that in the limit $n_{max}\to\infty$
\begin{equation}
\frac{\mu C_0^{(0)}(n_{max},\omega)\Lambda}{2\pi}\to -\frac{\pi}{2},
\end{equation}
just like in the continuum calculations \cite{pionless,*pionless2}. This limit is independent upon the value of the scattering length, as illustrated in Fig. 1 of Ref. \cite{Stetcu:063613}.

While the ground-state energy is thus fixed to the exact value in all model spaces, the remaining energies will depend upon the truncation parameter $n_{max}$ and HO frequency, $E^{(0)}_{i\ge 1}=E^{(0)}_{i\ge 1}(n_{max},\omega)$
satisfy Eq. (\ref{C0_LO}) and in general depend not only
on $\omega$ but also on $n_{max}$. These energies should converge as $n_{max}\to \infty$ to 
finite values $E^{(0)}_{i\ge 1}(\infty,\omega)$. However, for $\Lambda \sim 1/R$, the theoretical errors are dominated by the physics of the effective range $r_2$, not included in the LO. 

Physics left out at LO can be accounted for by considering higher order terms, in increasing perturbation theory order. Thus, at NLO, we include corrections as first-order perturbations on top of the LO wavefunction $\psi^{(0)}$. The NLO correction to the two-body potential in Eq. (\ref{taylorcoord}) comes with an unknown parameter $C_2$ that can be chosen so that a second energy level is fixed, let's say the first excited state. However, the NLO piece induces also a nonvanishing correction to the energy used to fix $C_0$ and hence requires a readjustment of $C_0$ so that at NLO one simultaneously reproduces the two observables (energy levels). While this is a valid avenue, we use a simpler method, motivated by the fact that one expects the NLO correction to be small. Thus, it is more convenient to split $C_0$ into a LO piece $C^{(0)}_0$, which remains unchanged, and a NLO piece $C^{(1)}_0$. The latter are treated in perturbation theory, like the ``genuine'' NLO term, and adjusted so that the level already fixed at LO does not change. According to Eq. (\ref{taylorcoord}), the NLO correction to the LO potential is 
\begin{equation}
V^{(1)}=C_0^{(1)} \delta(\vec{r})
  -C_2^{(1)} \left\{ \left[\nabla^2\delta(\vec{r})\right]
             +2\left[\vec{\nabla} \delta(\vec{r})\right]\cdot \vec{\nabla}
             +2\delta(\vec{r})\nabla^2\right\}.
\label{NLOpot}
\end{equation}
The first-order corrections to the energy, $E^{(1)}$, and to the wavefunction, $\psi^{(1)}(\vec{r})$, are obtained in first-order perturbation theory 
\begin{eqnarray}
\left(H^{(0)}-E^{(0)}\right) \psi^{(1)}(\vec{r})
&=&\left(E^{(1)}-V^{(1)}\right)\psi^{(0)}(\vec{r}).
\label{sch1}
\end{eqnarray}
The requirement that two energy levels have the correct values fixes the amount of change from LO, thus providing two equations that determine the unknown coupling constants $C_0^{(1)}$ and $C_2^{(1)}$ in each model space. In the case when the lowest level $E_0$ is already fixed to a given (experimental or theoretical) value, $E_0^{(1)}=0$.  For non-negligible range, on can alternatively choose that at LO $C_0^{(0)}$ be fixed to a level of the two-body spectrum without a range, while in NLO that level is shifted to the correct  position with range, thus requiring a non-vanishing $E_0^{(1)}$. These two alternatives are the HO-basis equivalent to fixing $C_0^{(0)}$ in a free-particle basis to, respectively, a known binding energy (such as the deuteron binding energy) or the scattering length.
In either case, if we take the first excited level $E_1(\omega)$ to be
reproduced at NLO in addition to the ground state, 
the two equations for the determination
of $C_0^{(1)}(n_{max}, \omega)$ and $C_2^{(1)}(n_{max}, \omega)$
can be written as
\begin{equation}
E_i^{(1)}(n_{max}, \omega) =E_i(\omega) - E_i^{(0)}(n_{max},\omega),
\qquad i=0,1.
\label{2eqs}
\end{equation}
The rest of the levels are prediction, and, as in the LO, converge to the exact values, with or without negligible range, when $n_{max}$ takes large enough values, with errors dominated by $1/\Lambda$. We will illustrate explicitly later that the magnitude of these errors is smaller than at LO, because more physics has been included in the theory. 

The next-to-next-to-leading order (N$^2$LO) correction to the potential  $V^{(2)}$ is given by
\begin{eqnarray}
V^{(2)}&=&C_0^{(2)} \delta(\vec{r})
  -C_2^{(2)} \left\{ \left[\nabla^2\delta(\vec{r})\right]
             +2\left[\vec{\nabla} \delta(\vec{r})\right]\cdot \vec{\nabla}
             +2\delta(\vec{r})\nabla^2\right\}
\nonumber\\
&&
+C_4^{(2)} \left\{\left[\nabla^4\delta(\vec{r})\right] 
         +4\left[\vec{\nabla} \nabla^2\delta(\vec{r})\right]\cdot \vec{\nabla}
         +4\left[\vec{\nabla} \vec{\nabla} \delta(\vec{r})\right]
                 \cdot \cdot \vec{\nabla}\vec{\nabla}
\right.\nonumber\\
&&\left. \qquad\quad
         +4\left[\vec{\nabla} \delta(\vec{r})\right]\cdot \vec{\nabla}\nabla^2
+2\delta(\vec{r})\nabla^4\right\}, 
\label{NNLOpot}
\end{eqnarray}
which introduces a new four-derivative parameter $C_4^{(2)}$. Additionally, similar to Eq. (\ref{NLOpot}), we have added in (\ref{NNLOpot}) perturbative shifts $C_0^{(2)}$ and $C_1^{(2)}$, which can be used to compensate the energy levels fixed at NLO. The three new parameters can be determined so that three energy levels (e.g., the ground-state energy and the first two excited levels) be fixed to the correct values. 

The correction $E^{(2)}$ to the energy is obtained using perturbation theory up to the second order.
In addition to the second-order correction from $V^{(1)}$ (\ref{NLOpot}), one has the first-order correction from $V^{(2)}$ in Eq. (\ref{NNLOpot}):

\begin{equation}
E^{(2)}= 
\langle \psi^{(0)}| V^{(2)} |\psi^{(0)}\rangle
+\frac{1}{2} \left\{\langle \psi^{(0)}| V^{(1)} |\psi^{(1)}\rangle
+\langle \psi^{(1)}| V^{(1)} |\psi^{(0)}\rangle\right\}.
\label{E2}
\end{equation}

Taking the lowest three levels $E_i(\omega)$, $i=0,1,2$, to be fixed, the system of three equations for the determination of $C_0^{(2)}(n_{max}, \omega)$, $C_2^{(2)}(n_{max}, \omega)$, and $C_4^{(2)}(n_{max}, \omega)$ is
\begin{equation}
E_i^{(2)}(n_{max}, \omega) =E_i(\omega) - E_i^{(0)}(n_{max},\omega)
- E_i^{(1)}(n_{max},\omega),
\qquad i=0,1,2.
\label{3eqs}
\end{equation}
The remaining levels are, again, prediction of the theory, and they eventually converge to the correct values.  If further precision is desired, one can continue the procedure to higher orders, although corrections beyond second-order perturbation theory are considerably more involved.

\begin{figure}[tb]
\begin{center}
\includegraphics*[scale=0.82]{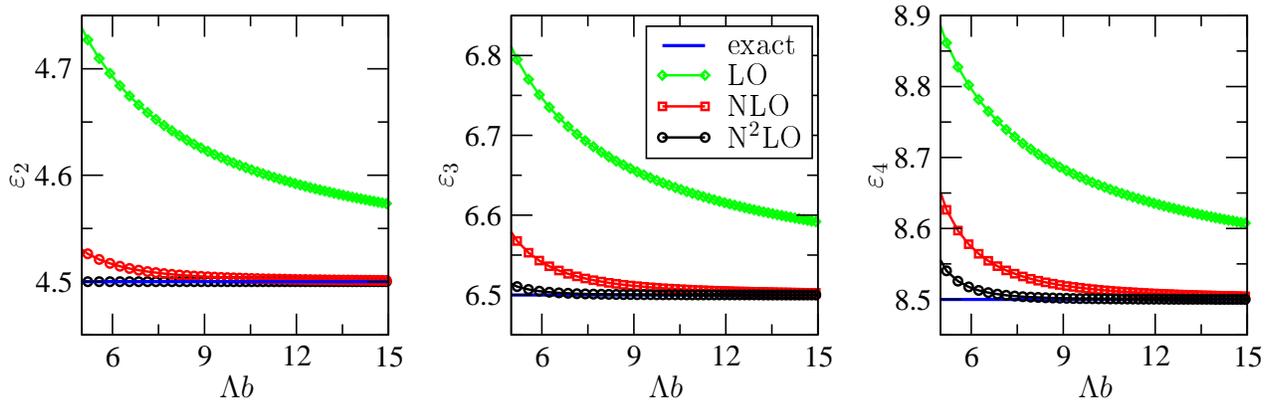}
\end{center}
\caption{The second ($\varepsilon_2$), third ($\varepsilon_3$), and fourth ($\varepsilon_4$) excited-state energies, in units of $\omega$, as a function of the dimensionless cutoff $\Lambda b$ for two trapped particles at unitarity. Results at LO (diamonds), NLO (squares), and  N$^2$LO (circles) are compared with the exact values (solid lines) given by Eq. (\ref{busch2}). The respective coupling parameters are fixed so that the ground-state,  ground- and the first excited-state, and ground-, the first and the second excited-state energies are reproduced in LO, NLO, and N$^2$LO, respectively. (Note that at N$^2$LO the second excited state in the leftmost panel is used as input and thus constant for all values of $\Lambda b$.) Figure reproduced from Ref. \protect\cite{NCSMeft_trap_2b}, with permission from Elsevier.}
\label{states_unitarity}
\end{figure}

In order to illustrate this approach, we consider now the case of two trapped particles at unitarity, characterized by infinite scattering length ($b/a_2=0$) and zero effective range ($r_2/b=0$). In the untrapped case, this situation can be realized by considering an attractive potential given by, for instance, a  square well with a fine-tuned depth. Asymptotically, the wave function at zero energy behaves as $1/r$ and the scattering length $a_2$ is then infinite. In the presence of the harmonic trap, the only scale is set by $b$, or, equivalently, $\omega$, so the solutions of Eq. (\ref{busch2}) have to be constants. The energy spectrum is given by the poles of the Gamma functions in the denominators, 
\textit{i.e.,}
\begin{eqnarray}
\varepsilon_n(\infty) = 2n + \frac{1}{2}, 
\label{trans_unit}
\end{eqnarray}
with $n\geq 0$ an integer. At each order we use a finite number of these energies to determine the interaction parameters in each model space: $\varepsilon_0$ at LO, $\varepsilon_0$ and $\varepsilon_1$ at NLO, and $\varepsilon_0$, $\varepsilon_1$, and $\varepsilon_2$ at N$^2$LO. We also calculate the remaining energies, which depend on $n_{max}$. Lowest excited states, from the second ($n=2$) to the fourth ($n=4$), are plotted in  Fig. \ref{states_unitarity} for different values of the cutoff $n_{max}$ in the dimensionless combination $\Lambda b$. All results change with $\Lambda b$ at LO and NLO. At N$^2$LO, the second excited-state energy is used as input to fix the coupling constants. The predicted energy levels converge as $n_{max}$ increases to the values given in Eq. (\ref{trans_unit}). In addition, Fig. \ref{states_unitarity} shows that the rate of convergence with respect to $n_{max}$ to the exact value increases when higher-order corrections are added in perturbation theory.

\begin{figure}[tb]
\begin{center}
\includegraphics*[scale=0.82]{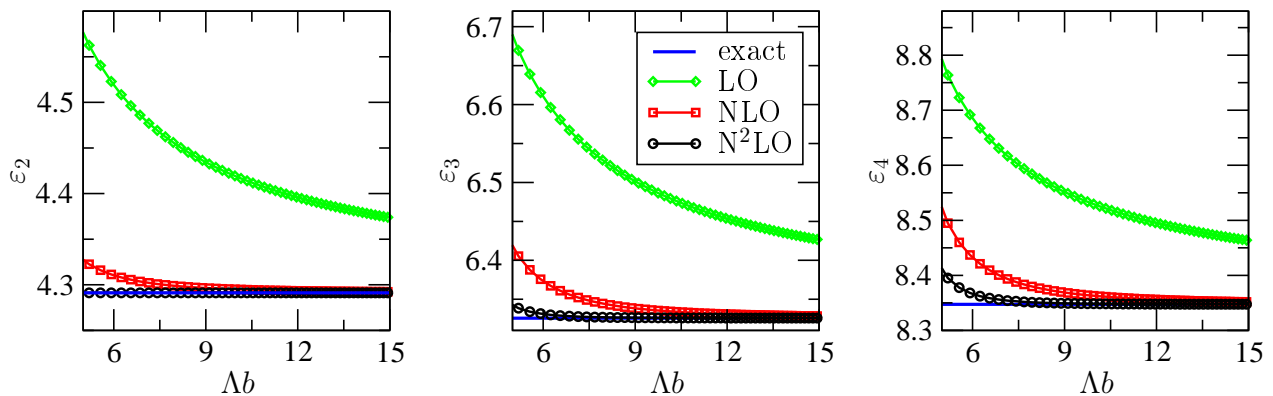}
\end{center}
\caption{Same as in Fig. \ref{states_unitarity}, but for $b/a_2=1$. Figure reproduced from Ref. \protect\cite{NCSMeft_trap_2b}, with permission from Elsevier.}
\label{states_b_over_a_1}
\end{figure}

No qualitative changes are expected with respect to the unitarity case for finite scattering length. Thus, in Fig. \ref{states_b_over_a_1} we consider two particles with the interaction characterized by $b/a_2=1$ and $r_2=0$. Like in the unitary case, the LO is iterated to all orders, whereas the higher-order corrections to the potential are treated in perturbation theory. The parameters at each order are adjusted so that the lowest levels satisfy Eq. (\ref{busch2}) exactly. The running of the energy for selected excited states is shown in Fig. \ref{states_b_over_a_1}. The exact values are now slightly lower than at unitarity, as the interaction is stronger and the states more tightly bound. The convergence of the energy levels to the exact values improves when higher corrections to the potential are added, and the difference between the truncated-space energy and the exact result is mitigated as more corrections are included.
 
Another case, of interest not only for nuclear physics but also for cold atoms, is the presence of a non-negligible range. Thus, in nuclear physics, the effective range plays an important role, while in atomic systems it should become relatively more important as one moves away from the resonance. A finite interaction range $R$ usually generates higher ERE parameters of the same magnitude, $|r_2|\sim R$, $|P_2| \sim R^3$, etc., even when there is fine-tuning that leads to $|a_2|\gg R$. As a last example, we account for a finite effective range $r_2$, for definitives choosing parameters $b/a_2=1$ and $r_2/b=0.1$. Regardless of the quantity chosen as input in LO,  the existence of range introduces errors that are energy dependent and can only be accounted for in subleading orders. As discussed earlier, we use as LO input the ground-state energy $\varepsilon_0$ given by Eq. (\ref{busch2}), with $b/a_2=1$, but with vanishing effective range $r_2$. Hence, the running of $C_0^{(0)}$ is the same as in the previous example. However, at NLO, we obtain $C_0^{(1)}$ and $C_2^{(1)}$ from the first two states of Eq. (\ref{busch2}) with $a_2$ {\it and} $r_2$ non-vanishing. Subsequently, the NLO is different from the previous section: it accounts not only for errors of order ${\cal O}(a_2 k^2/\Lambda)$, due to the explicit model space truncation, but also for implicit ones, ${\cal O}(a_2 R k^2)$, in the potential. The N$^2$LO corrections are  slightly more subtle \cite{pionless,*pionless2}. 
The introduction of range leaves an error that can be as big as
${\cal O}(a_2^2 r_2^2 k^4)={\cal O}(a_2^2 R^2 k^4)$.
Errors from the explicit truncation of the model space are now
${\cal O}(a_2^2 k^4/\Lambda^2)$ or ${\cal O}(a_2^2 r_2 k^4/\Lambda)$,
and, as in general, are smaller than errors from the truncation of
the expansion once $\Lambda \ge 1/R$.
These types of errors are one order
in $kR$ or $k/\Lambda$ from NLO, which requires $V^{(2)}$ for control.
In contrast, errors from the shape parameter are only 
${\cal O}(a_2 P^3 k^4)={\cal O}(a_2 R^3 k^4)$, 
two orders in $kR$ down from NLO.
Thus, at N$^2$LO we determine $C_0^{(2)}$, $C_2^{(2)}$, and $C_4^{(2)}$
from the lowest three levels of Eq. (\ref{busch2}),
still with non-vanishing $a_2$ and $r_2$ and neglecting
all higher-order ERE parameters. As an illustration, we take $r_2/b= 0.1$.
Qualitatively, the only change in energies with respect to previous
subsections is the finite jump from LO
to NLO due to the effective range.
\begin{figure}[tb]
\begin{center}
\includegraphics*[scale=0.82]{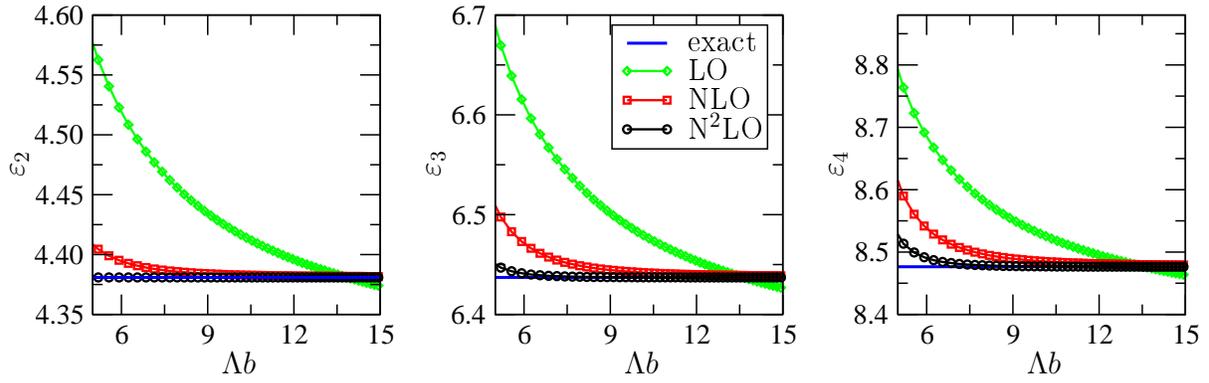}
\end{center}
\caption{Same as in Fig. \ref{states_unitarity}, 
but for $b/a_2=1$ and $r_2/b=0.1$. Figure reproduced from Ref. \protect\cite{NCSMeft_trap_2b}, with permission from Elsevier.}
\label{states_b_over_a_1_point_one}
\end{figure}

The energy levels for three excited states for two trapped particles with the interaction characterized by $b/a_2=1$ and $r_2/b=0.1$ are shown in Fig. \ref{states_b_over_a_1_point_one}. The lack of effective range information in LO translates into an asymptotic behavior that misses the correct value. However, both the NLO and N$^2$LO results converge as $n_{max}$ increases, to the values given by Eq. (\ref{busch2}) with the range, as they should. Moreover, the inclusion of N$^2$LO corrections speeds up convergence considerably: for a fixed value of the cutoff $n_{max}$ the results at N$^2$LO are much closer to the exact value.

\begin{figure}[tb]
\begin{center}
\includegraphics*[scale=0.82]{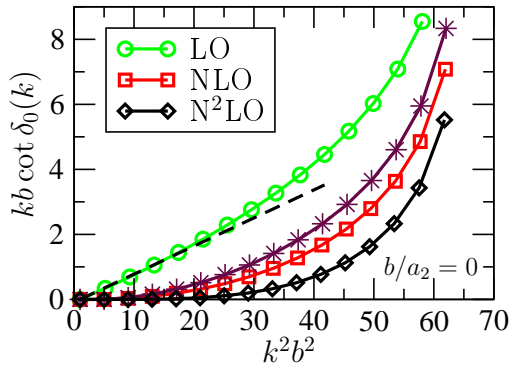}
\end{center}
\caption{Two particles at unitarity: prediction for S-wave scattering phase shifts $kb\cot\delta_0(k)$ as function of $k^2b^2$ in the finite model space characterized by $n_{max} = 15$. The points at LO (circles), NLO (squares), and N2LO (diamonds) are obtained from calculated energies via Eq. (\ref{inv_busch}). The dashed line corresponds to a linear fit of the LO curve at small $kb$ values, which allows an estimate of the effective range introduced by truncation. For comparison, we also show results with the NLO potential fully diagonalized (black stars), which clearly illustrates the effectiveness of the perturbation theory treatement. Figure reproduced from Ref. \protect\cite{NCSMeft_trap_2b}, with permission from Elsevier.}
\label{Stetcu_fig:phasesh_unit}
\end{figure}

Equation (\ref{busch1}) has been hitherto used in one direction: to adjust the LECs so that one describes a set of observables (energy levels) in finite model spaces. The rest of the levels constitute prediction of the theory, and they approach the exact values, given again by Eq. (\ref{busch1}), in the limit of large cutoffs. If we now invert Eq. (\ref{busch1}), it is possible to determine the S-wave phaseshift:
\begin{equation}
k_n \cot\delta_0(k_n)=-2\frac{\Gamma\left(3/4-\varepsilon_n/2\right)}{\Gamma\left(1/4-\varepsilon_b/2\right)},
\label{inv_busch}
\end{equation}
where $k_n=\sqrt{2\varepsilon_n}/b$ is the momentum associated with the state $\varepsilon_n$. This allows us to translate the errors introduced by the finite HO basis in more familiar ERE parameters. As an example, let us consider two interacting particles at unitarity in the absence of an effective range parameter. Thus, any deviation from 
\begin{equation}
k \cot \delta_0(k) = 0,
\end{equation}
is an error introduced by $n_{max}$, and even if the initial interaction is be tuned to have zero effective range, the truncation introduces an effective range (and other higher ERE parameters), as it can be seen in Fig. \ref{Stetcu_fig:phasesh_unit}. The main purpose of NLO (and beyond) is to reduce that effective range (and the remaining ERE parameters). In the example considered, the model space is defined by $n_{max} = 15$. Except for the levels used in the fit, where also the phaseshift are correctly reproduced, $\varepsilon_n$ (and the phaseshift) deviate from the exact value in this model space.  In Fig. \ref{Stetcu_fig:phasesh_unit}, we plot $kb\cot\delta_0(k)$ as a function of $k^2b^2$ (it is more natural to express k in units of 1/$a_2$, but at unitarity $b$ provides the only length unit). At LO, $k \cot\delta_0(k)$ starts off linear in $k^2$ (for $k^2$ up to 20), and then shows higher powers of $k^2$ as one approaches the cutoff. Thus, the LO truncation generates an effective range of about $0.17 b$ in this particular model space, and one could also determine further parameters of the ERE. NLO and N$^2$LO corrections reduce the size of the ERE parameters, so that the results for the phaseshifts improve order by order, getting closer and closer to the horizontal axis. Since $\Lambda^2b^2 = 63$, at $k^2b^2 \sim 60$, the errors are dominated by the truncation errors, and therefore the lower orders do little to improve on the previous order(s). However, at low momentum, the results systematically improve, as expected. Finally, because in current approaches to the NN interactions one often iterates subleading orders together with LO, we show also the results where NLO is fully diagonalized, finding that, indeed, the phaseshifts are more acurately described when NLO is treated in perturbation theory.

In this section we have constructed an interaction within a EFT framework in the case of two particles trapped in a HO potential. We have shown a systematic improvement and acceleration of the convergence to the exact results with respect to the size of the model space as more corrections to the potential are added, with subleading orders treated in perturbation theory. This has been observed for  finite and infinite scattering length cases as well as cases where the range was vanishing or taking a finite value, like in the case of interacting nucleons. The goal that motivated this approach, i.e., a method to fix the LECs only to two-body observables, has been achieved and we are thus ready to consider systems with more than two particles. Before applying the same method to nuclear systems, we consider trapped atomic systems, were a host of methods have been used to obtain few-particle spectra that can be used to test the accuracy of the current approach.

\subsubsection{Few-body systems}
\label{Stetcu_sec:few_ferms_trap}

We now consider  a systems of $A$  two-component fermions confined by a HO potential. The Hamiltonian that describes such a system writes:
\begin{eqnarray}
H= H_0+\sum_{i<j}V_{ij}+ \ldots  ,
\label{hami_many}
\end{eqnarray}
where $H_0$ is the intrinsic HO potential, $V_{ij}$ the two-body interaction between particle $i$ and $j$ and ``$\ldots $" denote three- and more-body interactions. The Hamiltonian $H_0$ can be expressed in terms of the intrinsic Jacobi coordinates (\ref{Stetcu_eq:JacobiCoord}):

\begin{eqnarray}
H_0&=&\sum_{i=1}^{A-1} \left( \frac{{\vec{p}_{\xi_{i}}}^{\; 2}}{2m}
+\frac{m \omega^{2}}{2}\vec{\xi}_{i}^{\; 2} \right),
\end{eqnarray}
where $\vec{p}_{\xi_{i}}$ is the momentum canonically conjugated to 
$\vec{\xi}_{i}$. In finite model spaces, we use the EFT method discussed above in order to define the two-body interaction. Just as in the two-body case, the LO many-body Hamiltonian, containing the HO term and the LO contact two-body interaction, is iterated to all orders, whereas higher order corrections are treated as perturbations. RG analysis shows that three-body forces appear beyond N$^2$LO, which is the highest order we consider here.

Because we limit our investigations to systems of three and four particles, we use the Jacobi coordinates discussed in Sec. \ref{Stetcu_Sec:NCSM} in order to solve the Schr\"odinger equation. For example, in the case of a three-particle system we use the basis states given by Eq. (\ref{Stetcu_eq:basis3b}) to expand the three-body wave function (we use a similar approach for four particles, although the construction of the basis is more involved \cite{Navratil:1999pw}). Since the basis states (\ref{Stetcu_eq:basis3b}) are eigenstates of the unperturbed Hamiltonian $H_0$, they are characterized by the quantum number $N_3$ given by
\begin{eqnarray}
N_3=2n_1+l_1+2 n_2 + l_2.
\label{quanta_tot}
\end{eqnarray}
The energy of each three-body state can be written as $(N_3+3)\omega$, and, as discussed in Sec. \ref{Stetcu_Sec:NCSM}, is used to define the truncated model space. In general, for a system of $A$ particles, the truncation is done by introducing a cutoff $N^{max}_{A}$ defined as the largest number of quanta in the eigenstates of $H_0$ used to construct the $A$-body  basis. Hence, the model space will include all the eigenstates of $H_0$ with energy up to $(N^{max}_A+3(A-1)/2)\omega$. 

The truncation of the many-body space is set by $N_A^{max}$, but, as we have discussed in the previous section, the truncation that defines the two-body interaction in a HO basis  is defined by $N^{max}_2=2n_{max}$ and an exact match between the two is impossible. To some extent, we have already encountered this issue while discussing the cluster approximation in Sec. \ref{Stetcu_sec:cluster}. In that case, there was a mismatch between the $A$-body model space and the $a$-body cluster model space in which the effective interaction is calculated. In practical NCSM applications, one chooses the truncation in such a way that the many-body space is the minimal required to include completely the $a$-body space.  For example, if we consider just two-body interactions, in particular $S$-wave only, $N^{max}_2=N^{max}_3$ when one describes positive-parity states, and $N^{max}_2=N^{max}_3-1$ for negative-parity solutions. However, in our approach the renormalization of the interaction in the two-body system assumes the the states lying above the cutoff $N_2^{max}$ have been ``integrated out'' rather than simply discarded. Hence, their effects are implicitly included in the effective two-body interaction. When these two interacting particles are embedded in a system with a larger number of particles, the spectators will carry energies associated with the HO levels they occupy. For example, of the $(N_3+3)\omega$ total energy of one of the basis states (\ref{Stetcu_eq:basis3b}), $(2n_2+l_2+3/2)\omega$ is carried by the relative motion of the spectator. As such, the maximum energy available to the two-body subsystem is smaller than that allowed by the $A$-body cutoff $N^{max}_A$ and some of the states removed by the truncation will not be accounted for by the renormalization.  One way to correct for this is to use the interactions renormalized with a state-dependent two-body cutoff $N_2^{max}=N^{max}_3-( 2n_2+l_2)$, as first suggested within a conventional NCSM approach in Ref.~\cite{NCSMcuts}. However, the resulting state-dependent interaction is difficult to handle in Jacobi coordinates for systems with more than three particles, and cannot be incorporated in a Slater-determinant basis. In order to account for all the two-body physics beyond our cutoff without the use of such an interaction, we simply decouple the cutoff of the many-body problem from that of the subcluster defining any interaction. Such a prescription has some similarity to the truncation used in Ref. \cite{alhassid2008}. For the three-body system, we have compared our approach of decoupling the two- and many-body spaces and the one involving the state-dependent interaction, but found no evidence that the latter is more suitable.

Each calculation is then characterized by two cutoff parameters: $N^{max}_2$ for the two-body subsystem, and $N^{max}_A$ for the few-body system. To the order we work, no three-body forces appear  (this changes for few-nucleon systems, where, in the pionless theory, a three-body force appears already in LO), so we do not need to consider a separate cutoff for renormalization of a three-body subsystem, when considering larger systems. Actually, for the systems considered here, only $S$-wave interactions have to be included up to N$^2$LO. Our final results are obtained by performing a double-converging approach: we first increase $N^{max}_A$ at fixed $N^{max}_2$ until they converge, and then increase $N^{max}_2$ further. For two-body states with $N_2>N_2^{max}$, we simply set the interaction matrix elements to zero. As we increase $N^{max}_A$ from $N^{max}_A=N^{max}_2$ ($N^{max}_A=N^{max}_2+1$) at fixed $N_2^{max}$ we observe a rapid dependence on $N^{max}_A$ until it is somewhat larger than $N_2^{max}$, the difference reflecting the typical number of quanta carried by the spectators. For low-lying many-body states,  further enlarging $N^{max}_A$ makes little difference because in the many-body states added, the weight of two-body states below the interaction cutoff gets smaller and smaller as  $N^{max}_A$ increases. Having achieved results for any observable of interest that are stable with respect to $N^{max}_A$ for each $N_2^{max}$, we can then take the limit of those values for large $N_2^{max}$. We illustrate this approach by presenting explicit results for the low-lying energies of trapped three- and four-fermion systems. Our goal is to show convergence as we increase the UV cutoff, $N^{max}_2$, 
as well as systematic improvement as the order in the EFT increases.

\begin{figure}[t]
\begin{center}
\includegraphics[scale=0.8,angle=-90,clip=true]{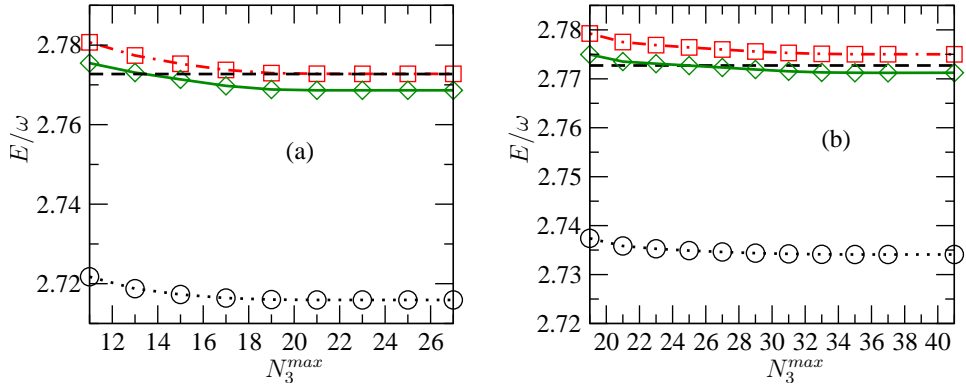}
\end{center} 
\caption{Energy in units of the HO frequency,
$E/\omega$, of the ground state $L^{\pi}=1^-$ of the $A=3$ system at 
unitarity, as function of the three-body model-space size, $N_3^{max}$:
(a) $N^{max}_2=10$; 
(b) $N^{max}_2=18$. 
Circles correspond to LO, squares to NLO, and diamonds to N$^2$LO. 
The dashed line marks the exact value \protect\cite{unitgas_prl,*unitgas_pra}. Figure from Ref. \protect\cite{NCSMeft_trap_34atoms}.}
\label{three_fermion_evol}
\end{figure}

\begin{figure}[t]
\begin{center}
\includegraphics[scale=0.45,angle=-90,clip=true]{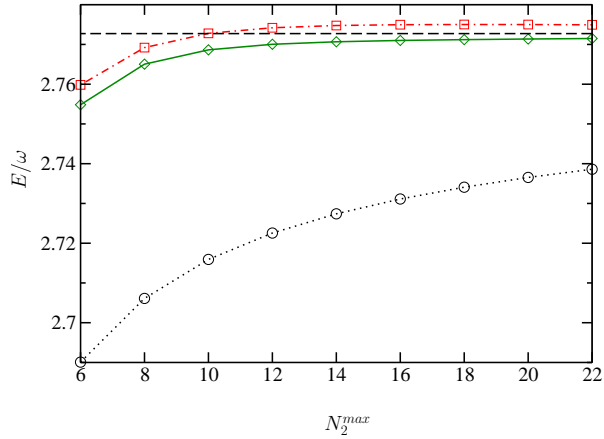}
\end{center} 
\caption{Energy in units of the HO frequency,
$E/\omega$, of the ground state $L^{\pi}=1^-$ of the $A=3$ system at 
unitarity, as function of the two-body cutoff, $N_2^{max}$.
Notation as in Fig. \ref{three_fermion_evol}. Figure from Ref. \protect\cite{NCSMeft_trap_34atoms}.}
\label{three_fermion_per_nmax}
\end{figure}

We consider first the three-fermion system  at unitarity ($b/a_2=0$ and $r_2/b=0$) where semi-analytical results  exist \cite{unitgas_prl}. The ground state is characterized by total angular momentum $L=1$ and negative parity. Figure \ref{three_fermion_evol} shows the convergence of the energy of this state with the truncation in the three-body model space, $N_3^{max}$, for two values of the  two-body UV cutoff: (a) $N_2^{max}=10$ and (b) $N_2^{max}=18$.  For fixed $N_2^{max}$, the Hamiltonian does not change, so that one expects a variational behavior of the ground-state energy with increasing the  three-body model space. Indeed, as shown in Fig. \ref{three_fermion_evol}, the energy decreases until convergence is reached for a large enough three-body model space. The value of $N_3^{max}$ for which the convergence is obtained depends on the particular value of 
$N_2^{max}$.  Thus, for $N_2^{max}=10$ the energy of the three-body ground state (in LO and corrections) does not 
change by more than $10^{-4}$ once $N_3^{max}\ge 19$, while for $N_2^{max}=18$ convergence at this level is achieved for $N_3^{max} \ge 31$. Even though for fixed $N^{max}_2$ the errors induced by the three-body cutoff are eliminated, the errors induced by the truncation in the two-body sector, where the interaction is defined, can be eliminated either by taking $N^{max}_2$ to large values or by adding corrections that take into account physics left out by the truncation to a certain order, or by combination of the two. Figure \ref{three_fermion_per_nmax} shows the convergence with respect to $N^{max}_2$ for the ground-state energy at unitarity.  The LO calculation converges to the exact result \cite{unitgas_prl}. Moreover, the result improves significantly when corrections are added to the potential: at NLO the agreement with the exact calculation is achieved faster  than at LO, and improves still at N$^2$LO. Subleading orders thus provide  systematic improvement over LO results, with the minimal additional overhead of using many-body perturbation theory to different orders rather than direct diagonalization.

\begin{figure}[t]
\begin{center}
\includegraphics[scale=0.8,angle=-90,clip=true]{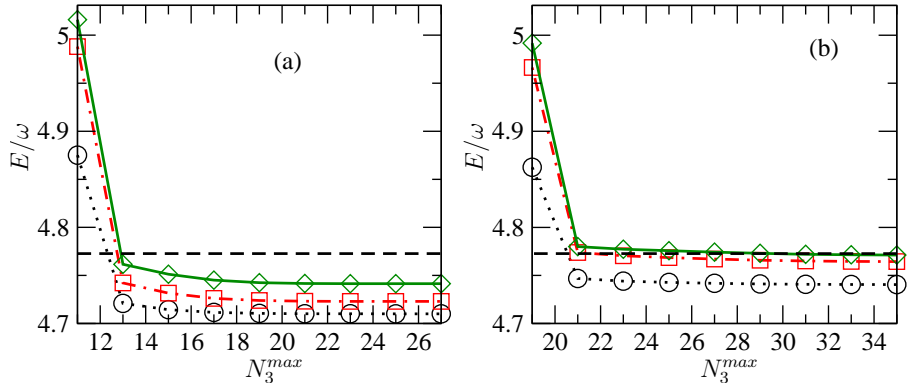}
\end{center} 
\caption{Same as Fig. \ref{three_fermion_evol}, but 
for the first excited state 
with $L^{\pi}=1^-$.  Figure from Ref. \protect\cite{NCSMeft_trap_34atoms}.}
\label{three_fermion_evol_1st_exc}
\end{figure}

\begin{figure}[t]
\begin{center}
\includegraphics[scale=0.4,angle=-90,clip=true]{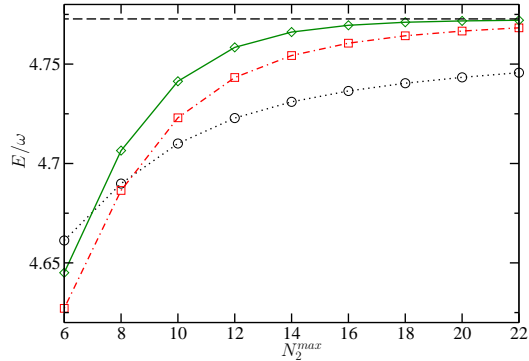}
\end{center} 
\caption{Same as Fig. \ref{three_fermion_per_nmax}, but 
for the first excited state with $L^{\pi}=1^-$. Figure from Ref. \protect\cite{NCSMeft_trap_34atoms}.}
\label{three_fermion_per_nmax_1st_exc}
\end{figure}

The direct diagonalization of the LO Hamiltonian provides not only the ground state, but also excited states. Furthermore, corrections beyond leading order can be carried out similarly. In Fig. \ref{three_fermion_evol_1st_exc} we show the running with the three-body cutoff of the energy of the first excited state with the same quantum numbers as the ground state, $L^{\pi}=1^-$, for the same two values of $N_2^{max}$ considered before, and in Fig. \ref{three_fermion_per_nmax_1st_exc} the convergence with $N_2^{max}$. The same $10^{-4}$ precision for the same two-body UV cutoffs considered before is achieved at somewhat larger three-body cutoffs, $N_3^{max}\ge 23$ and $N_3^{max}\ge 35$ respectively. Like for the ground state, for a fixed $N_2^{max}$ the values of energies at all orders decrease until convergence is reached. Note the sharp decrease of the energy as $N_3^{max}$ goes from $N_2^{max}+1$ to $N_2^{max}+3$, followed by small change as the three-body cutoff is further increased. This suggests that a small number of quanta is carried out by the spectator, so that most of the two-body physics can be accommodated by a relatively small three-body space. The importance of having two different cutoffs in the two- and many-body systems is evident in this case. Indeed, if $N_{3}^{max}$ is fixed at $N_{2}^{max}+1$ one can see from Fig. \ref{three_fermion_evol_1st_exc} that as corrections to the potential are added, results get worse: for both values of $N_{2}$ the energy at NLO and N$^2$LO is farther away from the exact value than the value obtained at LO. As Fig. \ref{three_fermion_per_nmax_1st_exc} shows, once $N_3^{max}$ is decoupled from $N_{2}^{max}$, the corrections to the potential again improve the energy systematically (except at very low two-body cutoff, {\it {i.e}} $N^{max}_2\le 10$). Agreement with the exact value \cite{unitgas_prl} $E/\omega=4.7727$ is very good: for $N^{max}_2=22$, we find at LO $E/\omega=4.7457$, slightly below the value $E/\omega=4.8554$ in Ref. \cite{Stetcu:063613}; at N$^2$LO, $E/\omega=4.7721$. 
The qualitative features of convergence with $N^{max}_3$ and $N_2^{max}$ are similar for states in other channels although details vary. In general, there is systematic improvement as $N_2^{max}$ increases, improvement accelerated by the inclusion of higher-order interactions \cite{NCSMeft_trap_34atoms}, as expected.

\begin{figure}[t]
\begin{center}
\includegraphics[scale=0.4,angle=-90,clip=true]{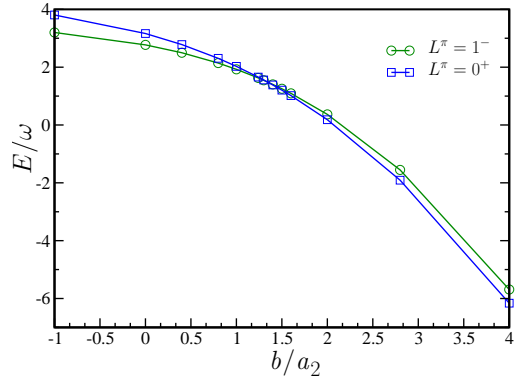} 
\end{center} 
\caption{Energy in units of the HO frequency, $E/\omega$, 
of the lowest states with $L^{\pi}=1^-$ (circles) 
and $L^{\pi}=0^+$ (squares) at N$^2$LO, as a function of the ratio $b/a_2$
between the HO length and the two-body scattering length,
when $r_2/b=0$.
At $b/a_2 \simeq 1.34$ there is an inversion of ground state.  Figure from Ref. \protect\cite{NCSMeft_trap_34atoms}.}
\label{energy_vs_b_a_plot}
\end{figure}

The approach presented here is not limited to the unitarity regime. As we have seen in the two-body case, it can be extended to finite scattering length, as well as finite effective range. Because the same overall features are exhibited for finite effective range, in this paper we limit ourselves to the case when $r_2/b=0$; concrete examples of applications to non-vanishing effective range can be found in Ref. \cite{NCSMeft_trap_34atoms}. Results for the lowest $L=0$ and $L=1$ states (calculated at N$^2$LO) as a function of $b/a_2$ are shown in Fig. \ref{energy_vs_b_a_plot}. In the limit $b/a_2\to -\infty$, the trapped system is non-interacting and the lowest state has angular momentum $L=1$, corresponding to a configuration with two particles in the first $S$ state, and the third in the first $P$ level. As interaction becomes stronger, the splitting between the two three-body states decreases, and at $b/a_2 \approx 1.5$, the two levels become degenerate, and the $L=0$ becomes the ground state, as pointed out in Ref. \cite{Stetcu:063613} and confirmed in Ref. \cite{dishonest}. A more precise N$^2$LO calculation puts the inversion point at $b/a_2 \simeq 1.34$ \cite{NCSMeft_trap_34atoms}. Increasing even further the interaction strength, the ground-state energy converges to $-1/2\mu a_2^2$, suggesting a configuration with one particle moving in an $S$-wave around a bound state of the other two (dimer). Thus, allowing the dimer to form inside a wide-enough trap, $b\gg a_2$, the low-lying three-body spectrum can be associated with the spectrum of two particles (one composite) in a trap. If the atom-dimer momenta is smaller than $1/a_2$, Eq. (\ref{busch2}) can be used to determine the ERE parameters for atom-dimer scattering. This method has been applied to the computation of the atom-dimer scattering in Refs. \cite{NCSMeft_trap_34atoms,PhysRevA.77.043619}, and will be used below to calculate the quartet neutron-deuteron scattering.


%
%

\begin{figure}[t]
\centering\includegraphics[scale=0.8,angle=-90,clip=true]{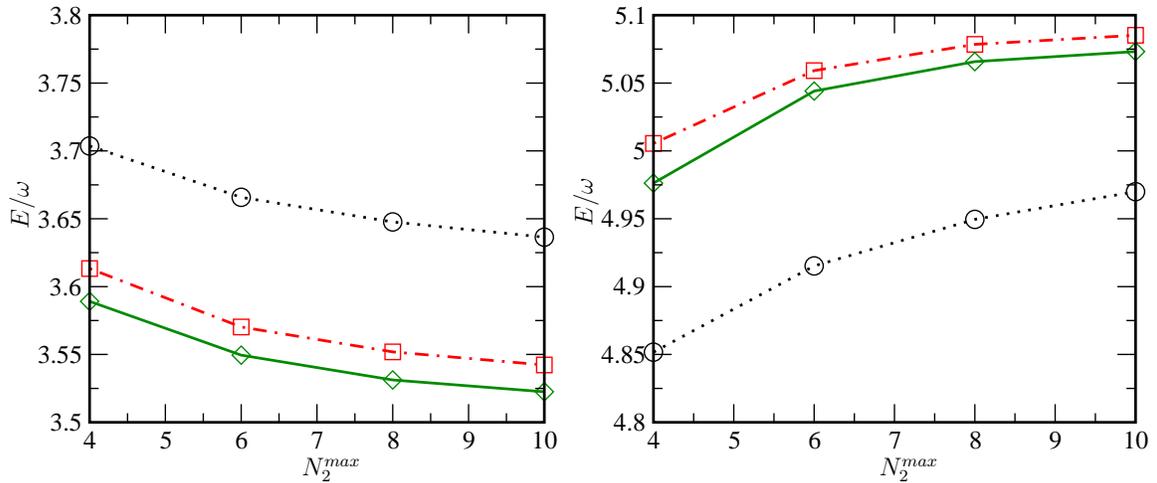} 
\caption{Same as Fig. \ref{three_fermion_per_nmax}, but 
for the ground state (left panel) and first excited state (right panel)
with $L^{\pi}=0^+$ for
the $A=4$ system at unitarity.
Notation as in Fig. \ref{three_fermion_evol}. Figure from Ref. \protect\cite{NCSMeft_trap_34atoms}.}
\label{four_fermion}
\end{figure}

The good agreement with semi-analytical solutions is not limited to the trapped three-body system. Comparable precision can be achieved for the $A=4$ system.  As before, we fix the value of the two-body cutoff $N_2^{max}$ and increase the size of the many-body model space, defined here as $N_4^{max}$. We show only results at unitarity, where other numerical solutions exist \cite{alhassid2008,chang2007,vonStecher:090402,*Blume:233201}, but finite scattering length and effective range can be entertained in the same framework. Figure \ref{four_fermion} shows the convergence of the ground and first-excited states as a function of the two-body ultraviolet cutoff $N^{max}_2$. In LO, the ground-state energy for $N^{max}_2=10$ is $E/\omega =3.64$, to be compared with $E/\omega =4.01$ obtained in Ref. \cite{Stetcu:063613}. Corrections up to N$^2$LO change the prediction to $E/\omega = 3.52$, which is in good agreement with previous calculations where the ground-state energy was found to be $3.6 \pm 0.1$ \cite{chang2007}, $3.551 \pm 0.009$ \cite{vonStecher:090402,*Blume:233201}, and $3.545 \pm 0.003$ \cite{alhassid2008}. Like in the case of two- and three-particle systems, improvement is significant and systematic.

In this section, we have considered physical systems composed of spin-$1/2$ fermions in external HO traps. In order to describe such systems, we have derived effective interactions in finite model spaces in a EFT framework, with corrections up to N$^2$LO.  To this order, the interactions are purely two-body and determined by the two-body scattering length $a_2$ and effective range $r_2$. We demonstrated both good agreement with existing solutions at unitarity \cite{alhassid2008,unitgas_prl,chang2007,vonStecher:090402,*Blume:233201}, as well as systematic improvement in the EFT expansion. In the last part of the paper, we illustrate how the same methods can be extended to solving the nuclear few-body problem.

\subsection{The trapped nuclear systems and the continuum limit}

The successful application of EFT principles to the derivation of effective interactions in finite model spaces in the case of trapped atomic systems motivates us to pursue a similar approach for nuclear systems. However, there is a major difference  between the two systems. In the case of atomic systems, the trap is physical and influences the long-range behavior of the wave function.  On the other hand, the nuclei are selfbound in the absence of the trap. The systems become very similar at the centre of the trap, where the wave function has no knowledge of the trapÕs existence.

The harmonic trap constitutes an essential element for the renormalization of the interaction. Therefore, in order to proceed with the same approach to the description of the nuclear system, one needs to place it in an external harmonic trap. In such a case, the spectrum of the trapped two-body nuclear system can be related to the nuclear phaseshifts through Eq. (\ref{busch1}), like for the trapped atomic systems presented before. But because in this case the HO trap represents an auxiliary tool, one needs to take the continuum limit when the trap vanishes, similar to the procedure discussed in Sec. \ref{Stetcu_Sec:EFT_NCSM}.  

Another significant difference for the nuclear case as compared to the spin-$1/2$ fermion case is the role of three-body forces. In the absence of the trap, in order to achieve RG invariance one needs to introduce a three-nucleon force in the pionless EFT at LO, as discussed in Sec. \ref{Stetcu_Sec:EFT_NCSM}. This feature is not affected by the presence of the trap \cite{hanstrap}, as expected from the short-distance character of renormalization. 

In this section, we present application to two- and three-nucleon systems. In order to renormalize the interaction, we trap the system. All the calculations are performed at finite HO frequency $\omega$, but the final results are obtained in the limit $\omega\to 0$ (the only exception is the calculation of two-body phaseshifts, which can be obtained in the presence of the trap, assuming zero range interactions). In the case of many-nucleon systems, we also introduce the three-body force in the appropriate three-body channels.

\subsubsection{Trapped two-nucleon system}
\label{Stetcu_sec:NNtrap}

Let us first consider two nucleons in a harmonic trap. In this case, the two-body interaction $V$ is constructed as in Sec. \ref{Stetcu_sec:two_ferms_trap}  from the exact spectrum in the trap (\ref{busch2}). For exemplification, we take the $^3$S$_1$ channel, although the treatment of the $^1$S$_0$ channel is similar. At LO the Schr\"odinger equation is solved exactly in a finite model space characterized by the cutoff $n_{max}$ and the coupling constant $C_0^0$ is adjusted such that the ground state corresponds to the lowest energy given by Eq. (\ref{busch2}) with only the scattering length in the triplet channel $a_{2t}=5.425$ fm on the right hand side. Beyond LO, corrections are treated as pertubations. At NLO, the coupling constants appearing at this order, $C_{0}^1$ and $C_2^1$ are fixed such that the two lowest energy correspond to the lowest energies given by Eq. (\ref{busch2}) with $a_{2t}$ and the effective $r_{2t}=1.749$ fm. 
 
\begin{figure}
 \centering \includegraphics*[scale=0.3,angle=-90]{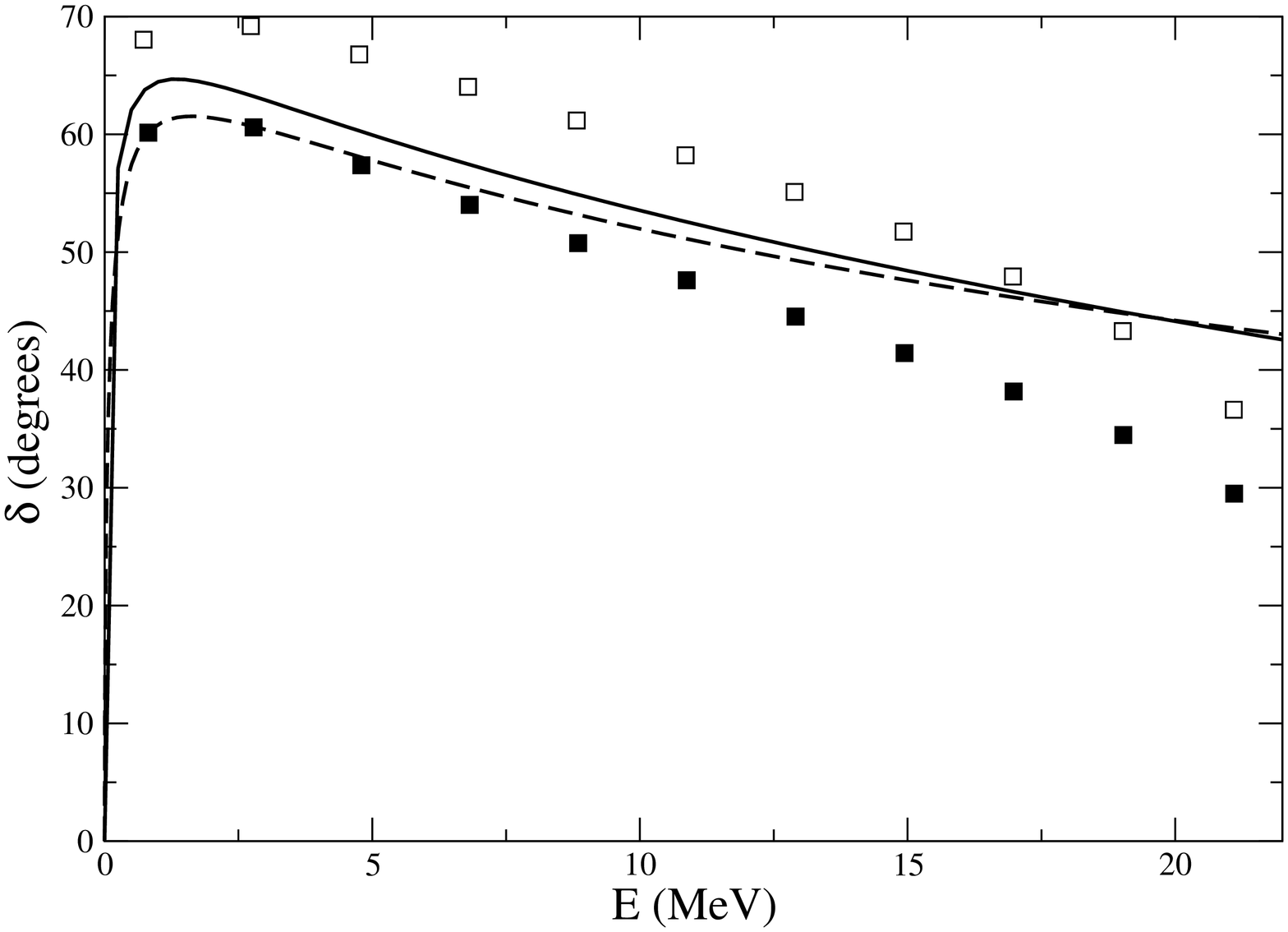}\includegraphics*[scale=0.3,angle=-90]{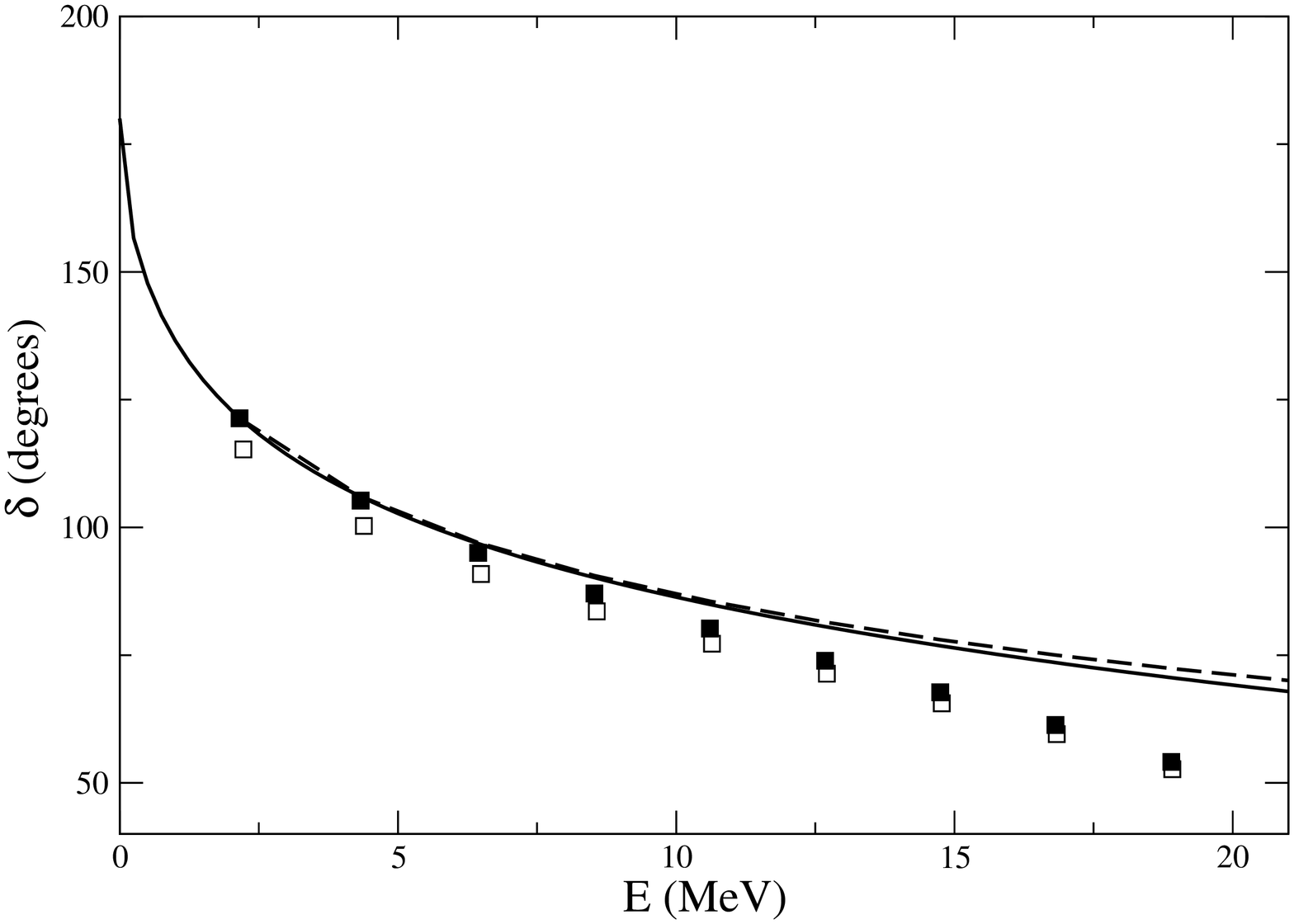}
\caption{The two-nucleon phaseshifts as a function of relative energy in the $^1$S$_0$ (left panel) and $^3$S$_1$ (right panel) channels. The ERE is marked by the continuous curves, while the dashed curves mark the ERE up to the effective range. The EFT results, displayed by open and filled symbols at LO and NLO respectively, are in reasonable agreement with ERE, especially at low energies. The calculations are performed at $\omega$ = 1 MeV and $N_2^{max}$ = 20.  Figure from Ref. \protect\cite{NCSMeft_trap_23nucl}.}
 \label{fig:phaseshiftNLO}
 \end{figure}

 As discussed in Sec. \ref{Stetcu_sec:two_ferms_trap} once the coupling constants are fixed in each order, one can calculate the scattering phaseshift for a certain cutoff \cite{NCSMeft_trap_2b} (see Fig. \ref{Stetcu_fig:phasesh_unit}). In Fig. \ref{fig:phaseshiftNLO}, we plot the phaseshift for both the singlet and triplet configurations. At low energies, as expected, one obtains good agreement with the experimental data, which worsens as the energy increases, but improves systematically order by order, as long as the momentum of the state is well below the cutoff imposed by the model space.

\begin{figure}
 \centering \includegraphics[scale=0.7,clip]{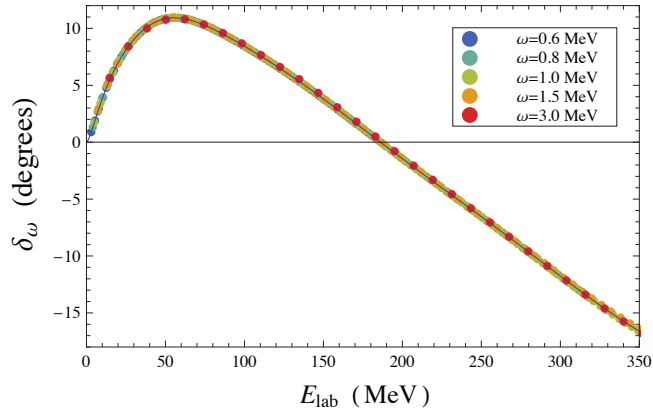}
\caption{The two-nucleon phaseshifts, as function of the lab energy, for the $^1$P$_0$ channel (discrete points) computed using Eq. (\ref{busch1_gen}) and compared with a continuum calculation (continuous line). Figure taken from Ref. \protect\cite{luu_2N_trap}, courtesy T. Luu.}
 \label{fig:luu_pwave}
 \end{figure}

 In nuclear physics, in order to achieve good accuracy it is important to include more than S waves. Actually, P and higher waves do come up in higher orders in pionless EFT (even though we do not consider those orders in this paper). But a generalization of Eq. (\ref{busch1}) to partial waves coupled to arbitrary angular momentum $l$ has been derived \cite{trap_higher_pws1,*trap_higher_pws2}:
 \begin{equation}
 \frac{\Gamma\left(\frac{2 l + 3}{4}-\frac{\varepsilon}{2}\right)}{\Gamma\left(\frac{1-2l}{4}-\frac{\varepsilon}{2}\right)}=(-1)^{l+1}\left(\frac{kb}{2}\right)^{2l+1}\cot\delta_l(k),
 \label{busch1_gen}
 \end{equation}
where the same conventions for $b$, $\varepsilon$ and $k$ as in Eq. (\ref{busch1}) were employed. Luu et. al. \cite{luu_2N_trap} have used the generalization (\ref{busch1_gen}) to extract the two-nucleon phaseshifts produced by the JISP16 NN potential \cite{JISP16_0,*JISP16_1}, constructed via an inverse scattering approach to reproduce the low-energy NN scattering data with high accuracy. Figure \ref{fig:luu_pwave} illustrate the approach for the $^1$P$_0$ two-nucleon channel, where a good agreement between the continuum and Eq. (\ref{busch1_gen}) results can be observed.
  
\begin{figure}
 \centering \includegraphics[scale=0.5,clip,angle=-90]{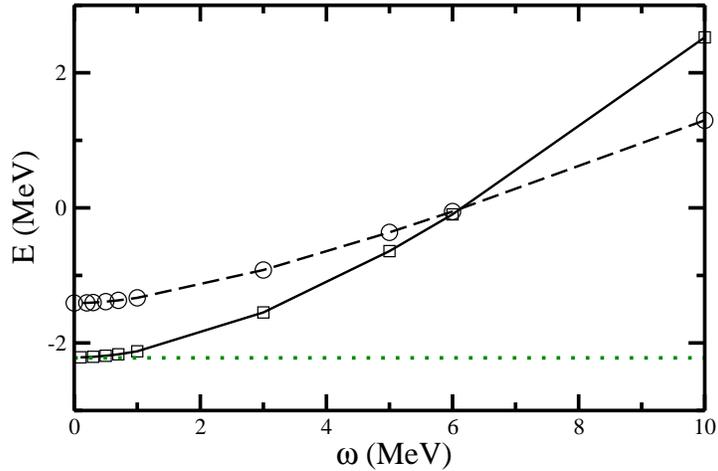}
\caption{The ground-state energy of the trapped two-nucleon system in the $^3S_1$ channel as a function of the HO frequency. At LO (dashed curve) the contact parameter is fixed to the triplet scattering length, and thus converges to $1/(m_N a_{2t}^2)$, while the range corrections introduced at NLO (continuous curve) bring the value very close to the experimental deuteron energy, denoted by the dotted line.  Figure from Ref. \protect\cite{NCSMeft_trap_23nucl}.}
 \label{fig:deuteron_limit}
 \end{figure}

If the phaseshifts  can be easily calculated in the presence of the harmonic trap, its removal is essential if one desires the description of the two-nucleon bound state in the triplet $^3$S$_1$ channel (the deuteron). The trap influences the long-range behavior of the two-body wavefunction and thus the energy of the state. As noted before, the trap is used to determine the LECs in each order, so all the  calculations are performed at finite $\omega$ which is then taken smaller and smaller. In Fig. \ref{fig:deuteron_limit} we illustrate the dependence of the lowest state of the trapped two-nucleon system as a function of the HO frequency. In the limit $\omega\to 0$, this state converges to the deuteron energy (marked by the dotted line), as it can be easily seen at NLO, while the excited states will approach a continuum spectrum. Because in LO we adjust the coupling constant to the triplet scattering length $a_{2t}$, the lowest energy converges to $1/(m_N a_{2t}^2)$, as in the continuum. 

The current approach is perhaps not the best suited to solve the two-body problem. However, the main motivation is the further application of the same techniques to the description of few-nucleons systems that we approach in the next section. 
  
\subsubsection{The three-nucleon system}

In order to describe a system of three nucleons in a trap, we turn to the Jacobi coordinate approach discussed in Sec. \ref{Stetcu_Sec:NCSM}, and consider first the system with total spin and isospin $J^{\pi}=\frac{3}{2}^{+}$ and $T=1/2$, respectively. In this case, while the three-nucleon force is not forbidden, in continuum it appears at higher orders than what is considered here \cite{triton_eft,*triton_eft2,*triton_eft3,*PhysRevC.69.034010}, and as a consequence, the properties of the three-nucleon system are determined by the two-nucleon input. The same is expected to hold for the trapped system.

\begin{figure}[t]
\begin{center}
\includegraphics[scale=0.33,angle=-90,clip=true]{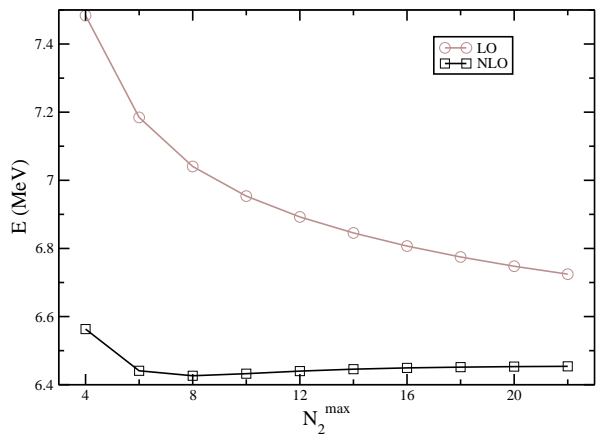}
\end{center} 
\caption{Ground-state energy of the trapped three-nucleon system with quantum numbers $T=1/2$ , $J^{\pi}=3/2^+$
as a function of $N_2^{max}$, for $\omega= 3$ MeV at LO (circles) and NLO (squares).  Figure from Ref. \protect\cite{NCSMeft_trap_23nucl}.}
\label{three_3_half_LO_NLO}
\end{figure}

The two-body interactions are constructed as described in the previous section and then proceed with the three-body problem. Like in the case of trapped few-body atoms, one must first fix the truncation of the two-body space ($N_2^{max}$) in which one performs the renormalization, while the three-body model-space size 
defined by $N_3^{max}$ is increased until the ground state energy reaches a converged value. Representative results for the ground-state energy as a function of the two-body model space size $N_2^{max}$ at $\omega =3$ MeV are summarized in Fig. \ref{three_3_half_LO_NLO}, while qualitative features are the same for other states and frequencies. Clearly the energy converges to a finite value as the two-body cutoff increases. This convergence is displayed in Fig. \ref{three_3_half_LO_NLO} and constitutes a confirmation that, as in the case of a free system, no three-nucleon force is needed at these orders, in this channel, to renormalize the three-nucleon system in the trap.

\begin{figure}
\centering{\includegraphics[scale=0.4,angle=0,clip=true]{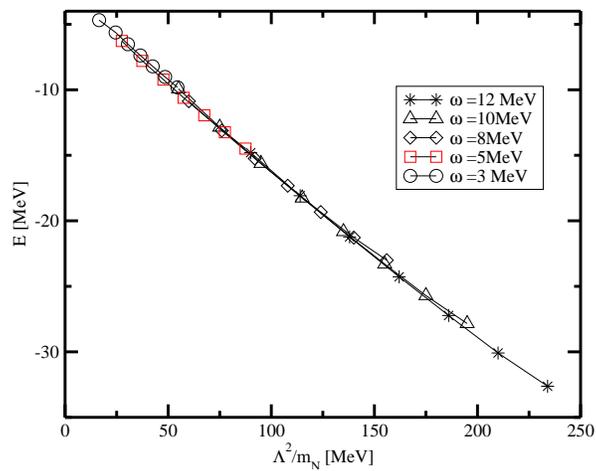}} 
\caption{Ground-state energy of the trapped three-nucleon 
system with
$T=1/2$ and $J^{\pi}=1/2^{+}$
as a function of $\Lambda_2^2/m_N$,
for different frequencies $\omega$.  
Calculations are performed at LO but {\it without} a three-nucleon force.   Figure from Ref. \protect\cite{NCSMeft_trap_23nucl}.}
\label{triton_no_3b}
\end{figure}

The situation changes for three nucleons in the triton channel, characterized by total isospin $T=1/2$ and total angular momentum $J^{\pi}=1/2^{+}$. In this case, the role of the three-body forces is similar to that of three trapped bosons \cite{hanstrap}. Since the renormalization concerns UV momenta, it is not expected to be affected by the trap. In order to check this assumption, one can perform a LO calculation in the trap, \textit{without} the inclusion of the three-body term. As before, for a fixed two-body cutoff $\Lambda_2$ the three-body model space cutoff is increased until convergence is reached. Figure  \ref{triton_no_3b} shows the ground-state energy of the trapped three-nucleon system as a function of $\Lambda_2^2/m_N$, which confirms its collapse as $\Lambda_2$ increases. Results for the same two-body cutoff and different values of $\omega$  are close to each other, which is a sign of the fact that the short-range two-nucleon interaction is much stronger than the long-range HO potential. Indeed, this illustrates the collapse of the three-nucleon system in this channel when only a two-nucleon force is included in the pionless EFT \cite{triton_eft,*triton_eft2,*triton_eft3,*PhysRevC.69.034010}.

The strong cutoff dependence demonstrates that short-distance physics has not been accounted for properly, independent of the existence of the trap. The proper renormalization of the system  can be achieved by including the non-derivative three-nucleon potential already at LO. Formally, the Hamiltonian is composed from the free space Hamiltonian, given by Eq. (\ref{ham}), and the relative harmonic potential term. Like in Sec. \ref{Stetcu_Sec:EFT_NCSM}, one can choose to determine the three-body contact parameter $D_0$ so that the lowest energy of the three nucleons in the trap be fixed at the experimental value of the triton binding energy, $E_{t}=-8.482$ MeV. Obviously, by doing so one makes an error because of the existence of the trap, but this error decreases when the HO frequency is taken smaller and smaller. In Ref. \cite{NCSMeft_trap_23nucl}, the behavior of the $D_0$ parameter was investigated and its behavior points to what looks like the beginning of the limit cycle, expected in the continuum \cite{triton_eft,*triton_eft2,*triton_eft3,*PhysRevC.69.034010}. However, the maximum cutoff for which the calculations were possible  ($\Lambda_2^2/m_N\sim 230$ MeV) was only approximately half the value where the second branch of the limit cycle appears in the continuum calculations \cite{triton_eft,*triton_eft2,*triton_eft3,*PhysRevC.69.034010}.

An alternative to fixing the triton energy to the experimental value in the presence of the trap is to consider three-body scattering observables. We have demonstrated that the presence of the trap allows us to determine two-body scattering phaseshifts and hence we can use the same methods in order to determine the three-body parameter. In the $\omega\to 0$ limit, the lowest state does converge to the triton.  The excited states in the trap at small $\omega$ correspond to three-body ``discretized" continuum states. Because in free space there is no bound excited state, the lowest energy states in the trap correspond to the scattering of a neutron on the deuteron, which can form inside the trap. Some states correspond to $S$-wave scattering. This is similar to the formation of the atom-dimer system in the limit $b/a_2\to 0$, discussed in Sec. \ref{Stetcu_sec:few_ferms_trap}. Thus, the three-body parameter can be adjusted so that the nucleon-deuteron (nd) phaseshift extracted from the lowest excited state of the trapped system [see Eq. (\ref{inv_busch})] reproduces exactly the experimental data.  

In order to avoid complications with the three-body forces and the bound-state, we illustrate how the calculation of neutron-deuteron scattering properties proceeds in the $T=1/2$, $J=3/2^+$ NNN channel. The procedure is similar to the extraction of the atom-dimer scattering parameters from the trapped three-body spectrum \cite{NCSMeft_trap_34atoms,PhysRevA.77.043619}.  Thus, the requirement that the deuteron forms inside the harmonic trap is that its size be much smaller than the HO length parameter, $a_{2t}\ll b$, or, equivalently, $\omega \ll 1/(m_Na_{2t}^2)\simeq 1.8$ MeV. Because the deuteron is placed already in the trap, and thus its long range behavior is modified, we also consider slightly larger frequencies, up to 3 MeV. The smaller the frequency, the larger the errors associated with the fact that smaller UV cutoff are accessible numerically. At fixed $\omega$, we calculate the energy spectra $E_{3;n}$ of the three-body system, as discussed at the beginning of this section. Assuming the separation between the NN bound state and a third nucleon, in the limit of zero range interaction, one can compute the nd phaseshift proceeding in a similar manner with Eq. (\ref{inv_busch}):
\begin{equation}
 \frac{\Gamma(3/4-(E_{3;n}-E_{d})/2\omega)}
      {\Gamma(1/4-(E_{3;n}-E_{d})/2\omega)}
=-\sqrt{\left(E_{3;n}-E_d\right)/2\omega} 
\; \cot\delta_3\left(\sqrt{2\mu_{nd} \left(E_{3;n}-E_d\right)}\right),
\label{eq:scat_3b}
\end{equation}
where $E_d$ is the energy of the trapped deuteron, $\mu_{nd}$ is the neutron-deuteron reduced mass. If the phaseshift $\delta_3(k)$ is given by an ERE expansion
\begin{equation}
k \, \cot \delta_3(k) =-\frac{1}{a_3}+\frac{r_3}{2}k^{2}+ \ldots
\label{ERE3}
\end{equation}
in terms of $nd$ ERE parameters $a_3$, $r_3$, {\it etc.}, it is a matter of simple algebra to obtain these directly from the spectrum of the trapped three-body system. The two-body energy $E_d$ is obtained at LO as the ground state solution in the $^3S_1$ channel of Eq. (\ref{busch2}) with the ERE truncated to the first term only, whereas at NLO, $E_d$ is the solution of the same equation, but includes range corrections. However, we have to impose limitations on which states from the spectrum we can reliably use in the calculations, as Eq. (\ref{eq:scat_3b}) holds only in the limit of zero range interactions. The range of the nd interaction is about the size of the deuteron itself, $\mathcal{R}_{nd}\sim a_{2t}$, while the relative neutron-deuteron momenta associated with the three-nucleon state $E_{3;n}$ is $p_{nd}=\sqrt{2\mu_{nd} \left(E_{3;n}-E_d\right)}$. From the zero-range interaction condition, $\mathcal{R}_{nd}p_{nd} \ll 1$, one can estimate how many states one can include in the calculation. In practice, we  select the two lowest eigenstates and extract the value of the quartet scattering length $a_{3q}$ using Eq. (\ref{ERE3}).

\begin{figure}[t]
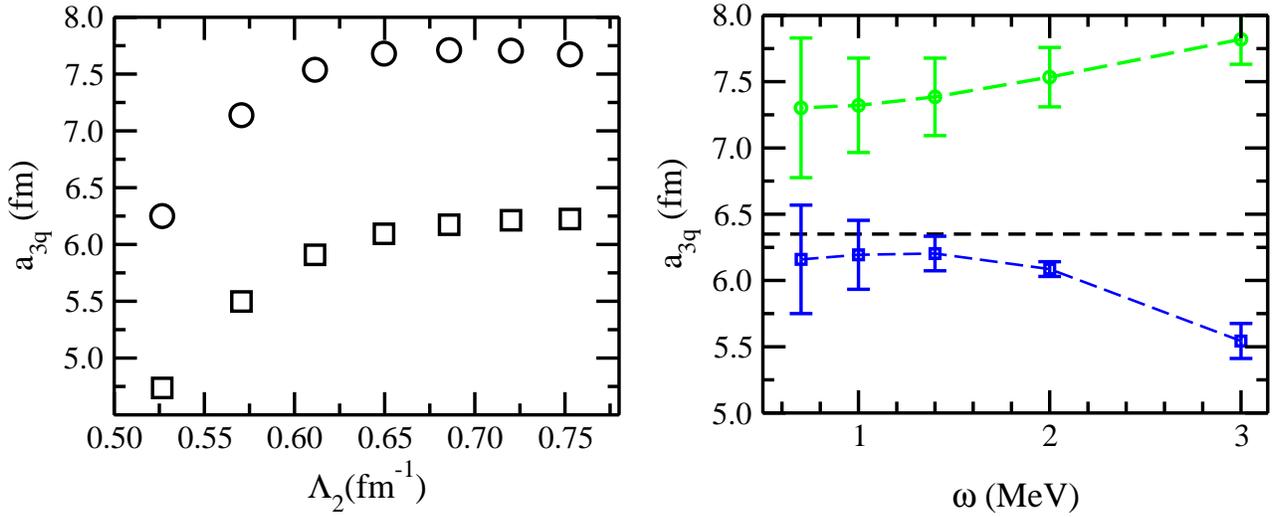

\begin{center}
\includegraphics[scale=0.7,angle=0,clip=true]{scat_length_LO_NLO_hw_1}\hspace{4mm}\includegraphics[scale=0.7,angle=0,clip=true]{scat_length_new}
\end{center} 
\caption{Left panel: Scattering length $a_{3q}$ extracted from the spectrum of the
 trapped three-nucleon system in the channel $T=1/2$, $J=3/2$
as function of the UV cutoff $\Lambda_2$, for $\omega=1$ MeV:
LO (circles) and NLO (squares). Right panel: 
Extrapolated values $a_{3q}({\infty})$ of the 
quartet scattering length  
for different values of $\omega$: LO (circles) and NLO (squares).
The error bars correspond to the standard fitting error.
The horizontal dotted line marks the experimental value \protect\cite{d_n_exp}.   Figure from Ref. \protect\cite{NCSMeft_trap_23nucl}.}
\label{scat_a3q}
\end{figure}

The values of $a_{3q}$ obtained for $\omega=1$ MeV at LO and NLO as a function of the two-body cutoff $\Lambda_2$ are shown in the left panel of Fig. \ref{scat_a3q}. Results for other values of the HO frequency are similar. At a fixed $\omega$, the scattering length $a_{3q}$ should converge as $\Lambda_2$ is increased since the energies of the three-nucleon system converge. As $\omega$ gets smaller, large  values for $N_{2}^{max}$ should be considered in order to reduce the errors associated with the cutoff $\Lambda_2$. Since in practice there are stringent limitations in the values of $N_{2}^{max}$ one can handle, an extrapolation for each value of $\omega$ is needed to extract the scattering length $a_{3q}(\infty)$ which would correspond to $\Lambda_2 \rightarrow \infty$.
For this purpose the trial extrapolation function was proposed \cite{NCSMeft_trap_23nucl}
\begin{eqnarray}
\frac{1}{a_{3q}}&=& \frac{1}{a_{3q}({\infty})}
+\frac{\alpha_1}{\Lambda_2^{p_1}}
+\frac{\alpha_2}{\Lambda_2^{p_2}}, 
\label{fit2} 
\end{eqnarray}
where $p_{1,2}$ and $\alpha_{1,2}$ are parameters, which are fit to the six values of the scattering length obtained at the largest cutoffs.

Results of the extrapolation are displayed in the right hand panel of Fig. (\ref{scat_a3q}), where $a_{3q}({\infty})$ is plotted as a function of $\omega$. For HO frequencies from about 0.4 MeV to about 2 MeV the scattering length is such that $7.30~\mathrm{fm} \leq a_{3q}^{LO}(\infty) \leq 7.53 ~\mathrm{fm}$ and $6.08~\mathrm{fm} \leq a_{3q}^{NLO}(\infty) \leq 6.16~\mathrm{fm}$ for the trial function (\ref{fit2}).

While larger traps are closer to the continuum limit, the error in the scattering length increases. First,
as $\omega$ gets smaller, the imprecision on the value of $a_{3q}$ stemming from the imprecision of the energy ({\it{i.e.}}, the difference between the values for finite $N_2^{max}$ and $N_2^{max}\to \infty$) are enhanced. This can be understood by noticing that in Eq. (\ref{eq:scat_3b}) the energy appears with $\omega$ in the denominator. Second, numerical imprecision also arises as $\omega$ gets smaller since the extrapolation to $a_{3q}({\infty})$ is performed in these cases from data at lower $\Lambda_2$. With all errors taken into account one can conclude that the results at NLO are in good agreement with the experimental value $a_{3q}=6.35 \pm 0.02 ~\rm{fm}$ \cite{d_n_exp} and with previous EFT calculations \cite{triton_eft,*triton_eft2,*triton_eft3,*PhysRevC.69.034010}. 

The same approach can be extended to other processes, e.g. neutron-alpha scattering. Its application requires only the computation of the many-body spectra in a harmonic trap, without the difficulties associated with the treatment of the antisymmetrization between the scattering components. However, if for two particles the errors are rather small for a large range of HO frequencies, when composed particles are involved it is important that the range of the interaction be smaller than the HO length, so that the cluster states can be properly identified. In the case of neutron-deuteron scattering it was important to allow for the deuteron to form inside the trap. Furthermore, the relative momenta between clusters has to be small enough so that Eq. (\ref{busch1}) and its equivalents, derived under the assumption of zero-range interactions, remain valid.

Finally, Eq. (\ref{ERE3}) can be used, as advertised, in order to fix the three-body contact parameter $D_0$. Thus, $D_0$ can be adjusted so that for all model spaces, defined now by $N_2^{max}$ and $N_3^{max}$, the first excited state of the three-body system reproduces correctly the phaseshift at the corresponding momentum. All the other restrictions, like the size of the HO length parameter, remain in place. Once $D_0$ is thus adjusted, one can apply the same procedure to more complex systems.
\section{EFTs and the Gamow Shell Model for exotic systems}
\label{eft_gsm}

Nuclei located far away from the valley of $\beta$-stability display features that do not occur for well bound systems. The strong coupling to the continuum manifests in the existence of halo configurations, in which some nucleons orbit far away from a core of more tightly bound nucleons, and of Borromean systems, where removal of one nucleon is accompanied by at least one more nucleon. The neutron-rich Helium isotopes $^6\rm{He}$ and $^8\rm{He}$ are two examples of such nuclei: both are Borromean halos that have no bound excited states.  They also exhibit the ``binding-energy anomaly'', {\it i.e.}, higher one- and two-neutron emission thresholds in $\rm{^8He}$ than in $\rm{^6He}$.

A microscopic description of weakly bound/unbound nuclei requires taking into account the interplay between bound states, scattering states, and resonances. In other words, these systems have to be described as open quantum systems (OQSs), in contradistinction with well-bound nuclei, which are nearly isolated from the environment of scattering states and decay channels (``closed quantum systems'').  
A shell-model realization OQSs is the so-called Gamow Shell Model (GSM) \cite{Michel:2002,*Michel:2003,*Hagen:2005,*Rotureau:2006Lett,*Papadimitriou:2011R,Michel:2009,*Michel:2010,GSM6he_v_low_k,GSM6he_v_low_k2}. 
GSM is based on the Berggren basis \cite{berg}, which consists of bound, resonant and  scattering single-particle wave functions generated by a finite-depth potential, and it provides the mathematical foundation for unifying 
bound and resonant states -- the poles of the $T$ matrix -- in the context of the Schr\"odinger equation. 

In Ref. \cite{Michel:2002,*Michel:2003,*Hagen:2005,*Rotureau:2006Lett,*Papadimitriou:2011R}, GSM has been used to describe exotic nuclei as systems of valence nucleons above a core using phenomenological potentials for the nucleon-core potential and schematic or phenomenological nucleon-nucleon interactions
for the valence particles. However, even when low-momentum interactions obtained from realistic NN potentials are employed in order to describe the interactions among valence nucleons \cite{GSM6he_v_low_k,GSM6he_v_low_k2}, the approach lacks proper RG invariance. In order to address this issue, a EFT derivation of effective interactions in GSM model spaces, similar to the one discussed in the previous section has been recently proposed \cite{Rotureau:2012fk}. Furthermore, because halo configurations are characterized by large nuclear radii compared to the size of the tightly bound core or, equivalently, by small nucleon separation energy compared to the core binding energy, the physics of halo nuclei is a perfect arena for the application of EFT which exploit separation of scales in physical systems in order to perform systematic, model-independent calculations for a large range of observables \cite{Hammer:2011ye}. By using potentials derived from EFT for the core-nucleon interaction and the nucleon-nucleon interaction for the valence particles, one do not need to assume a particular form for the interactions among constituents. Moreover, RG invariance is  automatically satisfied, and 
the ultraviolet cutoff, originally introduced as an  arbitrary separation between physics kept explicit in the theory and physics treated as short ranged, can be varied in a consistent manner. In the following, we briefly discuss application of halo EFT with the GSM to the $J^{\pi}=0^+$ ground state of $^6{\rm He}$, and for more details we refer the interested reader to the original publication \cite{Rotureau:2012fk}.

Because the first excited state of $^4$He is very high in energy, the $J^{\pi}=0^+$ ground state in $^6$He 
can be described as a three-body system $n+n+\alpha$. The neutrons in the halo interact with the alpha particle
via a two-body interaction $V_{n\alpha}$ and with each other via a potential $V_{nn}$. We denote the neutron (core) mass by $m_N$ ($M_c$) and the neutron-core reduced mass by $\mu_{Nc}=m_N M_c/(m_N+M_c)$. The potential between the alpha core and a neutron is constructed with EFT  as described in Ref. \cite{Bertulani:2002,*Bedaque:2003}. The small relative momentum means that neutron and alpha particle see each other, in a first approximation, as elementary objects. At LO there is only one contribution, which is in the P$_{3/2}$ channel, and the ``dimeron'' potential projected onto this channel can be written in the momentum space as
\begin{equation}
V_{n\alpha}(k',k,k_0)=\frac{k' k}{A+B k_0^2},
\label{Vnaunreg}
\end{equation}
where $\vec{k}$ ($\vec{k'}$)  is the incoming (outgoing) relative momentum and $k_0=\sqrt{2 \mu_{Nc} E_{n\alpha}}$ in terms of the total energy $E_{n\alpha}$ of the $n\alpha$ subsystem. $A$ and $B$ are LECs  to be determined from matching to physical observables. Since this interaction is singular, a regularization procedure is introduced in form of an ultraviolet cutoff $\Lambda_{n\alpha}$. Somewhat different from the procedure followed in the previous section, this is here achieved by introducing a smooth regulator function 
\begin{equation}
F(x)=\exp(-x),
\end{equation}
whose role is to suppress the high-energy contributions 
of the potential. We thus replace the potential in Eq. 
(\ref{Vnaunreg}) by
\begin{equation}
V_{n\alpha}(k',k,k_0;\Lambda_{n\alpha})=
\frac{k' k}{A(\Lambda_{n\alpha})+B(\Lambda_{n\alpha}) k_0^2} 
\, F\left(k'^2/\Lambda_{n\alpha}^2\right) 
\, F\left(k^2/\Lambda_{n\alpha}^2\right).
\label{pot_n_a}
\end{equation}
As discussed in Sec. \ref{Stetcu_Sec:NCSMasEFT}, in order for observables to be RG invariant, the LECs $A(\Lambda_{n\alpha})$ and $B(\Lambda_{n\alpha})$ must depend on the ultraviolet cutoff $\Lambda_{n\alpha}$. At LO,  $A(\Lambda_{n\alpha})$ and $B(\Lambda_{n\alpha})$ can be fixed such that the scattering volume and effective ``range" in the ERE in the P$_{3/2}$ channel exactly match the experimental values $a_{n\alpha}=-62.951$ fm$^3$ and $r_{n\alpha}=-0.8819$ fm$^{-1}$ \cite{param}, respectively.

The two neutrons in the halo have sufficiently low relative momentum that meson exchange can be considered a short-range force. The neutron-neutron potential is thus taken from the pionless EFT \cite{eftreview,pionless,*pionless2,rupak}. At LO, the potential acts only in the $\rm{^1S_0}$ channel and in momentum space it is simply a constant $C$.
As before, the potential requires regularization,
for which we continue to use the function $F(x)$,
but now in terms of the relative momentum between 
the two neutrons and an $nn$ cutoff $\Lambda_{nn}$:
\begin{equation}
V_{nn}(k',k;\Lambda_{nn})=
C(\Lambda_{nn}) 
\, F\left(k'^2/\Lambda_{nn}^2\right)
\, F\left(k^2/\Lambda_{nn}^2\right).
\end{equation}
Like in Sec. \ref{Stetcu_sec:NNtrap}, the LEC $C$ can be fixed to the neutron-neutron scattering length.

\begin{figure}[t]
\begin{center}
\includegraphics[scale=0.4,angle=-90,clip=true]{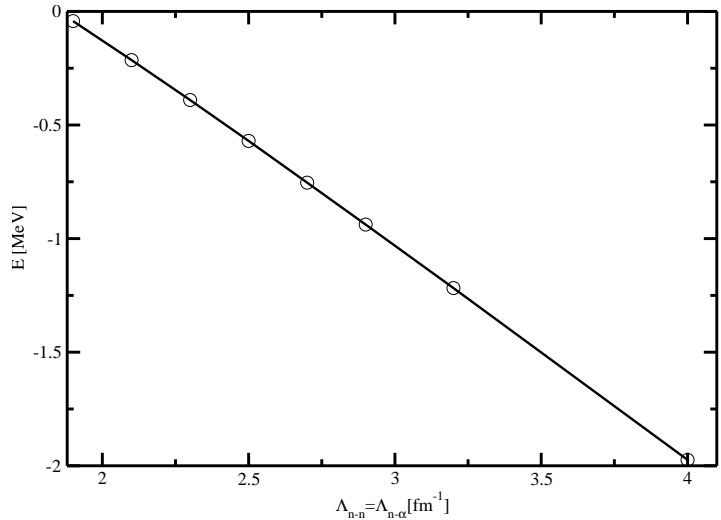}
\end{center} 
\caption{The running of the ground-state energy of $\rm{^6He}$ as a function of the ultraviolet cutoff $\Lambda_{nn}=\Lambda_{n\alpha}$. Only two-body forces have been included in the calculation.}
\label{energy_same_cut}
\end{figure}

The Hamiltonian of the $n+n+\alpha$ systems reads,
\begin{equation}
H=\sum_{i=1}^{2}
\left[\frac{\vec{p}_i^{\, 2}}{2 \mu}
+V_{n\alpha}(k_{0i};\Lambda_{n\alpha})\right]
+V_{nn}(\Lambda_{nn})+\frac{\vec{p}_1\cdot  \vec{p}_2}{M_c},  
\label{ham_3b}
\end{equation}
where $\vec{r}_{i}$ is the position of neutron $i=1,2$
relative to the $\alpha$ core,
and $\vec{p}_i$ the corresponding momentum.
This Hamiltonian is translationally invariant, 
the recoil term 
$\vec{p}_1 \cdot  \vec{p}_2/M_c$   
stemming from the choice of coordinates. 
The three-body equation is solved using a complete set of eigenstates of the LO potential $V_{n\alpha}(k_0;\Lambda_{n\alpha})$. In this particular case of the ground state 
of $^6$He, it is sufficient to consider the set of continuum states along the real axis since $^6$He is bound.
We show in Fig \ref{energy_same_cut} results of the calculations for the ground state of $^6$He as a function of the cutoffs $\Lambda_{nn}=\Lambda_{n\alpha}$.
The nearly linear dive of the ground state seen in
Fig. \ref{energy_same_cut}
is reminiscent of the behavior
observed with LO two-body forces in EFTs for systems of
three bosons or three-or-more-component fermions
\cite{Bedaque:1998kg,*Bedaque:1998km,*Platter:fk}.
There, the dive is even faster, more like quadratic
in the cutoff, stemming from the 
strong s-wave interactions among the three pairs.
In either case, what we see is a collapse
of the ground state under short-range two-body interactions
similar to the one first observed by Thomas \cite{thom}.
It is 
an indication that the three-body problem has not been 
properly renormalized with only two-body interactions 
\cite{Bedaque:1998kg,*Bedaque:1998km,*Platter:fk}. 
The cutoff dependence does not decrease as the cutoffs increase, as one would expected from residual cutoff dependence in a renormalized system that has been truncated correctly, but instead increases with positive powers of the cutoffs. The solution to this problem has to be found outside the two-body subsystems, which are perfectly well defined and well described by the EFT. Indeed, this problem disappears by  adding at LO, a three-body force to renormalize the three-body problem \cite{Rotureau:2012fk}. 

This novel approach thus provides a unique RG invariant approach to $\rm{^6He}$ and paves the way for more comprehensive studies of halo nuclei with EFT. More extensive investigation of $\rm{^6He}$, including higher-order corrections and calculation of other observables (such as the ground-state radius and the first excited-state energy) can be carried out. At the cost of more computational resources, other members of the $\rm{He}$ isotope family could be investigated as well, along the lines of Ref. \cite{Michel:2002,*Michel:2003,*Hagen:2005,*Rotureau:2006Lett,*Papadimitriou:2011R}. More generally, EFT provides a consistent and systematic derivation of the interactions to be used in GSM calculations and should become a valuable tool in the study of other three-body resonant states,
such as the Hoyle state in $\rm{^{12}C}$.


\section{Conclusions and outlook}
\label{Stetcu_Sec:concl}

NCSM remains today, about two decades after its introduction, a powerful many-body method for solving the nuclear many-body problem, and not only. Its applications include nuclear spectra, electro-weak transitions, reaction observables, and even physics beyond standard model. Trapped atomic systems were also studied within its framework. 

The method is similar to the phenomenological shell model and is based on a numerical diagonalization in a finite many-body basis constructed with HO wave functions. The inherent truncation to finite model spaces requires the use of effective interactions, which take into account contributions from the excluded space, whether or not the the excluded physics is known. The main purpose of the paper is a review of the effective interactions and  operators in finite model spaces, and not a comprehensive overview of all the current NCSM applications. Hence, although we have discussed selected applications to energy spectra and electromagnetic observables, we have not presented more recent developments like the importance truncated NCSM or the implementation of resonating group methods within NCSM framework, with applications to nucleon-nucleus scattering or fusion observables. These are important developments that extend the applicability of the NCSM and will be discussed in detail in an upcoming review paper \cite{Navratil_PPNP}.

We have discussed in detail the derivation of effective interactions and non-scalar operators within the unitary transformation approach. The starting point is a given Hamiltonian, which includes, besides the kinetic term, two- and three-body interactions defined in a continuum basis. The $A$-body Hilbert space is spit in two complementary subspaces: the model space, in which the many-body calculation is performed, and an excluded space, whose influence is taken into account by constructing a unitary transformation designed in such a way that the transformed Hamiltonian does not connect the model and excluded spaces. All the initial symmetries of the original Hamiltonian are preserved and, because of the unitarity condition, observables are not affected. The main shortcoming of this approach is that constructing an exact transformation is as difficult as solving the many-body problem. Therefore, the cluster approximation, in which the transformation is calculated for two or three interacting particles, and then the effective interaction used in the many-body calculation, has been introduced in practical applications. In this approximation, induced higher-body interactions are neglected, even though there is no guaranty that they are small, while contributions that initially are small or inexistent can be enhanced. Because the lowest approximation, i.e., the two-body cluster, integrates out mainly the short-range part of the interaction, short-range operators, such as the relative kinetic energy, are strongly renormalized, while long-range operators, e.g., quadrupole, are weakly renormalized. Implementation of the unitary approach to general tensor operators beyond the lowest cluster approximation becomes extremely challenging because the number of matrix elements explodes. All these shortcomings can be minimized by increasing the model space, as by construction the effective interaction approaches the starting interaction when the model space approaches the full Hilbert space, and so do the other operators.

One of the major advantages of using NCSM is its potential to provide an \textit{ab initio} description of heavier nuclei. Thus, the importance truncated NCSM, not discussed in detail here, can push the limits of applicability to medium mass nuclei. Another approach, described in detail in this paper, is more similar to the phenomenological shell model. However, instead of using phenomenologically fitted interactions, single-shell effective interactions can be derived from realistic NN and (possibly) NNN interactions in free space using a secondary unitary transformation. In addition, all the symmetries of the original Hamiltonian are preserved, even if part of the nucleons are frozen like in the phenomenological shell model. The potential of such an approach has been demonstrated in the case of p-shell nuclei, and could be improved by a more targeted design of the unitary transformation, as proposed for trapped atomic systems. However, its applicability beyond the p-shell is challenging because of the numerical power required to converge in the conventional NCSM approach closed shell nuclei, as well as those with one, two and possibly three nucleons outside closed shell.

We have also discussed in great detail a more recent alternative to the traditional unitary transformation approach, in which effective interactions in finite NCSM spaces are derived using principles of EFT.  Problems such as neglecting higher order clusters in the effective interaction, or difficult renormalization of general low-momentum operators can be mitigated if one formulates the problem as an EFT in a discrete basis. In such an approach, the form of effective interaction remains the same, set by the operators included in the EFT expansion to a certain order, while the strengths of the LECs are adjusted to reproduce experimental observables. Unlike in the conventional NCSM approach, in which the effective interaction converges to an underlying ÒbareÓ interaction with increasing the size of the model space, in the EFT framework it is important the behavior of observables with the variation of the UV cutoff. RG invariance is desired, and, at a certain order, the variation of observables with the UV cutoff puts a lower limit on the errors induced by neglected physics in that order. The errors are reduced systematically by adding correction terms which include relevant physics or correct for truncation errors.

We have illustrated the approach using an effective theory without pions, but the same general principles can be applied to other many-body techniques, as well as to the chiral EFT. Two alternative procedures of fixing the LECs have been illustrated. As a first application, we have considered the nuclear system in a pionless EFT approach. In LO, there are three LECs, two coupling constants in the NN sector and one three-body parameter in the triton channel. The strength of the contact interaction in the $^3$S$_1$ channel has been fixed to reproduce the deuteron binding energy in all the model spaces. Because the deuteron is the only two-body bound state, the strength of the two-body contact term in the $^1$S$_0$ and the three-body parameters have been determined so that one simultaneously reproduces the triton and $^4$He binding energies. Following an additional procedure of removing the infrared cutoff $\omega$ artifact, inherent to the HO basis, the energy of the first $(0^+;0)$ state in the alpha particle has been determined in very good agreement with the experiment. The ground-state energy of $^6$Li is predicted with less accuracy \cite{Stetcu:2006ey}, but still within 30\% of the experimental value. Despite the reasonable agreement for a LO calculation, in order to demonstrate systematic improvement one needs to include terms beyond LO. 

The lack of a larger number of bound states in the NN system makes the application of this procedure difficult for subleading orders, or for the EFT with explicit pions. This has motivated the search for a more flexible procedure, in which all the two-body parameters are fixed in the two-body system alone. The harmonically trapped systems present the advantage that all the states are bound, and the natural basis states are the HO wave functions used in NCSM calculations. Moreover, it has been shown that the spectrum of two particles trapped in a HO potential is entirely determined by the ERE parameters in the continuum \cite{NCSMeft_trap_2b,HT,DFT_sr,mehen}. The energy levels in the trapped two-body system can be used to determine the LECs in a finite oscillator basis \cite{Stetcu:063613,NCSMeft_trap_2b}. In this approach, the same procedure as in continuum was applied: the physics at the LO was iterated in all orders, while the subleading orders were treated in perturbation theory. Furthermore, application to the three-body system has shown that the exact solution at unitarity \cite{unitgas_prl,*unitgas_pra} can be reproduced: the errors associated with the truncation can be eliminated either by increasing the cutoff \cite{Stetcu:063613}, or by adding corrections beyond the LO \cite{NCSMeft_trap_34atoms}. The procedure was extended to arbitrary $b/a_2$ ratios, as well as small values of $r_2/b$ ratios or other ERE parameters \cite{NCSMeft_trap_2b,NCSMeft_trap_23nucl}, of great interest for applications to the nuclear systems.

Despite similarities between the atomic and nuclear systems, the procedure devised for trapped cold atoms cannot be applied directly to the nuclear many-body problem. In addition to subtleties related to the power counting, the nuclear system is self-bound in the absence of the trap, which modifies the long-range behavior. The trap plays an essential role in the renormalization procedure of the two-body interaction, as it connects scattering observables to bound-state energy spectra. Hence, in order to take advantage of the renormalization technique devised for cold atoms, the nuclear system is placed in a trap and all calculations are performed at finite HO frequency $\omega$. In addition, one has to take the limit of $\omega\to 0$, which corresponds to the continuum. Description of the three-nucleon system properties has been limited to the nucleon-deuteron quartet scattering up to NLO, with results that are in good agreement with the experimental value and previous EFT calculations. Also, the collapse of the triton ground state in the presence of the trap has been investigated. In order to attack systems with more nucleons, the three-body force has to be included properly in the calculation.

The approach can be extended in several directions. First, the calculations of Ref. \cite{Stetcu:2006ey} could be repeated in order to asses whether the reasonable agreement (in LO) with experimental $^6$Li binding energy was accidental or could be improved at NLO, thus testing the limits of the pionless EFT with increasing the number of nucleons. Second, the application of the same techniques presented for the extraction of the quartet nucleon-deuteron scattering length  could be extended to other systems in order to extract scattering information from  bound-state physics. The nucleon-$\alpha$ scattering process is particularly well suited to such an approach because of the lack of a bound-state system in the five-body system, as well as the compact size of $^4$He. Such an approach would provide an alternative to other methods under development \cite{Quaglioni:2007eg,PhysRevC.79.044606,*PhysRevC.82.034609,*PhysRevC.83.044609,*PhysRevLett.108.042503,LanczCoef,Nollett:2007}. Third, the investigation of general operators is in order, as this constitutes a strong motivation for going to an EFT inspired approach. Finally, the same techniques could be extended to the EFT with pions or to other many-body methods such as the Gamow shell model \cite{Rotureau:2012fk}.

\section*{Acknowledgments}
We would like to thank our close collaborators B. R. Barrett, and U. van Kolck for their support and contribution to the project. The work of I.S. was partially supported by the UNEDF SciDAC Collaboration under DOE grant DE-FC02-07ER41457 and was partly performed at Los Alamos National Lab under the auspices of US DOE. J.R. acknowledges support by the European Research Council (ERC StG 240603) under FP7, the US NSF under grant PHY-0854912 , and the US DOE under grant DE-FG02-04ER41338.


\begin{thebibliography}{100}

\bibitem{Wiringa:1995}
R.B. Wiringa, V.G.J. Stocks and R. Schiavilla,
\newblock Phys. Rev. C 51 (1995) 38.

\bibitem{QMC_rev}
S.C. Pieper and R.B. Wiringa,
\newblock Ann. Rev. Nucl. Part. Sci. 51 (2001) 53.

\bibitem{cdbonn}
R. Machleidt, F. Sammarruca and Y. Song,
\newblock Phys. Rev. C 53 (1996) R1483.

\bibitem{weinberg1990}
S. Weinberg,
\newblock Phys. Lett. B251 (1990) 288.

\bibitem{weinberg1991}
S. Weinberg,
\newblock Nucl. Phys. B363 (1991) 3.

\bibitem{ordonez1992}
C. Ord\'{o}\~{n}ez and U. van Kolck,
\newblock Phys. Lett. B291 (1992) 459.

\bibitem{eftreview}
P.F. Bedaque and U. van Kolck,
\newblock Ann. Rev. Nucl. Part. Sci. 52 (2002) 339.

\bibitem{Carlson:1986}
J. Carlson,
\newblock Phys. Rev. C 36 (1987) 2026.

\bibitem{Carlson:1997qn}
J. Carlson and R. Schiavilla,
\newblock Rev. Mod. Phys. 70 (1998) 743.

\bibitem{Schiavilla:1993tk}
R. Schiavilla, R.B. Wiringa and J. Carlson,
\newblock Phys. Rev. Lett. 70 (1993) 3856.

\bibitem{Nogga199719}
A. Nogga et~al.,
\newblock Phys. Lett. B409 (1997) 19.

\bibitem{Yakubovsky:1966ue}
O.A. Yakubovsky,
\newblock Sov. J. Nucl. Phys. 5 (1967) 937.

\bibitem{Gloeckle:1993mx}
W. Gloeckle and H. Kamada,
\newblock Phys. Rev. Lett. 71 (1993) 971.

\bibitem{Kamada1992205}
H. Kamada and W. Gl\"ockle,
\newblock Nuclear Physics A 548 (1992) 205 .

\bibitem{PhysRevLett.85.944}
A. Nogga, H. Kamada and W. Gl\"ockle,
\newblock Phys. Rev. Lett. 85 (2000) 944.

\bibitem{Kievsky:1992um}
A. Kievsky, S. Rosati and M. Viviani,
\newblock Nucl. Phys. A551 (1993) 241.

\bibitem{Barnea:2000be}
N. Barnea, W. Leidemann and G. Orlandini,
\newblock Phys. Rev. C 61 (2000) 054001.

\bibitem{Navratil:2000gs}
P. Navr\'atil, J.P. Vary and B.R. Barrett,
\newblock Phys. Rev. C 62 (2000) 054311.

\bibitem{Navratil:2009ut}
P. Navratil et~al.,
\newblock J. Phys. G36 (2009) 083101.

\bibitem{benchmark:4he_gs}
H. Kamada et~al.,
\newblock Phys. Rev. C 64 (2001) 044001.

\bibitem{PhysRevLett.92.132501}
K. Kowalski et~al.,
\newblock Phys. Rev. Lett. 92 (2004) 132501.

\bibitem{Hagen2007169}
G. Hagen et~al.,
\newblock Phy. Lett. B 656 (2007) 169 .

\bibitem{PhysRevC.76.044305}
G. Hagen et~al.,
\newblock Phys. Rev. C 76 (2007) 044305.

\bibitem{PhysRevC.82.034330}
G. Hagen et~al.,
\newblock Phys. Rev. C 82 (2010) 034330.

\bibitem{Gandolfi:2006bt}
S. Gandolfi and F. Pederiva,
\newblock Eur. Phys. J. 35 (2008) 207.

\bibitem{PhysRevLett.99.022507}
S. Gandolfi et~al.,
\newblock Phys. Rev. Lett. 99 (2007) 022507.

\bibitem{latticeEFT}
E. Epelbaum et~al.,
\newblock Phys. Rev. Lett. 106 (2011) 192501.

\bibitem{Okubo:1954}
S. Okubo,
\newblock Prog. Theor. Phys. 12 (1954) 603.

\bibitem{DaProvidencia:1964}
J. Da~Providencia and C.M. Shakin,
\newblock Ann. of Phys. 30 (1964) 95.

\bibitem{Suzuki:1980}
K. Suzuki and S. Lee,
\newblock Prog. Theor. Phys. 64 (1980) 2091.

\bibitem{Suzuki:1994ok}
K. Suzuki and R. Okamoto,
\newblock Prog. Theor. Phys. 92 (1994) 1045.

\bibitem{Suzuki:1982}
K. Suzuki,
\newblock Prog. Theor. Phys. 68 (1982) 246.

\bibitem{lattQCD}
S.R. Beane et~al.,
\newblock Phys. Rev. Lett. 97 (2006) 012001.

\bibitem{Stetcu:2006ey}
I. Stetcu, B.R. Barrett and U. van Kolck,
\newblock Phys. Lett. B 653 (2007) 358.

\bibitem{Stetcu:063613}
I. Stetcu et~al.,
\newblock Phys. Rev. A 76 (2007) 063613.

\bibitem{NCSMeft_trap_2b}
I. Stetcu et~al.,
\newblock Annals of Physics 325 (2010) 1644.

\bibitem{NCSMeft_trap_34atoms}
J. Rotureau et~al.,
\newblock Phys. Rev. A 82 (2010) 032711.

\bibitem{NCSMeft_trap_23nucl}
J. Rotureau et~al.,
\newblock Phys. Rev. C 85 (2012) 034003.

\bibitem{bertsch_core_pol}
G.F. Bertsch,
\newblock Nucl. Phys. 74 (1965) 234.

\bibitem{ring}
P. Ring and P. Schuck,
\newblock The nuclear many-body problem, 1st ed. (Springer-Verlag, New York,
  1980).

\bibitem{kuo74}
T.T.S. Kuo,
\newblock Ann. Rev. Nucl. Sci. 24 (1974) 101.

\bibitem{wildenthal}
B.H. Wildenthal,
\newblock Prog. Part. Nucl. Phys. 11 (1984) 5.

\bibitem{KB3}
T.T.S. Kuo and G.E. Brown,
\newblock Nucl. Phys. A 114 (1962) 235.

\bibitem{KB3-2}
A. Poves and A.P. Zuker,
\newblock Phys. Rep. 70 (1981) 562.

\bibitem{PhysRevC.74.034315}
B.A. Brown and W.A. Richter,
\newblock Phys. Rev. C 74 (2006) 034315.

\bibitem{PhysRevC.65.061301}
M. Honma et~al.,
\newblock Phys. Rev. C 65 (2002) 061301.

\bibitem{NCSMcore}
A.F. Lisetskiy et~al.,
\newblock Phys. Rev. C 78 (2008) 044302.

\bibitem{NCSMcore_op}
A.F. Lisetskiy et~al.,
\newblock Phys. Rev. C 80 (2009) 024315.

\bibitem{Johnson2010:pei}
C.W. Johnson,
\newblock Phys. Rev. C 82 (2010) 031303.

\bibitem{Navratil:1999pw}
P. Navr\'atil, G.P. Kamuntavicius and B.R. Barrett,
\newblock Phys. Rev. C 61 (2000) 044001.

\bibitem{Lipkin_CM}
H.J. Lipkin,
\newblock Phys. Rev. 109 (1958) 2071.

\bibitem{Navratil:1993plb}
P. Navr\'atil, H.B. Geyer and T.T.S. Kuo,
\newblock Phys. Lett. B315 (1993) 1.

\bibitem{Navratil:1993NPA}
P. Navr\'atil and H.B. Geyer,
\newblock Nucl. Phys. A556 (1993) 165.

\bibitem{Haxton:2001}
W.C. Haxton and T. Luu,
\newblock Nucl. Phys. A 690 (2001) 15.

\bibitem{Haxton:2002kb}
W. Haxton and T. Luu,
\newblock Phys. Rev. Lett. 89 (2002) 182503.

\bibitem{Haxton:2008}
W.C. Haxton,
\newblock Phys. Rev. C 77 (2008) 034005.

\bibitem{luu:103202}
T. Luu and A. Schwenk,
\newblock Phys. Rev. Lett. 98 (2007) 103202.

\bibitem{ViazVary}
C.P. Viazminsky and J.P. Vary,
\newblock J. Math. Phys. 42 (2001) 2055.

\bibitem{Krenciglowa}
E.M. Krenciglowa and T.T.S. Kuo,
\newblock Nucl. Phys. A235 (1974) 171.

\bibitem{Navratil:1996_effint}
P. Navr\'atil and B.R. Barrett,
\newblock Phys. Rev. C 54 (1996) 2986.

\bibitem{Navratil:1998_pshell}
P. Navr\'atil and B.R. Barrett,
\newblock Phys. Rev. C 57 (1998) 3119.

\bibitem{Barnea:clusterExpansion}
O. Mintkevich and N. Barnea,
\newblock Phys. Rev. C 69 (2004) 044005.

\bibitem{nonHermOp}
B.R. Barrett et~al.,
\newblock Journal of Physics A: Mathematical and General 39 (2006) 9983.

\bibitem{Hayes:2003ni}
A.C. Hayes, P. Navr\'atil and J.P. Vary,
\newblock Phys. Rev. Lett. 91 (2003) 012502.

\bibitem{Caurier:2005rb}
E. Caurier and P. Navr\'atil,
\newblock Phys. Rev. C 73 (2006) 021302.

\bibitem{Nogga:2005hp}
A. Nogga et~al.,
\newblock Phys. Rev. C 73 (2006) 064002.

\bibitem{Navratil:2007}
P. P.~Navr{\'a}til et~al.,
\newblock Phys. Rev. Lett. 99 (2007) 042501.

\bibitem{Stetcu:polarizHe}
I. Stetcu et~al.,
\newblock Phys. Rev. C 79 (2009) 064001.

\bibitem{forssen:2p}
C. Forss{\'e}n, R. Roth and P. Navr{\'a}til,
\newblock (2011), arXiv:1110.0634v2.

\bibitem{PhysRevLett.106.202502}
P. Maris et~al.,
\newblock Phys. Rev. Lett. 106 (2011) 202502.

\bibitem{Quaglioni:2007eg}
S. Quaglioni and P. Navr\'atil,
\newblock Phys. Lett. B652 (2007) 370.

\bibitem{PhysRevC.79.044606}
S. Quaglioni and P. Navr\'atil,
\newblock Phys. Rev. C 79 (2009) 044606.

\bibitem{PhysRevC.82.034609}
P. Navr\'atil, R. Roth and S. Quaglioni,
\newblock Phys. Rev. C 82 (2010) 034609.

\bibitem{PhysRevC.83.044609}
P. Navr\'atil and S. Quaglioni,
\newblock Phys. Rev. C 83 (2011) 044609.

\bibitem{PhysRevLett.108.042503}
P. Navr\'atil and S. Quaglioni,
\newblock Phys. Rev. Lett. 108 (2012) 042503.

\bibitem{Stetcu:2008vt}
I. Stetcu et~al.,
\newblock Phys. Lett. B665 (2008) 168.

\bibitem{PhysRevC.84.065501}
J. de~Vries et~al.,
\newblock Phys. Rev. C 84 (2011) 065501.

\bibitem{PhysRevC.77.024301}
C. Forss\'en et~al.,
\newblock Phys. Rev. C 77 (2008) 024301.

\bibitem{Feldmeier:1997zh}
H. Feldmeier et~al.,
\newblock Nucl. Phys. A 632 (1998) 61.

\bibitem{Bogner:2001yi}
S. Bogner, T.T.S. Kuo and L. Coraggio,
\newblock Nucl. Phys. A684 (2001) 432.

\bibitem{Bogner:2002yw}
S. Bogner et~al.,
\newblock Phys. Rev. C 65 (2002) 051301.

\bibitem{Bogner:2006pc}
S.K. Bogner, R.J. Furnstahl and R.J. Perry,
\newblock Phys. Rev. C 75 (2007) 061001.

\bibitem{PhysRevLett.103.082501}
E.D. Jurgenson, P. Navr\'atil and R.J. Furnstahl,
\newblock Phys. Rev. Lett. 103 (2009) 082501.

\bibitem{PhysRevLett.107.072501}
R. Roth et~al.,
\newblock Phys. Rev. Lett. 107 (2011) 072501.

\bibitem{PhysRevC.83.034301}
E.D. Jurgenson, P. Navr\'atil and R.J. Furnstahl,
\newblock Phys. Rev. C 83 (2011) 034301.

\bibitem{Coraggio:2008in}
L. Coraggio et~al.,
\newblock Prog. Part. Nucl. Phys. 62 (2009) 135, 0809.2144.

\bibitem{PhysRevC.83.064317}
H. Hergert, P. Papakonstantinou and R. Roth,
\newblock Phys. Rev. C 83 (2011) 064317.

\bibitem{Roth:2011fk}
R. Roth et~al.,
\newblock (2011), 1112.0287v1.

\bibitem{N3LO}
D.R. Entem and R. Machleidt,
\newblock Phys. Rev. C 68 (2003) 041001(R).

\bibitem{vanKolck:1994}
U. van Kolck,
\newblock Phys. Rev. C 49 (1994) 2932.

\bibitem{Epelbaum:2002}
E. Epelbaum et~al.,
\newblock Phys. Rev. C 66 (2002) 064001.

\bibitem{Osnes}
P.J. Ellis and E. Osnes,
\newblock Rev. Mod. Phys. 49 (1974) 777.

\bibitem{Navratil:1996jq}
P. Navr\'atil, M. Thoresen and B.R. Barrett,
\newblock Phys. Rev. C 55 (1997) 573.

\bibitem{Stetcu:2004bk}
I. Stetcu et~al.,
\newblock Int. J. Mod. Phys. E14 (2005) 95.

\bibitem{Stetcu:2004wh}
I. Stetcu et~al.,
\newblock Phys. Rev. C 71 (2005) 044325.

\bibitem{Stetcu:2006zn}
I. Stetcu et~al.,
\newblock Phys. Rev. C 73 (2006) 037307.

\bibitem{PhysRevLett.88.152502}
P. Navr\'atil and W.E. Ormand,
\newblock Phys. Rev. Lett. 88 (2002) 152502.

\bibitem{PhysRevC.65.054003}
A. Nogga et~al.,
\newblock Phys. Rev. C 65 (2002) 054003.

\bibitem{Weinberg1992}
S. Weinberg,
\newblock Physics Letters B 295 (1992) 114.

\bibitem{Kaplan1998}
D.B. Kaplan, M.J. Savage and M.B. Wise,
\newblock Physics Letters B 424 (1998) 390 .

\bibitem{Fleming2000}
S. Fleming, T. Mehen and I.W. Stewart,
\newblock Nucl. Phys. A 677 (2000) 313.

\bibitem{TMprime99}
S.A. Coon and H.K. Han,
\newblock Few Body Syst. 30 (2001) 131.

\bibitem{PhysRevC.82.054001}
E.R. Anderson et~al.,
\newblock Phys. Rev. C 82 (2010) 054001.

\bibitem{Navratil_FBS}
P. Navr\'atil,
\newblock Few Body Syst. 41 (2007) 117.

\bibitem{PhysRevC.79.021303}
C. Forss\'en, E. Caurier and P. Navr\'atil,
\newblock Phys. Rev. C 79 (2009) 021303.

\bibitem{Roth:2007sv}
R. Roth and P. Navratil,
\newblock Phys. Rev. Lett. 99 (2007) 092501, 0705.4069.

\bibitem{PhysRevLett.98.162503}
T. Dytrych et~al.,
\newblock Phys. Rev. Lett. 98 (2007) 162503.

\bibitem{PhysRevC.76.014315}
T. Dytrych et~al.,
\newblock Phys. Rev. C 76 (2007) 014315.

\bibitem{Dytrych:2008}
T. Dytrych et~al.,
\newblock Journal of Physics G: Nuclear and Particle Physics 35 (2008) 095101.

\bibitem{INOY:2003}
P. Doleschall et~al.,
\newblock Phys. Rev. C 67 (2003) 064005.

\bibitem{INOY:2004}
P. Doleschall,
\newblock Phys. Rev. C 69 (2004) 054001.

\bibitem{kaplan}
D.B. Kaplan,
\newblock (2005), nucl-th/0510023.

\bibitem{pionless}
U. van Kolck,
\newblock Workshop on Chiral Dynamics 1997, Theory and Experiment, edited by A.
  Bernstein, D. Drechsel and T. Walcher, Springer-Verlag, 1997, hep-ph/9711222.

\bibitem{pionless2}
U. van Kolck,
\newblock Nucl. Phys. A645 (1999) 273.

\bibitem{nogga}
A. Nogga, R.G.E. Timmermans and U. van Kolck,
\newblock Phys. Rev. C 72 (2005) 054006.

\bibitem{Bedaque:1998kg}
P.F. Bedaque, H.W. Hammer and U. van Kolck,
\newblock Phys. Rev. Lett. 82 (1999) 463.

\bibitem{Bedaque:1998km}
P.F. Bedaque, H.W. Hammer and U. van Kolck,
\newblock Nucl. Phys. A646 (1999) 444.

\bibitem{Platter:fk}
L. Platter and D.R. Phillips,
\newblock Few Body Syst. 40 (2006) 35.

\bibitem{rupak}
J.W. Chen, G. Rupak and M.J. Savage,
\newblock Nucl. Phys. A653 (1999) 386.

\bibitem{triton_eft}
P.F. Bedaque, H.W. Hammer and U. van Kolck,
\newblock Nucl. Phys. A 676 (2000) 357.

\bibitem{triton_eft2}
P.F. Bedaque et~al.,
\newblock Nucl. Phys. A 714 (2003) 589.

\bibitem{triton_eft3}
L. Platter,
\newblock Phys. Rev. C 74 (2006) 037001.

\bibitem{PhysRevC.69.034010}
I.R. Afnan and D.R. Phillips,
\newblock Phys. Rev. C 69 (2004) 034010.

\bibitem{Platter:2004zs}
L. Platter, H.W. Hammer and U.G. Meissner,
\newblock Phys. Lett. B607 (2005) 254.

\bibitem{hammerrev}
E. Braaten and H.W. Hammer,
\newblock Phys. Rept. 428 (2006) 259.

\bibitem{lattice}
H.M. M\"uller et~al.,
\newblock Phys. Rev. C 61 (2000) 044320.

\bibitem{Coon:2012ab}
S.A. Coon et~al.,
\newblock (2012), arXiv:1205.3230.

\bibitem{haxtonSMasEF}
W.C. Haxton and C.L. Song,
\newblock Phys. Rev. Lett. 84 (2000) 5484.

\bibitem{HT}
T. Busch et~al.,
\newblock Found. Phys. 28 (1998) 549.

\bibitem{luscher}
M. L\"uscher,
\newblock Nucl. Phys. B 354 (1991) 531.

\bibitem{atomexpt}
M. K\"ohl et~al.,
\newblock Phys. Rev. Lett. 94 (2005) 080403.

\bibitem{stoferle:030401}
T. St\"oferle et~al.,
\newblock Phys. Rev. Lett. 96 (2006) 030401.

\bibitem{beth}
H.A. Bethe,
\newblock Phys. Rev. 76 (1949) 38.

\bibitem{DFT_sr}
A. Bhattacharyya and T. Papenbrock,
\newblock Phys. Rev. A 74 (2006) 041602(R).

\bibitem{mehen}
T. Mehen,
\newblock Physical Review A 78 (2008) 013614.

\bibitem{alhassid2008}
Y. Alhassid, G.F. Bertsch and L. Fang,
\newblock Phys. Rev. Lett. 100 (2008) 230401.

\bibitem{NCSMcuts}
P. Navr\'atil and B.R. Barrett,
\newblock Phys. Lett. B 369 (1996) 193.

\bibitem{unitgas_prl}
F. Werner and Y. Castin,
\newblock Phys. Rev. Lett. 97 (2006) 150401.

\bibitem{unitgas_pra}
F. Werner and Y. Castin,
\newblock Phys. Rev. A 74 (2006) 053604.

\bibitem{dishonest}
J.P. Kestner and L.M. Duan,
\newblock Phys. Rev. A 76 (2007) 033611.

\bibitem{PhysRevA.77.043619}
J. von Stecher, C.H. Greene and D. Blume,
\newblock Phys. Rev. A 77 (2008) 043619.

\bibitem{chang2007}
S.Y. Chang and G.F. Bertsch,
\newblock Phys. Rev. A 76 (2007) 021603.

\bibitem{vonStecher:090402}
J. von Stecher and C.H. Greene,
\newblock Phys. Rev. Lett. 99 (2007) 090402.

\bibitem{Blume:233201}
D. Blume, J. von Stecher and C.H. Greene,
\newblock Phys. Rev. Lett. 99 (2007) 233201.

\bibitem{hanstrap}
S. T\"olle, H.W. Hammer and B.C. Metsch,
\newblock Comptes Rendus Physique 12 (2011) 59.

\bibitem{trap_higher_pws1}
S.K. Yip,
\newblock Phys. Rev. A 78 (2008) 013612.

\bibitem{trap_higher_pws2}
A. Suzuki, Y. Liang and R.K. Bhaduri,
\newblock Phys. Rev. A 80 (2009) 033601.

\bibitem{luu_2N_trap}
T. Luu et~al.,
\newblock Phys. Rev. C 82 (2010) 034003.

\bibitem{JISP16_0}
A. Shirokov et~al.,
\newblock Physics Letters B 644 (2007) 33 .

\bibitem{JISP16_1}
P. Maris, J.P. Vary and A.M. Shirokov,
\newblock Phys. Rev. C 79 (2009) 014308.

\bibitem{d_n_exp}
W. Dilg, L. Koester and W. Nistler,
\newblock Physics Letters B 36 (1971) 208.

\bibitem{Michel:2002}
N. Michel et~al.,
\newblock Phys. Rev. Lett. 89 (2002) 042502.

\bibitem{Michel:2003}
N. Michel et~al.,
\newblock Phys. Rev. C 67 (2003) 054311.

\bibitem{Hagen:2005}
G. Hagen, M. Hjorth-Jensen and J.S. Vaagen,
\newblock Phys. Rev. C 71 (2005) 044314.

\bibitem{Rotureau:2006Lett}
J. Rotureau et~al.,
\newblock Phys. Rev. Lett. 97 (2006) 110603.

\bibitem{Papadimitriou:2011R}
G. Papadimitriou et~al.,
\newblock Phys. Rev. C  (2011) 051304(R).

\bibitem{Michel:2009}
N. Michel et~al.,
\newblock J. Phys. G 36 (2009) 013101.

\bibitem{Michel:2010}
N. Michel et~al.,
\newblock J. Phys. G 37 (2010) 064042.

\bibitem{GSM6he_v_low_k}
G. Hagen, M. Hjorth-Jensen and N. Michel,
\newblock Phys. Rev. C 73 (2006) 064307.

\bibitem{GSM6he_v_low_k2}
K. Tsukiyama, M. Hjorth-Jensen and G. Hagen,
\newblock Phys. Rev. C 80 (2009) 051301.

\bibitem{berg}
T. Berggren,
\newblock Nucl. Phys. A 109 (1968) 265.

\bibitem{Rotureau:2012fk}
J. Rotureau and U. van Kolck,
\newblock Few Body Syst.  (2012), 1201.3351v1,
\newblock in press.

\bibitem{Hammer:2011ye}
H.W. Hammer and D.R. Phillips,
\newblock Nucl. Phys. A 865 (2011) 17.

\bibitem{Bertulani:2002}
C.A. Bertulani, H.W. Hammer and U. van Kolck,
\newblock Nucl. Phys. A 712 (2002) 37.

\bibitem{Bedaque:2003}
P.F. Bedaque, H.W. Hammer and U. van Kolck,
\newblock Phys. Lett. B 569 (2003) 159.

\bibitem{param}
R.A. Arndt, D.L. Long and L.D. Roper,
\newblock Nucl. Phys. A 209 (1973) 429.

\bibitem{thom}
L.H. Thomas,
\newblock Phys. Rev. 47 (1935) 903.

\bibitem{Navratil_PPNP}
P. Navr\'atil, B.R. Barrett and J.P. Vary,
\newblock to be published, 2012.

\bibitem{LanczCoef}
M.A. Marchisio et~al.,
\newblock Few Body Syst. 33 (2003) 259.

\bibitem{Nollett:2007}
K.M. Nollett et~al.,
\newblock Phys. Rev. Lett. 99 (2007) 022502.

\end{thebibliography}

\end{document}